\documentclass[a4paper,11pt]{article}
\usepackage[a4paper,margin=2.0cm]{geometry}
\usepackage[utf8]{inputenc}
\usepackage[T1]{fontenc}
\usepackage{graphicx}
\usepackage{epstopdf}
\usepackage{amsmath,amssymb}

\usepackage{bm}
\usepackage{dcolumn}
\usepackage{psfrag}
\usepackage{xstring}
\usepackage{xcolor}
\usepackage{array}
\usepackage{tabularx}
\usepackage{booktabs}
\usepackage{caption}
\usepackage{placeins}
\usepackage{orcidlink}
\usepackage{authblk}
\usepackage[numbers,sort&compress]{natbib}

\title{Models and Measures of Statistical Physics and Sociophysics for Fracture Mechanics and Earthquake Dynamics: An Introduction}

\author[1]{Sudip Sarkar\,\orcidlink{0009-0009-7832-1211}\thanks{\texttt{sudip\_sarkar@srmap.edu.in}}}
\author[1]{Soumyajyoti Biswas\,\orcidlink{0000-0002-0729-0587}\thanks{\texttt{soumyajyoti.b@srmap.edu.in}}}
\author[2,3]{Bikas K. Chakrabarti\,\orcidlink{0000-0001-9004-7221}\thanks{\texttt{bikask.chakrabarti@saha.ac.in}}}

\affil[1]{Department of Physics, SRM University - AP, Amaravati, Andhra Pradesh - 522240, India}
\affil[2]{Saha Institute of Nuclear Physics, Kolkata - 700064, India}
\affil[3]{Indian Statistical Institute, Kolkata - 700108, India}
\date{}
\begin{document}

\maketitle

\begin{abstract}
The breaking or fracture of materials and damages or devastations due to
earthquakes have caught our attention since the earliest stage of human
civilization. In fact, at around 1500, Leonardo da Vinci made the pioneering observation that, unlike elastic constants (formalized later by Hooke in 1678) of the materials, the tensile (breaking) strengths of nominally identical iron wires decrease drastically with increasing their length, indicating the vanishing breaking strength of materials in the large-size limit, suggesting the impossibility of constructing arbitrarily large structures. This was the first observation that, unlike the elasticity of disordered materials, which remains defined in the
thermodynamic limit (self-averaging statistics), the breaking properties of the disordered materials have extreme (non-self-averaging) statistics. In spite of major and some precise developments in statistical physical modelling and characterizing the fracture mechanics (starting with Alan Griffith's crack nucleation theory in 1921 and the weakest link failure Fibre Bundle Model of Frederick Thomas Peirce in 1926) and of earthquake dynamics (starting with Robert Burridge and Leon Knopoff's continuum dynamical model in 1967 and later other discrete or lattice train models or fractal overlap models of stick-slip earthquakes), no graduate-level textbook in condensed matter physics or in statistical physics introduces the basic models and their properties. And now computer scientists and social scientists (coming with their social inequality measures applied to the avalanche statistics in precursor failures) are joining their efforts to explore the new precursory inequality measures for big avalanches or failures. This introductory review is designed to fill this gap with a view to introducing these to this wider audience and to take them to the current frontiers in these fields.
\end{abstract}


\bigskip

\section{Introduction}
\label{sec:introduction}

The problem of fracture or breakdown properties of solids is perhaps one of the oldest commonly encountered problems in the physical and engineering sciences, and it remains intriguing, vibrant, and very much an active topic of research. Robert Hooke (1635-1703) defined the linear elastic response of solids \cite{hooke1678}. After the discovery of crystal structures and interatomic potentials, the estimates of the elastic constants of different (pure) solids or crystals could be extracted and compared with the estimates obtained from the velocity measurements of different acoustic (or lattice dynamical) modes in the solids \cite{born1954}. For disordered solids, one could again define such linear response elastic constants in the thermodynamic limit of the crystal structures. Their critical behavior near the critical disorder or percolation point of the lattice was established, for example, by Pierre-Gilles de Gennes \cite{gennes1976} and others (for a recent account, see e.g., \cite{StaufferAharony2018}). For the nonlinear and irreversible behavior of solids, like fracture, the story (for similar estimates for the fracture strength of disordered solids), however, could not be closed yet. Extending the Griffith's crack nucleation theory \cite{griffith1921} for the weakest crack in the percolation model of the disordered solids, \cite{RayChakrabarti1985} obtained the Weibull-type and \cite{duxbury1986} obtained the Gumbel-type  breaking statistics \cite{ChakrabartiBenguigui1997} for a review on these extreme breaking statistics of disordered solids. It may be mentioned that these studies could identify the breaking or fracture point of disordered solids as critical points where avalanche size fluctuations tend to diverge. A brief introduction to these developments will be given here in sections \ref{sec:crack_nucleation}, \ref{sec:extension_of_griffith} and \ref{sec:Fracture_exponents_for_disorder_concentration}. The breaking statistics of the Fibre Bundle Models (FBMs), originally introduced by Frederick Thomas Peirce \cite{Peirce1926}, and  major analytical \cite{HemmerHansen1992} as well as computational advances were made (see \cite{Pradhan2010, HansenHemmerPradhan2015} for reviews). Again, a very brief introduction to these models and results, including some Self-Organized Critical (SOC) FBM \cite{Biswas_LLSFBM_2013} will be  discussed here in section \ref{sec:fbm_fracture}.

As discussed earlier, the emergence of criticality—manifested as the divergence of avalanche-size fluctuations prior to complete failure—has been observed in both percolation-based fracture models and FBMs. In parallel, socio-statistical inequality measures, such as the Gini, Kolkata, and Hirsch indices, originally developed to quantify wealth, income, or scientific citation inequalities, have recently been shown to provide useful precursor indicators of impending failure \cite{Ghosh2022, ghosh_2022_h_index, gini_prl_2023, jordi}. These developments have stimulated an extensive body of literature on the use of inequality measures as failure precursors, which is introduced in sections \ref{sec:inequality_measures} and \ref{inequality_in_fracture_models}.

Since the availability of compelling empirical evidence for Gutenberg-Richter (GR) power-law statistics relating the earthquake frequencies with their magnitude \cite{gr1944} and the Omori-Utsu \cite{Utsu_1961, Utsu_1995} power-law for the time decay of aftershock frequency, they  got established, important statistical physics models, like the celebrated \cite{bkmodel} continuous stick-slip model and
other discrete lattice train \cite{biswas2013} and fractal overlap \cite{ChakrabartiStinchcombe1999, bhattacharyya2005, Bhattacharya_2011}, etc., models were extensively studied (see \cite{Carlson1994, kawamura2012} for detailed reviews). A brief introduction to these models and their statistics will be discussed here in section \ref{sec:eq_stat_models}. In all these models, the GR-type power laws and the associated SOC behavior \cite{BTW}, helped the successful applications  \cite{PRE_2026} of the socio-statistical avalanche inequality measures. These are again discussed at length in sections \ref{sec:forecasting_with_b}, \ref{sec:modern_forecasting} and \ref{inequality_measure_in_EQ}.

We finally conclude, after a brief discussion on the main results discussed in the paper, in section \ref{sec:discussion}.


\section{da Vinci's experiment: Extreme Statistics for Fracture Strength}
\label{sec:da_vinci}
    
The problem of fracture or breakdown properties of solids might be the oldest commonly encountered one in physical and engineering sciences,
which remain intriguing and very much alive and active still as research topics. Robert Hooke (1635-1703) defined the linear elastic
response of solids in 1678 (see \cite{hooke1678}). After the discovery of crystal structures and of the interatomic potentials, the estimates of the elastic constants of different (pure) solids or crystals could be extracted and compared with estimates from the velocity measurements of different acoustic (or lattice dynamical) modes in the solids (see e.g., \cite{born1954}) for a recent account; see e.g., \cite{kittel_2004} for defining such linear response constants of the solids even in the thermodynamic limit of the crystal. For disordered solids, one could again define such linear response elastic constants in the thermodynamic limit of the crystal structures. Their critical behavior near the critical disorder or percolation point of the lattice was established, for example, by Pierre-Gilles de Gennes (see \cite{gennes1976}) and others (for a recent account, see e.g., \cite{Stauffer2003}). For the nonlinear and irreversible behavior of solids, like fracture, the story (for similar estimates for the fracture strength of disordered solids), however, could not be closed yet.

\begin{figure}[h!]
\begin{center}
        \includegraphics[width=0.65\textwidth]{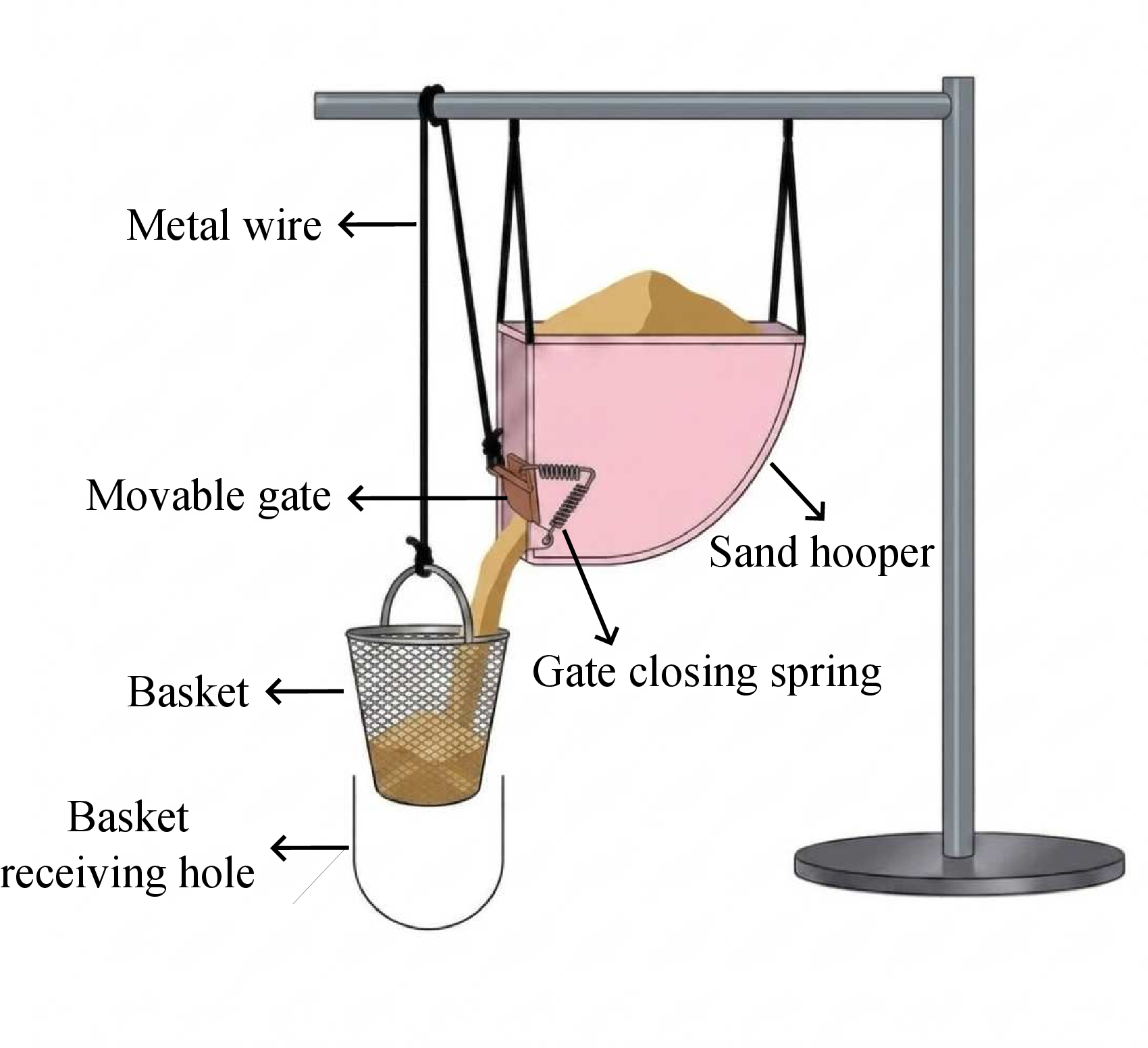}
\end{center} 
\caption{In this classic experiment, originally conceived by Leonardo da Vinci, an iron wire of fixed cross-sectional area and length was used to suspend a basket. The basket was gradually filled with fine sand released from a hopper positioned above it. As the load increased, the wire eventually fractured. At the moment of failure, a spring mechanism automatically closed the hopper, preventing any additional sand from entering the basket. Simultaneously, the basket dropped a short distance into a receiving hole, ensuring that its contents remained intact. The sand collected in the basket was then weighed to determine the tensile strength of the wire. The experiment was repeated for wires of different lengths while keeping the wire material, cross-sectional area, and all other experimental conditions unchanged. [Adopted from \cite{lund_byrne_2001}]}
    \label{fig:vinci}
    \end{figure}

Over 500 years ago Leonardo da Vinci (1452–1519) carried out experiments in his notebooks on the tensile strength of iron wires of various lengths and concluded that the shorter the wire the greater the load it could sustain. This is, of course, a classic example of the statistical size effect in fracture, which can be understood in terms of the weakest-link concept \cite{Irwin_1961, lund_byrne_2001}. Weibull went on to formalize this concept in a statistical framework and showed that under certain assumptions regarding the distribution of flaws and the loading conditions, the characteristic strength will decrease with increasing specimen size \cite{weibull_1939}. In the classical Weibull weakest-link model, the characteristic strength goes to zero in the infinite-size limit. Of course this is dependent on the model assumptions, and the result is not generally true for all materials and all failure mechanisms. Other mechanisms such as stress redistribution, stable crack growth, and fracture process zone effects can lead to different size effect scaling laws \cite{Bazant_1993}. Furthermore, a reduction in the mean fracture strength with increasing size does not in itself imply statistical non-self-averaging, which is a description of the behavior of the sample-to-sample fluctuations as the system size increases \cite{Wiseman_1995}. While size-dependent mechanical strength may be one of the factors that limits the size of biological structures, the size limits of trees and animals cannot be explained by fracture strength alone. The height and mass of trees and animals are also constrained by hydraulic transport, growth form, and physiological scaling, and mechanical stability may well be a further constraint depending on the structure and the loading conditions (see \cite{McMahon, Koch2004, Niklas_2004}). In Fig. \ref{fig:vinci} (adopted
from \cite{lund_byrne_2001}) we show da Vinci's  experimental setup. An English translation of Leonardo da Vinci's text says (see \cite{lund_byrne_2001}): `` The object of this test is to find the load an iron wire can carry. Attach an iron wire $2$ braccia long to something that will firmly support it, then attach a basket or similar container to the wire and feed into the basket some fine
sand through a small hole placed at the end of the hopper. A spring is fixed so that it will close the hole as soon as the wire breaks. The
basket is not upset while falling since it falls through a very short distance. The weight of sand and the location of the fracture of the wire
are to be recorded. The test is repeated several times to check the
results. Then a wire of $\frac{1}{2}$ the previous length is tested, and the
additional weight it carries is recorded; then a wire of $\frac{1}{4}$ length is
tested, and so forth, noting the ultimate strength and the location of the fracture. "This remarkable observation clearly indicated that unlike the case of linear responses of solids like elastic
modulus,  the  fracture strength of a disordered solid is dictated by the extreme fluctuation responsible for creating the weakest point (or defect) in the disordered solid, and it decreases with the sample
volume. After discussing Griffith's theory of nucleation of a micro-crack in an otherwise homogeneous elastic solid in the next section, we will discuss the origin of the typical forms of extreme statistics, which explains the  observation of da Vinci for a disordered solid, in the next-to-next section.
  
\section{Griffith's crack nucleation theory}
\label{sec:crack_nucleation}

Since da Vinci's above-mentioned observation, there was no progress with the science of fracture strength of solids for the next 300 years or so.
Alan Griffith (see \cite{griffith1921}) estimated how the crack nucleation stress for
an otherwise pure material decreases with the dimension of the single
defect. This was done assuming that the solid is still in the brittle
limit, when the stress–strain relationship remains linear until breaking.

\begin{figure}[h!]
    \begin{center}
         \includegraphics[width=0.25\textwidth]{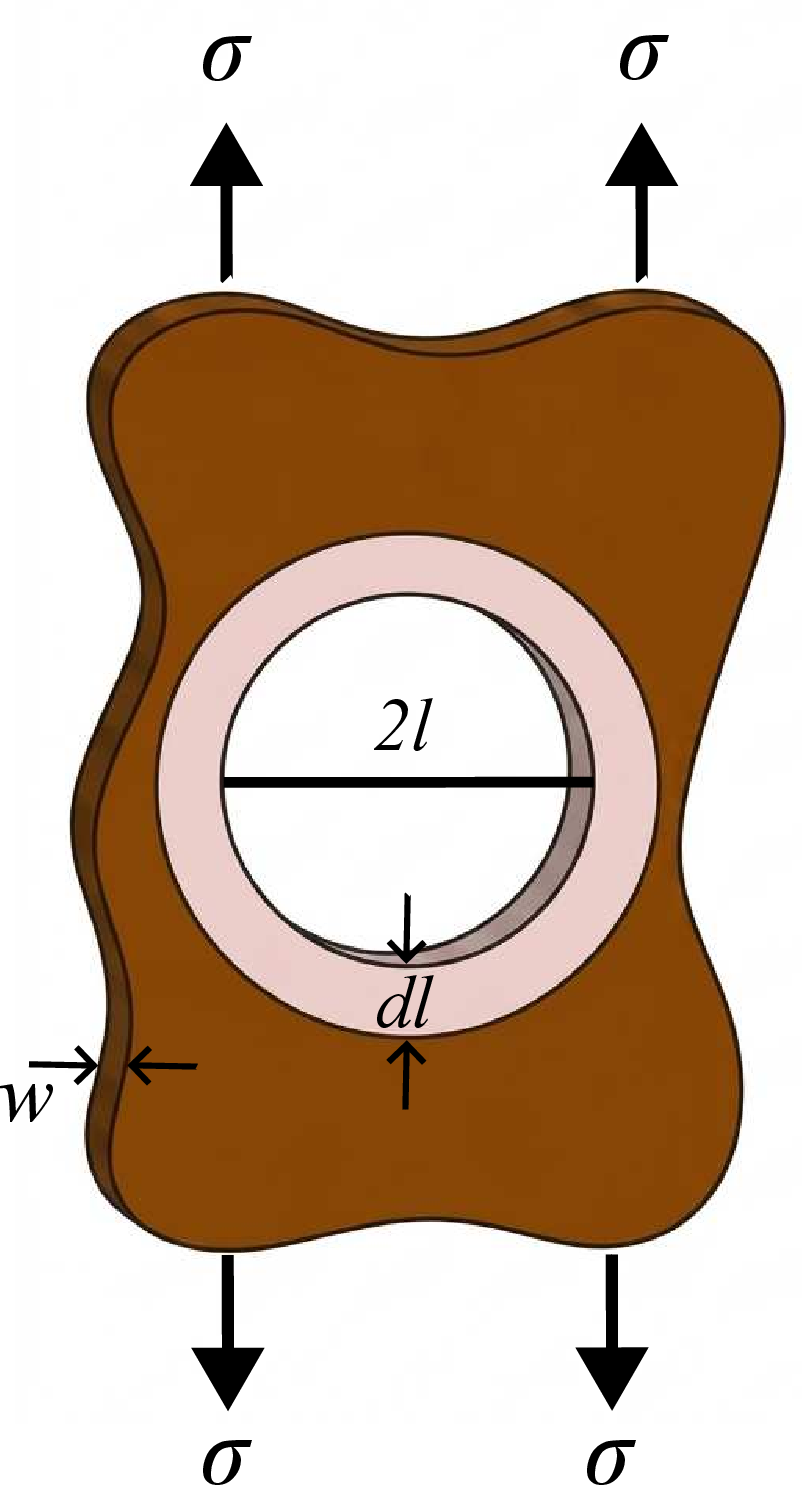}
    \end{center}
    \caption{A portion of an elastic plate of thickness $w$ is subjected to a tensile stress $\sigma$ (Mode I loading), containing a linear micro-crack of length $2l$ (indicated by a solid black line). The crack is assumed to extend symmetrically by an additional length $2dl$ (on both sides). The elastic energy released from the surrounding annular region of volume $2\pi ldlw$, with energy density $\frac{\sigma^2}{2Y}$, must be sufficient to supply the surface energy $4\Gamma wdl$ required to create the two new crack surfaces, where $Y$ is Young's modulus and $\Gamma$ is the surface energy per unit area.}
    \label{fig:g1}
    \end{figure}
    
For a mode I fracture  in an elastic slab of thickness $w$, containing a  micro-crack of length $2l$ perpendicular to the tensile stress $\sigma$, further extension of  the crack length by $2dl$ requires the elastic energy (having density $\frac{\sigma^2}{2Y}$ and the modulus of elasticity, or Young's modulus $Y$) of the annular region $2\pi ldlw \times \frac{\sigma^2}{2Y}$
to be sufficient to provide the surface energy $4\Gamma w dl$ for the new surfaces created (having surface energy density $\Gamma$) (see Fig. \ref{fig:g1}). Excess elastic energy, over the required surface energy,  will  contribute to the sound wave amplitude (energy) required for the crack-tip motion \cite{Mott1948}. The fracture nucleation stress ($\sigma_c$), given by the equality of the crack-released elastic energy and the required surface energy to open up the extended crack surfaces, is

\begin{equation}
\sigma_c \sim \sqrt{\frac{Y\Gamma}{l}}
\label{eq1}
\end{equation}

\noindent where $\sigma_c$ decreases inversely with the square root of the existing crack length $l$ for a brittle solid having a linear stress-strain relation up to the breaking point.

\section{Extension of Griffith Theory for Disordered Solids \& Extreme Statistics}
\label{sec:extension_of_griffith}

We considered in the previous section the fracture property of a stressed
solid, having a single defect (with the major lateral size in the
direction perpendicular to the stress in the sample). In the linear
responses like the (mechanical) elastic or (electrical) conductivity of
such solids (see e.g., \cite{gennes1976, Stauffer2003}), all the ``parallel" elastic or
conducting parts of the sample contribute with their ``respective share" to
the net elasticity or conductivity, leading to their self-averaging
behavior and statistics (well-defined in the thermodynamic limit).
However, the fracture or breakdown behavior  of such disordered solids are
determined  only by the  weakest (often the largest) defect or crack in
the entire solid. Except for some indirect effects, weak or small defects
or cracks in the solid do not determine the breakdown strength of the
sample. The fracture or breakdown statistics of a solid sample are
therefore determined essentially by the extreme statistics of the weakest
defect in the solid formed by fluctuations of the nominal micro-defects in
the sample.

\begin{figure}[h!]
    \begin{center}
         \includegraphics[width=0.25\textwidth]{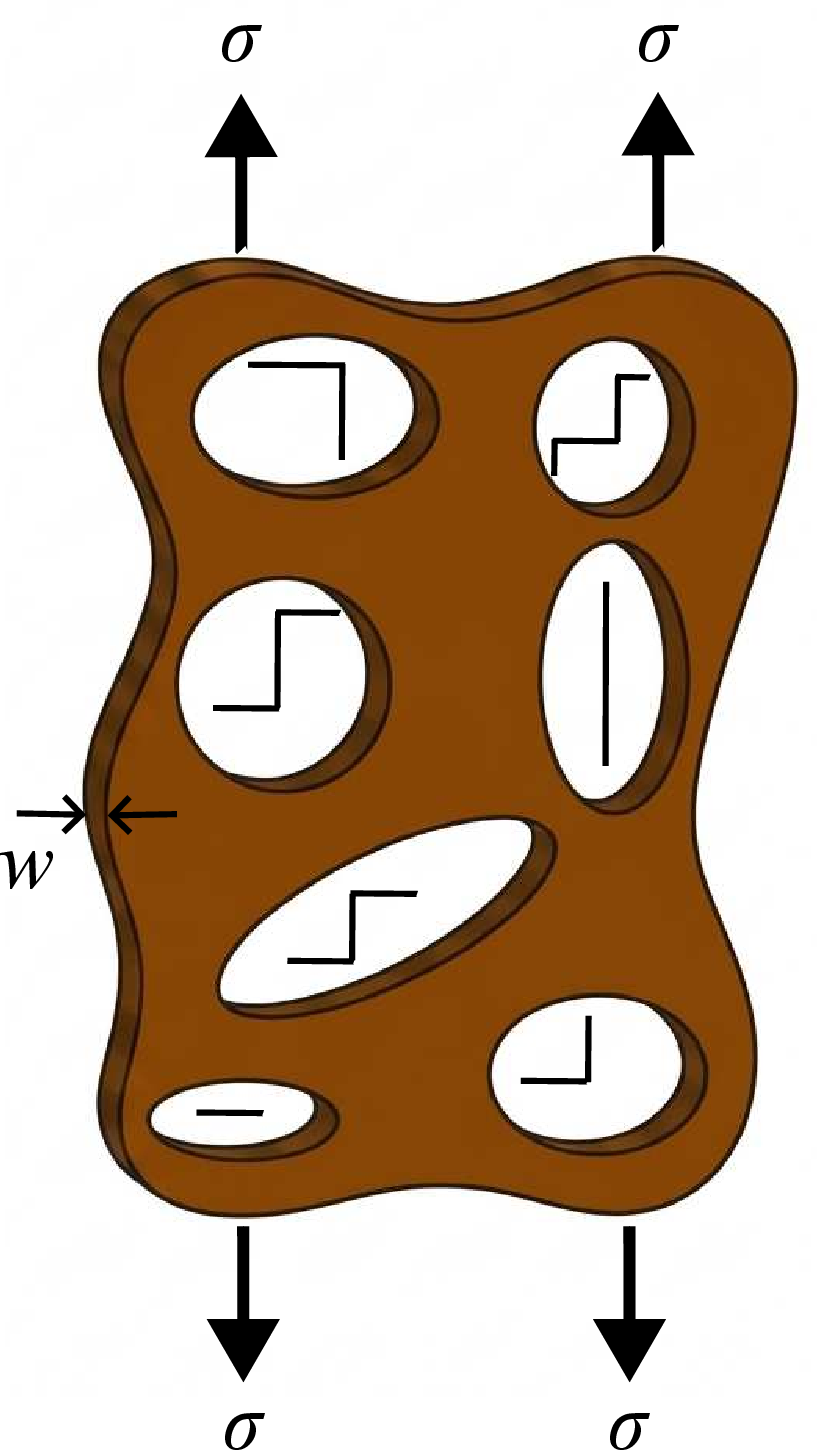}
    \end{center}
    \caption{A solid sample of thickness $w$ subjected to a tensile stress $\sigma$ (Model I loading), containing numerous micro-cracks of varying sizes, shapes, and orientations that arise from material disorder. The solid black lines represent the micro-cracks, while the surrounding white regions indicate the zones of elastic energy release associated with the formation and growth of these micro-cracks.}
    \label{fig:g2}
    \end{figure}

We now consider a solid sample containing many micro defects of various
orientations, where, due to fluctuations in disorder, all kinds of defect
sizes and shapes are produced. For simplicity, let us take a simple square
lattice model of the sample of size $\mathcal{L}$, with bond dilution (representing
the absence of elastic springs) with probability $p$ (see Fig. \ref{fig:g2})

If we assume the fluctuations have produced $\tilde{n}$ defects or micro-cracks
of different sizes and shapes, and each of these defects has their stress
released regions well separated, then consider the defects
independently. We then denote their failure probability under stress
$\sigma$ by $f_i(\sigma)$, $i = 1, 2, ..., \tilde{n}$. If we denote the failure
probability of the entire sample under stress $\sigma$ by $F(\sigma)$,
then for small values of $f_i(\sigma)$, we get

\begin{equation}
F(\sigma)
= \prod_{i=1}^{\tilde{n}} \left[1-f_i(\sigma)\right]
\simeq \exp\!\left[-\sum_{i=1}^{\tilde{n}} f_i(\sigma)\right]
= \exp\!\left[-\mathcal{L}^{d}\tilde{g}(\sigma)\right]
\label{eq2}
\end{equation}

\noindent where $\tilde{g} (\sigma)$ denotes the density of cracks in the sample of volume $\mathcal{L}^d$, which starts propagating at and above the stress level $\sigma$ on the sample.

In the percolation model of a solid, the defect cluster sizes $l$,  are
typically of the order of the correlation length, $\xi \sim |p -
p_c|^{-\nu_p}$, where $p_c$ is the percolation concentration and $\nu_p$ is the percolation correlation length exponent (see e.g.,\cite{Stauffer2003}). The pair-connectedness function (or pair correlation function) $\mathcal{G}(l)$, which gives the probability that two occupied sites separated by a distance $l$ belong to the same percolation cluster, can take the form

\begin{subequations}
\renewcommand{\theequation}{\theparentequation(\alph{equation})}
\label{eq3}

\begin{align}
\mathcal{G}(l) &\sim \exp\left(-\frac{l}{\xi}\right), \label{eq3a}\\
&\sim l^{-w}. \label{eq3b}
\end{align}

\end{subequations}

\noindent For $p \neq p_c$, $\mathcal{G}(l)$ decays exponentially with
distance, as given by Eq. (\ref{eq3a}), whereas at the critical point ($p=p_c$), it exhibits a power-law decay, as given by Eq. (\ref{eq3b}). To get $\tilde{g}(\sigma)$ in Eq. (\ref{eq2}) from $\mathcal{G}(l)$ the given expressions (\ref{eq3a}) and (\ref{eq3b}), we need to use Griffith's brittle solid fracture stress relation (\ref{eq1})
to transform the crack length $l \sim \sigma^{-2}$ to its corresponding nucleation stress $\sigma$. This finally gives the Gumbel and Weibull distributions, respectively, for the sample failure probability $F(\sigma)$  under stress
$\sigma$ (see \cite{RayChakrabarti1985, duxbury1986, BiswasRayChakrabarti2015}):

\begin{subequations}
\renewcommand{\theequation}{\theparentequation(\alph{equation})}
\label{eq4}

\begin{align}
F(\sigma)
&\sim 1-\exp\!\left[-\mathcal{L}^{d}\exp\!\left(-C\sigma^{-2}\right)\right],
\label{eq4a}\\
&\sim 1-\exp\!\left[-\mathcal{L}^{d}C'\sigma^{m}\right],
\label{eq4b}
\end{align}

\end{subequations}

\noindent where $C$ and $C^{\prime}$ are some constants determined by the disorder
in the sample and $m = 2w$ is called the Weibull modulus. In both cases,
the sample failure probability approaches unity as the stress 
$\sigma$ on
the sample becomes large or the volume of the sample $\mathcal{L}^d$ becomes large
(or both become large). Assuming it $f(\sigma)$ to be finite at $\sigma =\sigma_c$, one gets $\sigma_c \sim \frac{1}{\sqrt{\log \mathcal{L}}}$ for the Gumbel distribution (\ref{eq4a}), and $\sigma_c \sim \frac{1}{\mathcal{L}^{1/m}}$ for the Weibull distribution (\ref{eq4b}) of the sample failure probability $F(\sigma)$. This explains the da Vinci observation discussed in section \ref{sec:da_vinci}.

For details of these studies on fracture properties of disordered solids, see e.g., \cite{ChakrabartiBenguigui1997, Sahimi2003, BiswasRayChakrabarti2015, Herrmann_2014, Bertalan_2014,ray_2018} for reviews.

\section{Fracture exponents for disorder concentration near
percolation threshold}
\label{sec:Fracture_exponents_for_disorder_concentration}

If the sample size is large but finite and the disorder concentration
$p$ approaches the finite-size percolation threshold $p_c(\mathcal{L})$,
the correlation length $\xi$ becomes comparable to the system size
$\mathcal{L}$, i.e.,\[\xi \sim \mathcal{L} \sim |p_c(\mathcal{L})-p|^{-\nu},\] where $\nu$ is the correlation-length critical exponent (see \cite{Stauffer2003}). Like the critical exponents describing the linear response of percolating solids, such as the elasticity exponent $T_e$ and the
conductivity exponent, which were derived using the node-link-blob
model (see \cite{gennes1976,ray1988,Stauffer2003}), the fracture
exponent $T_f$ can also be obtained within the same framework. While
the elasticity exponent $T_e$ governs the critical scaling of the
elastic modulus,\[Y \sim |p_c(\mathcal{L})-p|^{T_e} \sim \xi^{-T_e/\nu},\]  the fracture exponent $T_f$ (for a fixed finite but large sample size; see discussions in the earlier  section \ref{sec:extension_of_griffith}) characterizes the
critical scaling of the fracture stress as the percolation threshold is approached. Combining the Griffith criterion (as in Eq. (\ref{eq1})) with the scaling relations
$Y\sim\xi^{-T_e/\nu}$, $l\sim\xi$, and $\Gamma\sim\xi^{d_B}$, where $l$ is the characteristic micro-crack
length, $\Gamma$ is the crack surface energy, and $d_B$ is the backbone fractal dimension (see, e.g., \cite{Stauffer2003,ray1988}), yields (for a fixed finite but large sample size)

\begin{equation}
\sigma_c \sim |p_c(\mathcal{L})-p|^{T_f},
\qquad
T_f=\frac{T_e+(d-d_B)\nu}{2}.
\label{eq5}
\end{equation}

For further extensions and modifications of relation (\ref{eq5}) see \cite{ChakrabartiBenguigui1997, Sahimi2003, BiswasRayChakrabarti2015}.

\section{ Fracture in fibre Bundle Models}
\label{sec:fbm_fracture}

The Fibre Bundle Model (FBM) was introduced by Frederick Thomas
Peirce \cite{Peirce1926} to estimate the strength of composite
materials. Despite its simplicity, the model captures many essential
features of fracture in heterogeneous solids. It consists of a
macroscopically large number of parallel Hookean fibres (or springs),
each having the same elastic constant and initial length but different
breaking thresholds. The fibres are attached between two horizontal
plates: a rigid upper plate and a lower plate supporting the external
load. As the applied load increases, fibres break permanently once their individual strength limits are reached. When a fibre fails, the extra load is distributed in the remaining intact fibres. Failures in the model are influenced by how this load is distributed. In the Local Load-Sharing (LLS) scheme, the extra load is shared among the nearest fibres. This causes high stress in the surrounding areas, leading to closely related failures that are difficult to model mathematically. As a consequence, the failure dynamics of LLS fibre bundles have been studied mainly through numerical simulations (see, however, \cite{Pradhan2010,HansenHemmerPradhan2015}).

At the other extreme is the Equal Load-Sharing (ELS) model, in which
both loading plates are assumed to be perfectly rigid. The load released
by a failed fibre is redistributed equally among all surviving fibres,
independent of their spatial locations. The ELS assumption makes the
model analytically tractable while retaining many of the universal
features of failure in disordered media. The strength of the ELS bundle
was first analyzed by Henry Ellis Daniels \cite{Daniels1945}. Although the model remained largely unnoticed by the
physics community for several decades, it attracted considerable
attention following the work of Sornette \cite{Sornette1989}. Since then, the FBM has become one of the standard statistical
models for studying fracture and breakdown phenomena in disordered
materials (see \cite{Kun_Hidalgo_Herrmann_2006, Stat_models_of_fracture_zapperi_alava_nakula, Pradhan2010,HansenHemmerPradhan2015} for details).

\subsection{Equal Load-Sharing Fibre Bundle Model (ELS-FBM)}
\label{subsec:ELSFBM}

\begin{figure}[h!]
    \begin{center}
         \includegraphics[width=0.60\textwidth]{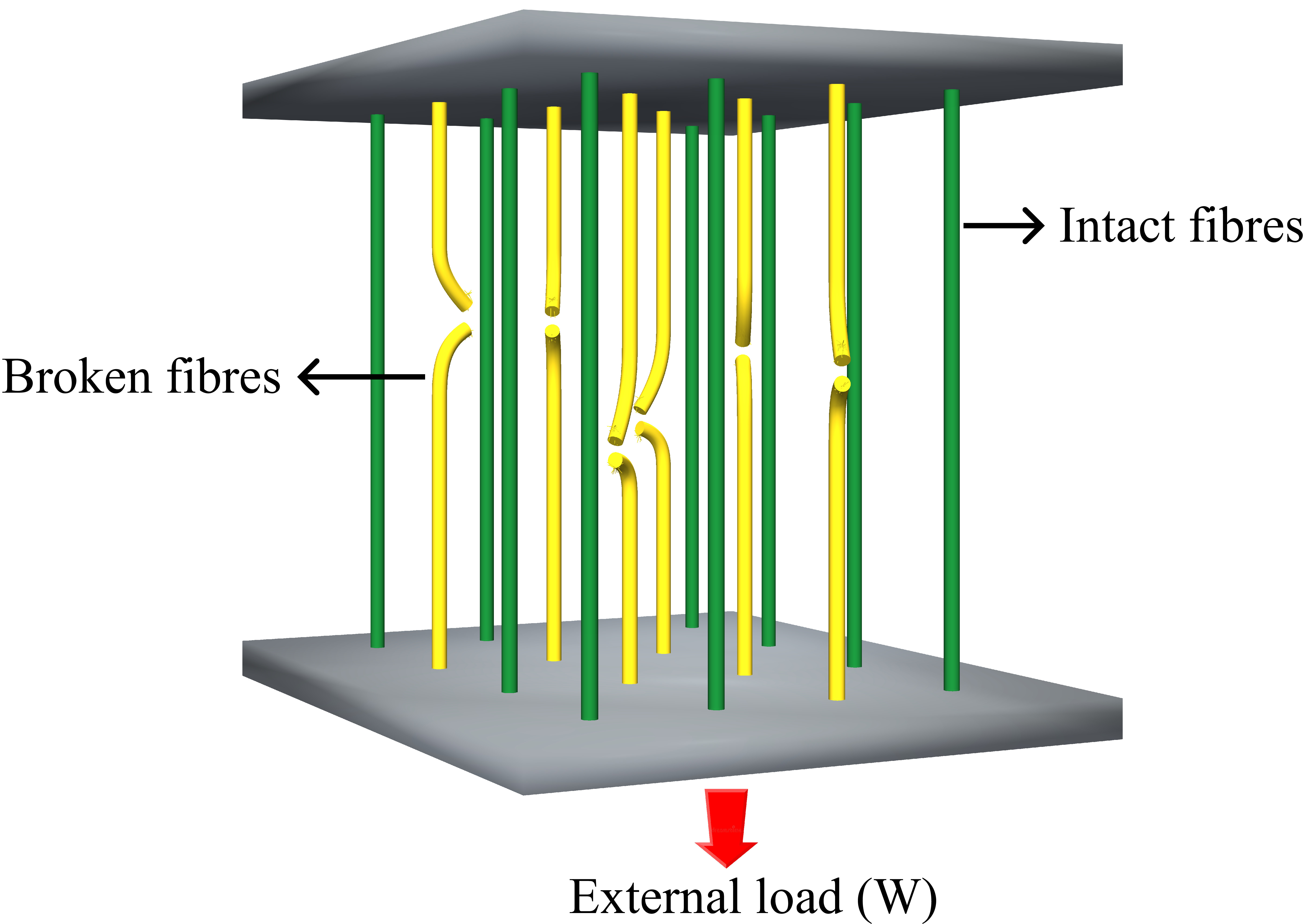}
    \end{center}
    \caption{The fibre bundle initially consists of $N$ parallel fibres clamped between two perfectly rigid plates: a fixed plate at the top and a movable plate at the bottom, from which a total load $W$ is suspended. The fibres are assumed to be linearly elastic and perfectly brittle, failing abruptly without exhibiting any plastic deformation or necking prior to fracture. In the ELS model, the applied load $W$ is redistributed uniformly among all the surviving (intact) fibres after each fibre failure so that every intact fibre carries the same load throughout the deformation process.}
    \label{fig:fbm}
    \end{figure}

We now briefly describe the mean-field-like failure behavior of the ELS-FBM and derive the corresponding critical exponents (see, e.g.,
\cite{Pradhan2003,Pradhan2010,HansenHemmerPradhan2015}). As discussed
earlier (see Fig. \ref{fig:fbm}), the bundle consists of $N$ parallel
Hookean fibres (or springs), each having the same elastic constant.
The bundle is subjected to a total external load,
\(W=N\sigma,\)
where $\sigma$ denotes the applied load per fibre. The individual fibres, however, possess different failure thresholds $\sigma_{th}$, which are drawn from a prescribed probability distribution
$\rho(\sigma_{th})$.

For the ELS model considered here, both the upper and lower loading plates are assumed to be perfectly rigid. As a result, no local deformation or stress concentration develops around the failed fibres. When a total external load $W=N\sigma$ is initially applied to the bundle, it is distributed equally among all $N$ fibres. Those with failure thresholds below the applied stress $\sigma$ fail
immediately, causing the load carried by the failed fibres to be redistributed equally among the surviving fibres. Consequently, the load per surviving fibre increases after each failure event, leading
to a sequence of successive failures. This failure dynamic can, therefore, be described by a recursion relation.

\subsubsection{ELS-FBM with uniform distribution of fibre strength:}
\label{sec:uniform_ELSFBM}

\begin{figure}[h!]
\begin{center}
    \includegraphics[width=0.75\textwidth]{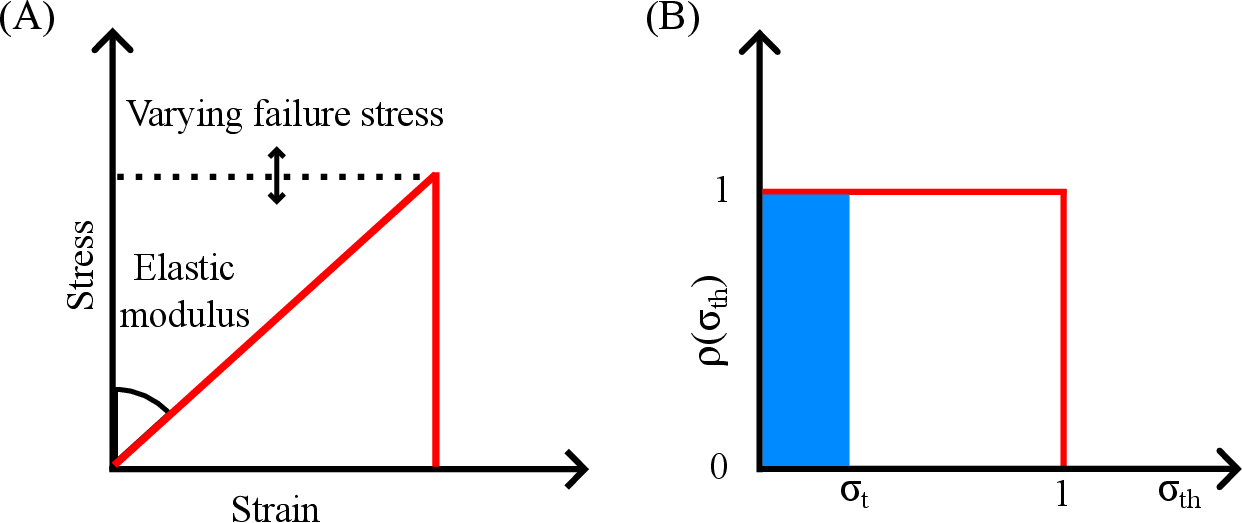}
\end{center}   
    \caption{\textbf{(A)} The stress-strain curves of all fibres have the same elastic modulus (slope), while the failure stress varies from fibre to fibre according to the strength (threshold) distribution $\rho(\sigma_{th})$. \textbf{(B)} The simple model considered here assumes uniform density $\rho(\sigma_{th})$ of the fibre strength distribution up to a cut-off strength (normalized to unity). At time $t$, when the load per surviving fibre is $\sigma_t$, a fraction $\sigma_t$ of the fibres fails, while the remaining fraction $1-\sigma_t$ survives.}
    \label{fig:fbm2}
    \end{figure}

For simplicity, first  we assume the distribution $\rho (\sigma_{th})$ of
the fibre strength thresholds $\sigma_{th}$ to be uniform in the range $(0,1)$ as shown in Fig. \ref{fig:fbm2}(B).

Let, $U_{t}(\sigma)$ denote the fraction of surviving fibres after the
$t$-th redistribution step, where $t=0$ corresponds to the instant when
the external load is applied to the bundle. If the total applied load
is $W=N\sigma$, then the stress carried by each surviving fibre at time
$t$ is $\sigma_{t}=\frac{W}{NU_{t}}=\frac{\sigma}{U_{t}}$. For the uniform threshold distribution,
\[
\rho(\sigma_{th})=
\begin{cases}
1, & 0\leq \sigma_{th}\leq 1\\
0, & \sigma_{th}>1
\end{cases}
\]
\noindent so the corresponding cumulative distribution function becomes $\rho_c(\sigma_t)=\int_{0}^{\sigma_t}\rho(\sigma_{th})\,d\sigma_{th}=\sigma_t$. The fraction of surviving fibres at the next redistribution step satisfies

\begin{equation}
U_{t+1}
=1-\rho_c(\sigma_t).
\label{ELSU_1}
\end{equation}

\noindent For a uniform threshold distribution Eq.~(\ref{ELSU_1}) becomes

\begin{equation}
U_{t+1}
=1-\sigma_t
=1-\frac{\sigma}{U_t}.
\label{ELSU_1a}
\end{equation}

\noindent The steady-state solution of the recursion relation is obtained by setting $U_{t+1}=U_t=U^*$. Substituting this condition into Eq.~(\ref{ELSU_1a}) gives $U^*
=\frac{1\pm\sqrt{1-4\sigma}}{2}$. For the fixed-point solution to be real, the discriminant must satisfy $1-4\sigma\ge0$, which gives $\sigma\le\frac{1}{4}$. Thus, the critical stress is $\sigma_c=\frac{1}{4}$. Using this we can get, $\sqrt{1-4\sigma}=2\sqrt{\sigma_c - \sigma}$. Substituting this into the quadratic solution, $U^*=\frac{1}{2}\pm\sqrt{\sigma_c-\sigma}$. Hence, the two fixed-point solutions are $U^*(\sigma)=\frac{1}{2}\pm\sqrt{\sigma_c-\sigma}$. Out of these

\begin{equation}
    U^*(\sigma)=\frac{1}{2}+\sqrt{\sigma_c-\sigma,}
    \label{ELSU_2}
\end{equation}

\noindent corresponds to the physically stable solution and represents the surviving fraction of fibres. The other solution $U^*(\sigma)=\frac{1}{2}-\sqrt{\sigma_c-\sigma}$ is unphysical (since the critical stress is $\sigma_c=\frac{1}{4}$, the stable fixed point approaches $U^*(\sigma)=0$ when $\sigma=0$). The fixed-point solution also enables us to define an order parameter $\mathcal{O}\equiv U^{*}(\sigma)-U^{*}(\sigma_c)$ where $\sigma\leq\sigma_c$, which measures the deviation of the surviving fraction of fibres from its critical value and characterises the transition from a partially failed state to complete breakdown. Using Eq.~(\ref{ELSU_2}) we obtain $\mathcal{O}=(\sigma_c-\sigma)^{1/2}$. Comparing the above result with the general scaling form of the order parameter

\begin{equation*}
\mathcal{O}
\sim
(\sigma_c-\sigma)^{\eta},
\qquad
\eta=\frac{1}{2}
\label{ELSU_3}.
\end{equation*}

\noindent To determine the relaxation dynamics, the recursion relation is approximated by a differential equation. Since $U_{t+1}-U_t\simeq\frac{dU}{dt}$, Eq.~(\ref{ELSU_1a}) can be written as

\begin{align}
-\frac{dU}{dt}
=
\frac{U^2-U+\sigma}{U}.
\label{ELSU_4}
\end{align}

\noindent To examine the stability of the fixed point, if we consider a small perturbation about the fixed-point solution, $U(t,\sigma)=U^*(\sigma)+\varepsilon(t)$, where $|\varepsilon(t)|\ll1$ and $U^*=\frac{1}{2}+\sqrt{\sigma_c-\sigma}$. Substituting these into Eq.~(\ref{ELSU_4}) and neglecting the second-order term $\varepsilon^2$, we obtain $-\frac{d\varepsilon}{dt}=\varepsilon\left(\frac{2U^*-1}{U^*}\right)=-2\varepsilon\sqrt{\sigma_c-\sigma}$ giving $\varepsilon(t)\sim e^{-t/\tau}$. Then the relaxation time ($\tau$) diverges as

\begin{equation*}
\tau
\sim (\sigma_c-\sigma)^{-\frac{1}{2}} \sim
(\sigma_c-\sigma)^{-\zeta},
\qquad
\zeta=\frac{1}{2}
\label{ELSU_5}.
\end{equation*}

\noindent At the critical point, we can write the solution as,

\begin{equation*}
U_t-\frac12
\sim
t^{-\lambda},
\qquad
\lambda=1
\label{ELSU_8}.
\end{equation*}

\noindent When the external load is increased quasi-statically, the avalanche size
($s$) defined as the number of fibres breaking between two successive load increments is given by $s \sim\frac{d(1-U^*)}{d\sigma}\sim \frac{1}{2\sqrt{\sigma_c-\sigma}}$. Hence, $\sigma_c-\sigma \propto s^{-2}$. If $P(s)$ denotes the avalanche-size distribution, then $P(s)\,ds\sim d\sigma$. It gives

\begin{equation*}
P(s)\sim
\frac{d\sigma}{ds}
\sim
s^{-\theta},
\qquad
\theta=3.
\label{ELSU_7}
\end{equation*}

\noindent The breakdown susceptibility ($\chi$) is defined as the response of the surviving fraction to an infinitesimal increase in the applied stress. So we can write using Eq. (\ref{ELSU_2})

\begin{align*}
\chi
&=
\left|
\frac{dU^*(\sigma)}{d\sigma}
\right|
\sim
(\sigma_c-\sigma)^{-\gamma},
\qquad
\gamma=\frac12.
\label{ELSU_6}
\end{align*}

\subsubsection{ELS-FBM with linearly increasing distribution of fibre strength:}
\label{sec:increasing_ELSFBM}

\begin{figure}[h!]
\begin{center}
    \includegraphics[width=0.75\textwidth]{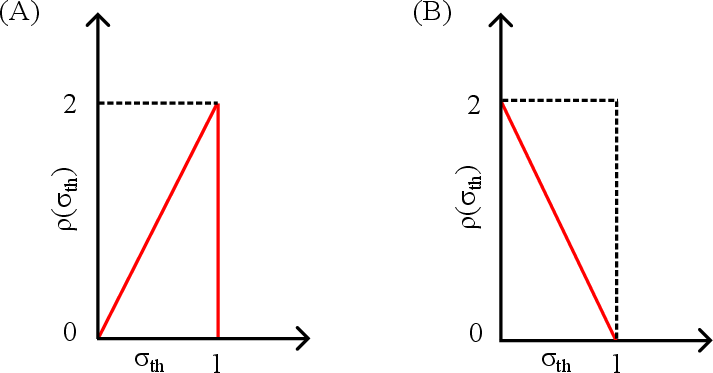}
    \end{center}
    \caption{The simplest model considered here assumes \textbf{(A)} the linearly increasing and \textbf{(B)} linearly decreasing density $\rho(\sigma_{th})$ of the fibre strength distribution up to a cutoff strength.}
    \label{fig:pf}
    \end{figure}   

For a linearly increasing distribution of fibre strength thresholds as shown in Fig.~\ref{fig:pf}(A), the
probability density is

\begin{equation*}
\rho(\sigma_{th})=
\begin{cases}
2\sigma_{th}, & 0\leq \sigma_{th}\leq1,\\
0, & \sigma_{th}>1,
\end{cases}
\end{equation*}

\noindent which satisfies the normalization condition $\int_0^1\rho(\sigma_{th})\,d\sigma_{th}=1$. The corresponding cumulative distribution function is $\rho_c(\sigma_t)=\int_0^{\sigma_t}\rho(\sigma_{th})\,d\sigma_{th}=\sigma_t^{\,2}$. From Eq.~(\ref{ELSU_1}) we can get
\begin{equation}
    U_{t+1}=1-\left(\frac{\sigma}{U_t}\right)^2
    \label{ELSI_1a}
\end{equation}.

\noindent At the fixed point, $U_{t+1}=U_t=U^*$ and hence
\begin{align}
(U^*)^3
-(U^*)^2
+\sigma^2
=
0.
\label{ELSI_1}
\end{align}

\noindent The critical point is obtained from the condition that the cubic has a
double root. Therefore, $\frac{d}{dU^*}\left[(U^*)^3-(U^*)^2+\sigma^2\right]=0$ which gives $U^*=0$ or $\frac{2}{3}$. Since the physically relevant stationary solution at the critical point is $U^*=U_c^*=\frac{2}{3}$. Then from Eq.~(\ref{ELSI_1}) we can get, $\sigma_c=\sqrt{\frac{4}{27}}$. Now, define the order parameter as the deviation of the stable fixed point from its critical value, $\mathcal{O}=U^{*}(\sigma)-U^{*}(\sigma_c)=U^{*}(\sigma)-\frac{2}{3}$. Substituting this expression into Eq.~(\ref{ELSI_1}), we obtain $\mathcal{O}^{3}+\mathcal{O}^{2}=\sigma_c^{2}-\sigma^{2}$. Close to the critical point, the order parameter is small ($\mathcal{O}\ll1$), so that the cubic term is negligible compared with the quadratic term. Hence, $\mathcal{O}^{2}\sim 2\sigma_c(\sigma_c-\sigma)$ (using the approximation $\sigma_c+\sigma\simeq2\sigma_c$). Therefore

\begin{equation*}
\mathcal{O}
\propto (\sigma_c-\sigma)^{1/2}
\sim
(\sigma_c-\sigma)^{\eta},
\qquad
\eta=\frac{1}{2}.
\label{ELSI_2}
\end{equation*}

\noindent To obtain the relaxation time ($\tau$), we can apply the continuum approximation $U_{t+1}-U_t\simeq \frac{dU}{dt}$ in Eq. (\ref{ELSI_1a}) and write it as 

\begin{equation}
    \frac{dU}{dt}=-\frac{U^3-U^2+\sigma^2}{U}.
    \label{ELSI_3}
\end{equation}

\noindent Let us consider a small perturbation about the fixed point solution $U(t,\sigma)=U^*(\sigma)+\varepsilon(t)$, where $|\varepsilon(t)|\ll1$ and $U^*=\frac23+\sqrt{\sigma_c-\sigma}$ then Eq. (\ref{ELSI_3}) becomes $\frac{d\varepsilon}{dt}=-\left.[\frac{3U^{*2}-2U^*}{U*}]\right.\varepsilon=-3\varepsilon\sqrt{\sigma_c-\sigma}$ giving $\varepsilon(t)\sim e^{-\frac{t}{\tau}}$. So, the relaxation time diverges as

\begin{equation*}
\tau
\sim
(\sigma_c-\sigma)^{-\frac{1}{2}}
\sim
(\sigma_c-\sigma)^{-\zeta},
\qquad
\zeta=\frac{1}{2}.
\end{equation*}

\noindent At the critical point, the first-order term of $\varepsilon$ vanishes. Therefore, $\frac{d\varepsilon}{dt}\simeq-\frac{3}{2}\varepsilon^2$, giving the solution as

\begin{equation*}
U_t-\frac23
\sim
t^{-\lambda},
\qquad
\lambda=1.
\end{equation*}

\noindent For avalanche-size distribution

\begin{equation*}
P(s)\sim \frac{d\sigma}{ds}
\sim
s^{-\theta},
\qquad
\theta=3.
\end{equation*}

\noindent If we differentiate Eq.~(\ref{ELSI_1}) with respect to $\sigma$ then, the breakdown susceptibility becomes

\begin{equation*}
\chi=\left|\frac{dU^*(\sigma)}{d\sigma}\right|\sim(\sigma_c-\sigma)^{-\gamma},
\qquad \gamma=\frac{1}{2}.
\end{equation*}

\subsubsection{ELS-FBM with linearly decreasing distribution of fibre strength:}
\label{sec:decreasing_ELSFBM}

For a linearly decreasing distribution of fibre strength thresholds as shown in Fig.~\ref{fig:pf}(B), the
probability density is

\begin{equation*}
\rho(\sigma_{th})=
\begin{cases}
2(1-\sigma_{th}), & 0\leq\sigma_{th}\leq1,\\
0, & \sigma_{th}>1,
\end{cases}
\end{equation*}

\noindent which satisfies the normalization condition $\int_0^1\rho(\sigma_{th})\,d\sigma_{th}
=\int_0^1 2(1-\sigma_{th})\,d\sigma_{th}=1$. The corresponding cumulative distribution function is

\begin{equation}
\rho_c(\sigma_t)
=
\int_0^{\sigma_t}
\rho(\sigma_{th})\,d\sigma_{th}
=
2\sigma_t-\sigma_t^2.
\label{ELSD_1}
\end{equation}

\noindent Substituting Eq.~(\ref{ELSD_1}) into the general recursion relation (\ref{ELSU_1}), we can get

\begin{equation}
U_{t+1}
=
1
-
2\left(\frac{\sigma}{U_t}\right)
+
\left(\frac{\sigma}{U_t}\right)^2.
\label{eq:recursion_decreasing}
\end{equation}

\noindent At the fixed point, $U_{t+1}=U_t=U^*$, the recursion relation (\ref{eq:recursion_decreasing}) becomes

\begin{align}
(U^*)^3
-
(U^*)^2
+
2\sigma U^*
-
\sigma^2
=
0.
\label{ELSD_2}
\end{align}

\noindent The critical point is obtained from the condition

\begin{align}
\frac{d}{dU^*}
\left[
(U^*)^3-(U^*)^2+2\sigma U^*-\sigma^2
\right]
=0
\Rightarrow
\sigma=
\frac{2U^*-3(U^*)^2}{2}.
\label{ELSD_3}
\end{align}

\noindent If we put the value of $\sigma$ in Eq. (\ref{ELSD_2}), then we can find  $U^*=0
$ or $\frac{4}{9}$. Since the physically meaningful solution corresponds to a finite surviving fraction of fibres, $U^*=U_c^*=\frac49$. So from Eq. (\ref{ELSD_3}), $\sigma_c =\frac{4}{27}$. The order parameter becomes $\mathcal{O}=U^*(\sigma)-\frac{4}{9}$. Now from Eq. (\ref{ELSD_2}) we can get $\mathcal{O}\sim \frac{4}{3}(\sigma_c-\sigma)^{\frac{1}{2}}$. Comparing the expression with the scaling relation

\begin{equation*}
\mathcal{O}
\sim
(\sigma_c-\sigma)^{\eta},
\qquad
\eta=\frac{1}{2}.
\end{equation*}

\noindent To obtain the relaxation time ($\tau$), we can apply the continuum approximation $U_{t+1}-U_t\simeq \frac{dU}{dt}$ in Eq. (\ref{eq:recursion_decreasing}) and write it as

\begin{equation}
\frac{dU}{dt}
=
-\frac{U^3-U^2+2\sigma U-\sigma^2}{U^2}.
\label{ELSD_4}
\end{equation}

\noindent Let us consider a small perturbation about the stable fixed-point solution
$U(t,\sigma)=U^*(\sigma)+\varepsilon(t)$, where $|\varepsilon(t)|\ll1$. Near the critical point, the fixed-point solution is given by $U^*(\sigma)=
\frac49+\frac43\sqrt{\sigma_c-\sigma}$. Substituting $U(t,\sigma)=U^*(\sigma)+\varepsilon(t)$ into Eq.~(\ref{ELSD_4}) and retaining the leading term in $\varepsilon$ gives $\frac{d\varepsilon}{dt} \simeq -\frac92\varepsilon\sqrt{\sigma_c-\sigma}$. Thus, $\varepsilon(t)\sim e^{-t/\tau}$, where the relaxation time diverges as

\begin{equation*}
\tau
\sim
(\sigma_c-\sigma)^{-1/2}
\sim
(\sigma_c-\sigma)^{-\zeta},
\qquad
\zeta=\frac12.
\end{equation*}

\noindent At the critical point, the first-order term $\varepsilon$ vanishes. Therefore, $\frac{d\varepsilon}{dt}\simeq-\frac{27}{16}\varepsilon^2$, giving the solution as

\begin{equation*}
U_t-\frac49
\sim
t^{-\lambda},
\qquad
\lambda=1.
\end{equation*}

\noindent For avalanche-size distribution

\begin{equation*}
P(s)\sim \frac{d\sigma}{ds}
\sim
s^{-\theta},
\qquad
\theta=3.
\end{equation*}

\noindent If we differentiate Eq.~(\ref{ELSD_2}) with respect to $\sigma$, the breakdown susceptibility becomes

\begin{equation*}
\chi
=
\left|
\frac{dU^*}{d\sigma}
\right|
\sim
(\sigma_c-\sigma)^{-\frac{1}{2}}
\sim
(\sigma_c-\sigma)^{-\gamma},
\qquad
\gamma=\frac{1}{2}.
\end{equation*}

\subsubsection{ Universality class of failure statistics in the ELS-FBM:}
\label{universility_in_ELSFBM}

We have thus shown that, although the critical stress $\sigma_c$ depends on the specific form of the fibre failure-threshold distribution $\rho(\sigma_{\mathrm{th}})$, the associated critical exponents remain unchanged for the class of threshold distributions considered under a given loading protocol. The order parameter ($\mathcal{O}$) is characterized by the exponent $\eta=\frac{1}{2}$, the breakdown susceptibility ($\chi$) by $\gamma=\frac{1}{2}$, the relaxation time ($\tau$) by $\zeta=\frac{1}{2}$, and the avalanche-size distribution $P(s)$ by $\theta=3$. Furthermore, at the critical point ($\sigma=\sigma_c$), the deviation of the surviving fraction from its critical value decays algebraically as $(U_t-U_c)\sim t^{-\lambda}$, with the critical relaxation exponent $\lambda=1$. These results demonstrate the robustness of the critical exponents against changes in the fibre-strength distribution within the specified loading protocol \cite{Zapperi_PRL_1997,Pradhan2001PRE,Pradhan2003,Pradhan2010,HansenHemmerPradhan2015}.

The avalanche-size exponent depends on how the external load is increased. The exponent $\theta=3$ applies to a protocol in which the load is increased in finite, uniform discrete increments that are not too small, with each increment potentially initiating an avalanche of fibre failures \cite{HemmerPradhan2007}. By contrast, under quasistatic loading, in which the load is increased until the next fibre reaches its failure threshold and the ensuing avalanche develops through load redistribution at fixed external load, the avalanche-size distribution exponent becomes $\frac{5}{2}$ \cite{HemmerHansen1992}. Thus, the exponents $\theta=3$ and $\frac{5}{2}$ describe event-by-event avalanche-size distributions under different loading protocols and should not be interpreted as exponents of a load-integrated distribution. The distribution of released energy is a distinct observable. For the ELS-FBM with Hookean fibres, the energy-burst distribution follows the asymptotic power law with exponent $\frac{5}{2}$, irrespective of the specific fibre strength distribution, under the assumptions of the model \cite{Pradhan2010, HansenHemmerPradhan2015}. However, this finding should not be confused with the size distribution of avalanches that results for finite load increments, when $\theta=3$. Combining the Rushbrooke relation, $\alpha+2\eta+\gamma=2$, with the hyperscaling relation, $2-\alpha=d\nu$, gives $2\eta+\gamma=d\nu$ \cite{Stanley1971}. Here, $\nu$ is the correlation-length exponent, and $d$ is the spatial dimension. Using $\eta=\gamma=\frac{1}{2}$ gives $d\nu=\frac{3}{2}$. Within the proposed Flory-Kolmogorov scaling framework the upper critical dimension is estimated as $d_u=6$ \cite{Biswas_flory_2020}. Taking $d=d_u=6$ in the proposed scaling relation yields $\nu=\frac{1}{4}$. These estimates depend on the assumptions of the proposed scaling framework and should not be regarded as direct consequences of the avalanche-size exponents alone.

\subsection{Local Load Sharing Fibre Bundle Model (LLS-FBM)}
\label{subsec:LLSFBM}

The ELS-FBM represents fracture within a mean-field approximation, in which the load released by a broken fibre is equally redistributed among all intact fibres. This mean-field picture can be generalized by introducing spatial dependence into the load-transfer mechanism. In particular, \cite{kun_herrmann_hidalgo_PRE_2002,Biswas_nucleation_2015} introduced a distance-dependent load-transfer function, allowing the effective interaction range to vary continuously between the global and local load-sharing (LLS) limits. A crossover from mean-field to short-range fracture behavior was observed as the interaction range was reduced, accompanied by changes in the ultimate strength and in the avalanche and cluster size distributions. Thus, the spatial range of load redistribution emerges as an important factor controlling the collective failure behavior of fibre bundles. The introduction of temporal dynamics provides a further extension of the LLS framework. The kinetics of failure and statistical properties of lifetimes in time-dependent LLS models are determined by both the kinetics of breakdown and the character of the stress redistribution process in the vicinity of the local breakdown event \cite{Newman_Phoenix_2001}. In this connection, high enough values of the breakdown exponent lead to formation of critical failure clusters that result in the brittle failure. Time-dependent failure in viscoelastic fiber bundles exhibits another qualitative difference between the global and local redistribution of load, with LLS demonstrating a sharp crossover to the global failure regime compared to ELS \cite{Hidalgo_Kun_Herrmann_CREEP}. All these studies provide the basis for the LLS-FBM model as the model of transfer of the stress released during the failure of a fiber to its neighbors. In this context, not only the magnitude but also the very way the loads are redistributed is capable of affecting the macroscopic behavior. Thus, one may see the redistribution rule as another factor responsible for the strength and criticality of the system. For instance, it was reported that the redistribution model based on the excess strength of the surviving fibres leads to a modification of the maximum strength of the bundle, as well as to changing the universality class of the failure transition \cite{SB_PS_PRL_2015}. The localization of the load redistribution is also able to lead to the collective dynamics going far beyond the critical failure. In particular, it has been reported that the LLS-FBM under slow driving evolves into a self-organized critical state with scale-invariant avalanche dynamics. This has been shown explicitly for the two-dimensional LLS-FBM \cite{Biswas_LLSFBM_2013}.

\subsubsection{SOC in two-dimensional LLS-FBM:}
\label{sec:soc_llsfbm}

\begin{figure}[h!]
\begin{center}
    \includegraphics[width=1.0\textwidth]{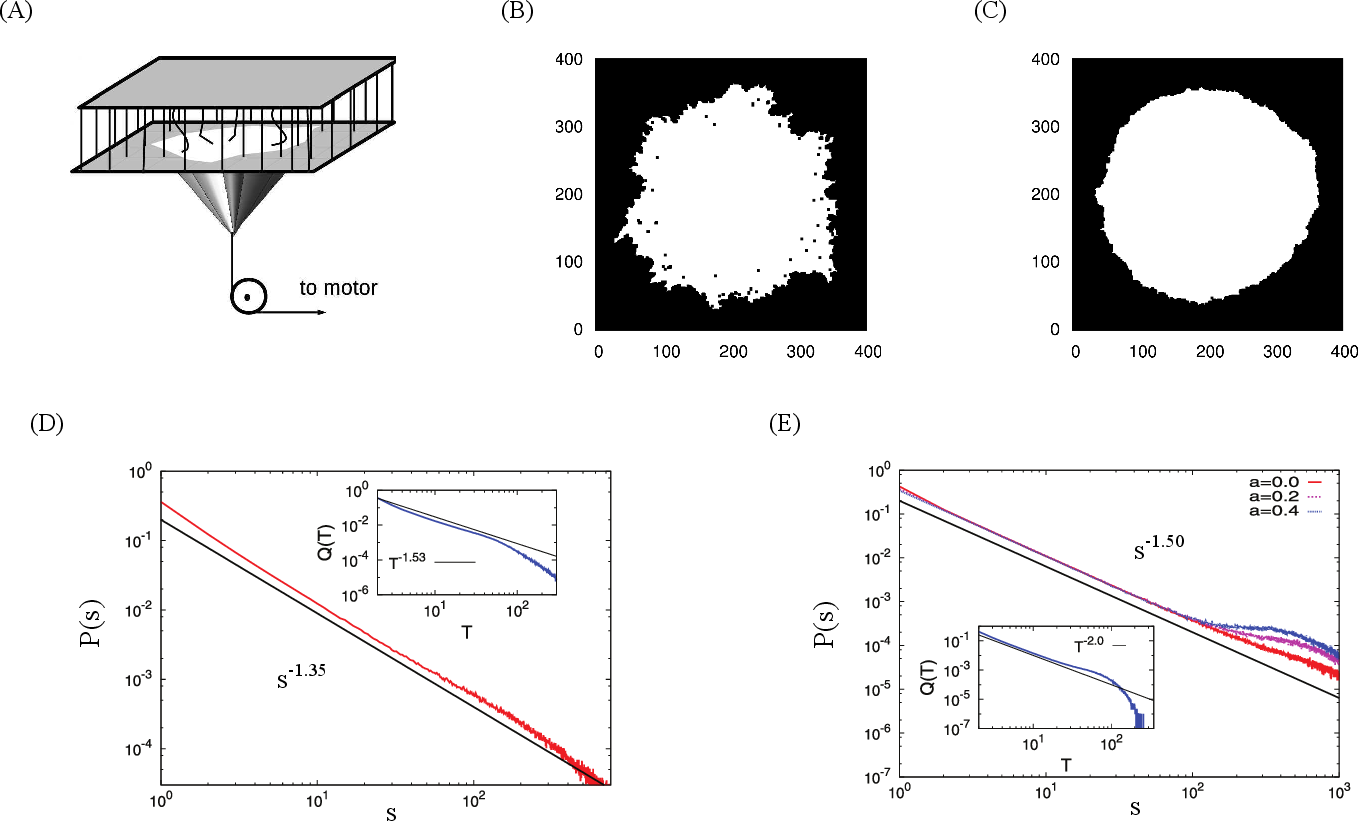}
    \end{center}
    \caption{\textbf{(A)} A schematic diagram of the model used in \cite{Biswas_LLSFBM_2013}. Load is slowly increased at a single point on the lower platform. The patch of broken fibres is indicated by the white portion in the lower platform. \textbf{(B)} A snapshot of the patch of broken fibres when the load redistribution is among the nearest surviving neighbors (Model I). The fibre failure thresholds are uniformly distributed between $[0:1]$. \textbf{(C)} A snapshot of the patch of broken fibres when the load redistribution is uniform along the broken patch boundary, i.e., Model II. The failure thresholds are uniformly distributed between $[0:1]$. [From \cite{Biswas_LLSFBM_2013}] \textbf{(D)} The distribution of the avalanche sizes for Model I (load redistribution is among the nearest surviving neighbors). The distribution is a power law with an exponent value of $1.35 \pm 0.03$. Inset: Distribution of the duration of an avalanche is shown, which is also a power-law decay with exponent value $1.53 \pm 0.02$.  \textbf{(E)} The distribution of the avalanche sizes is plotted for zero and finite lower cutoffs for Model II. The distribution function is a power law with an exponent value of $1.50 \pm 0.01$, which is also our estimate from scaling arguments. Inset: The distribution of the duration of an avalanche is plotted for Model II. This shows a power-law decay with an exponent value of $2.00 \pm 0.01$. [Adapted from \cite{Biswas_LLSFBM_2013}]}
    \label{fig:soc_fbm}
    \end{figure}

SOC can emerge in LLS-FBM without introducing explicit dissipation or fibre regeneration. Instead, the growth of the damaged region continuously increases the number of load-carrying boundary fibres, naturally balancing the slow external driving and producing stationary scale-free avalanche dynamics. In \cite{Biswas_LLSFBM_2013}, a modified two-dimensional LLS-FBM is introduced that exhibits SOC behavior through localized loading rather than the conventional uniform loading employed in standard LLS models, as shown in Fig. \ref{fig:soc_fbm}. In this model the external load is applied quasi-statically to a single fibre at the center of the lattice instead of uniformly across the system. Following the failure of the central fibre, the released load is redistributed according to two different load-sharing schemes. 

In Model I, the load of a broken fibre is shared equally among its nearest surviving neighbors, leading to the growth of a compact damaged patch while preserving local stress fluctuations. During an avalanche, the external load remains fixed, and successive failures are driven solely by load redistribution until a stable configuration is reached, see Figs. \ref{fig:soc_fbm}(B) and (D). As a simplified counterpart, Model II redistributes the load of a failed fibre uniformly among all surviving fibres located along the boundary of the damaged region. By suppressing local stress fluctuations, this boundary-based redistribution provides a mean-field approximation. Numerical simulations demonstrated that Model I exhibits SOC behavior with avalanche exponents close to the Manna universality class \cite{manna_1991}, whereas Model II reproduces the mean-field SOC limit with analytical predictions in excellent agreement with the simulation results, as shown in Figs. \ref{fig:soc_fbm}(C) and (E).

\section{Social Inequality Measures: Gini, Kolkata, and Hirsch Indices}
\label{sec:inequality_measures}

\begin{figure}[h!]
\begin{center}
    \includegraphics[width=1.0\textwidth]{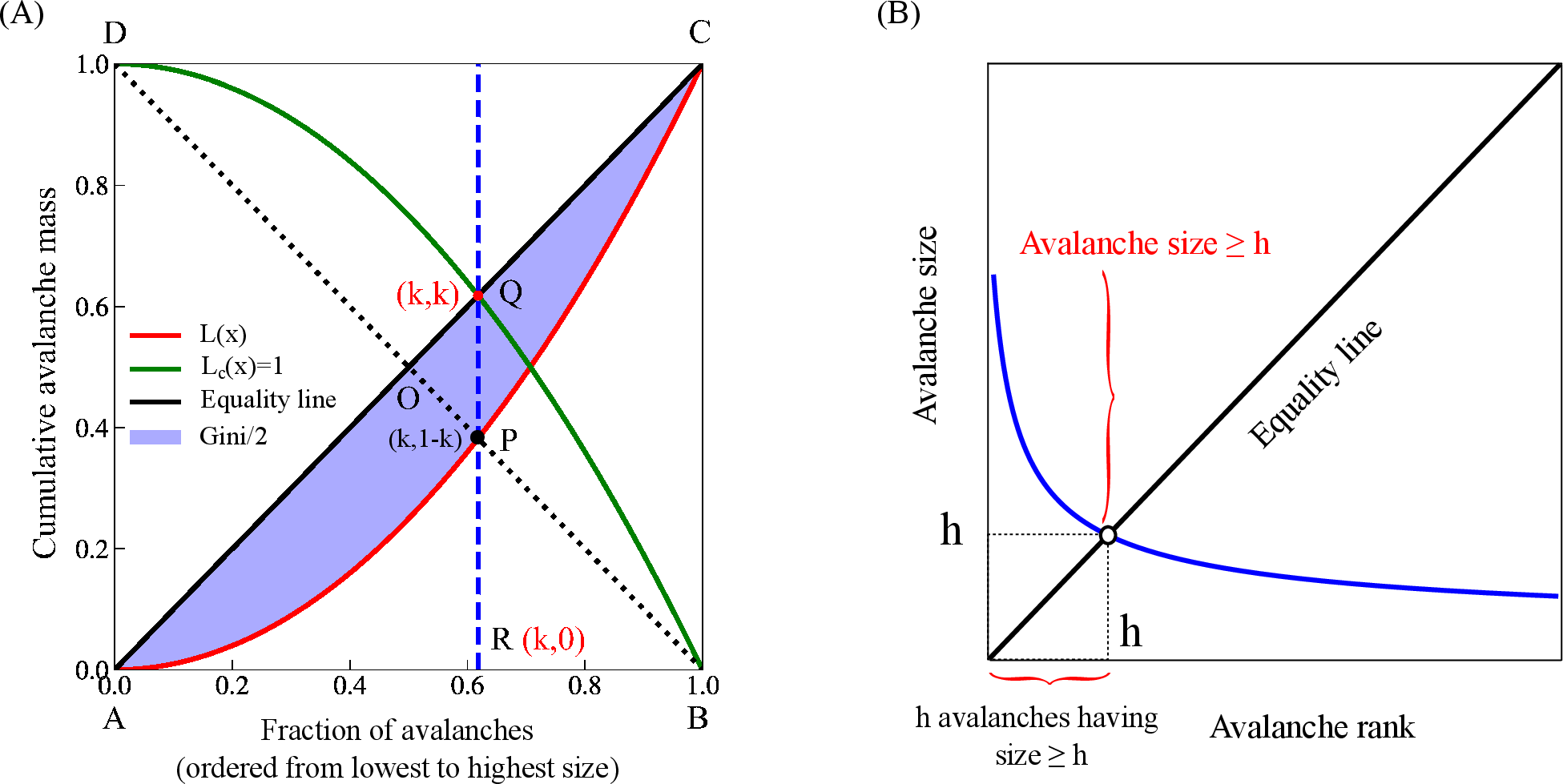}
    \end{center}
    \caption{\textbf{(A)} Schematic diagram of the Lorenz function $L(x)$, the line of perfect equality, and the shaded area used to define the Gini index. The Gini index ($g$) is given by the ratio of the shaded area APCOA to the triangular area ABCA, or equivalently, twice the area between the Lorenz curve (the red line) and the line of perfect equality (the black line). The Kolkata index ($k$) is defined by the intersection of the Lorenz curve with the anti-diagonal line, satisfying $L(k)=1-k$. \textbf{(B)} Avalanche events are arranged in descending order of their sizes and plotted against their rank. The black diagonal corresponds to the equality line. The Hirsch index ($h$) is defined by the intersection of the ranked avalanche-size curve with the equality line (where avalanche size $=$ avalanche rank) satisfying avalanche size $s_h\ge h$ for the largest $h$ avalanches. Equivalently, at least $h$ avalanche events have sizes not smaller than $h$, whereas the remaining avalanche events have sizes not exceeding $h$.}
    \label{fig:lorenz}
    \end{figure}

Social statistical measures \cite{sen_chakrabarti_sociophysics_2014} were originally developed in economics to quantify disparities in the distribution of wealth and income within a population. By providing a quantitative description of how unevenly a resource is distributed, these measures have become powerful tools for characterizing heterogeneity in complex datasets. The applicability of inequality measures extends far beyond economics. Over the past decade, they have been successfully employed in diverse fields such as sociology, finance, network science, statistical physics, etc., where they quantify the heterogeneity in the distribution of resources, interactions, or events. Inequality measures have recently emerged as powerful tools for investigating systems exhibiting critical phenomena and catastrophic failure. Some of the most used social statistical measures are the Gini index \cite{gini}, the Hirsch index \cite{h_index} and the Kolkata index \cite{kolkata}. 

The Gini ($g$) and Kolkata ($k$) inequality indices are derived from the Lorenz curve \cite{lorenz_1905}, $L(x)$, which represents the cumulative fraction of energy released by the smallest fraction $p$ of events arranged in increasing order of their released energy; see Fig. \ref{fig:lorenz}(A). By construction, the Lorenz curve satisfies $L(0)=0$ and $L(1)=1$. In the limiting case where all events release exactly the same amount of energy, the Lorenz curve coincides with the diagonal line joining $(0,0)$ and $(1,1)$, indicating perfect equality. Any deviation of the Lorenz curve below this diagonal reflects the degree of inequality in the energy distribution. The Gini index is defined as the normalized area enclosed between the line of perfect equality and the Lorenz curve. Mathematically,

\begin{equation*}
g = 2\int_{0}^{1}\left[x-L(x)\right]\,dx
  = 1 - 2\int_{0}^{1} L(x)\,dx.
\end{equation*}

\noindent The Gini index ranges from 0 to 1, where $g=0$ corresponds to perfect equality and $g=1$ corresponds to complete inequality. The Kolkata index $(k)$, defined as the fixed point of the Lorenz curve satisfying

\begin{equation}
    1-k=L(k).
    \label{eq:k}
\end{equation}

\noindent Equivalently, the largest $(1-k)$ fraction of events contributes a fraction $k$ of the total released energy, as shown in Fig. \ref{fig:lorenz}(A). The Kolkata index generalizes Pareto's well-known 80-20 law (see \cite{pareto}), according to which approximately $80\%$ of the total wealth is possessed by the richest $20\%$ of the population. In the context of fracture and avalanche dynamics, the Kolkata index quantifies the extent to which a small fraction of large events dominates the cumulative energy release.

To obtain an analytical relation between the Gini and Kolkata indices, we first consider a minimal (Landau-like; see e.g., \cite{joseph_2022}) polynomial approximation for the Lorenz function
$L(x)=Ax+Bx^2$, where $A,B>0$, such that $L(x)$ becomes a monotonically increasing function of $x$ and $A + B = 1$ to ensure $L(x) = 1$ for $x = 1$ and also $L(x) = 0$ at $x = 0$. One can then express the Gini index through

\begin{equation}
    g = 1-2\int_0^1 L(x)dx = 1-A-\frac{2B}{3}= \frac{B}{3}.
    \label{eq:g=B/3}
\end{equation}

\noindent Substituting $A=1-3g$ and $B=3g$ into the assumed polynomial expression for $L(x)$, and using Eq. (\ref{eq:k}), we obtain $1-k=(1-3g)k+3gk^2\;\Rightarrow\;3gk^2+(2-3g)k-1=0$ giving the Kolkata index as, \cite{ghosh_gkp, g_eq}

\begin{equation}
k = \frac{1}{2}+\frac{3g}{8}, \quad \text{for } g \to 0.
\label{eq:3g/8}
\end{equation}

\noindent While Eq. (\ref{eq:3g/8}) provides a leading-order relation valid in the weak-inequality limit, a more realistic description of fracture and avalanche dynamics can be obtained by considering the experimentally observed power-law distribution of event sizes. In many fracture and avalanche systems, event sizes follow a power-law distribution as, $P(s)\sim s^{-\beta}$ where, $s$ is the avalanche size. Of course, there is always a physical lower and upper cut-off.

Let the $r_0$-th event denote the largest event. If the events are arranged in ascending order of size, then the size of the $r$-th event scales as, $s_r \sim (r_0-r)^{-n}$, where the diverging/system-spanning event happens at $r=r_0$, with $n=\frac{1}{\beta-1}$ (see e.g., \cite{eswar}. For a system driven towards a failure/transition, the natural order is often close to the ascending order. For systems driven towards catastrophic failure, successive events often exhibit an increasing trend in released energy as failure approaches (see e.g., \cite{jordi}). Consequently, the chronological order is expected to be approximately consistent with the ascending order of event sizes over the precursory stage (see e.g., \cite{schmit}). This approximation is more appropriate for systems approaching catastrophic failure than for stationary systems. Considering a fraction $\mathcal{R}=\frac{r}{r_0}$ of the ordered sequence (assuming that the chosen sequence of events falls within that segment), the corresponding Lorenz function becomes

\begin{equation}
L(x,\mathcal{R},n)=
\frac{\displaystyle\int_{0}^{x\mathcal{R}r_0}(r_0-r)^{-n}\,dr}
{\displaystyle\int_{0}^{\mathcal{R}r_0}(r_0-r)^{-n}\,dr}
=
\frac{1-(1-x\mathcal{R})^{1-n}}
{1-(1-\mathcal{R})^{1-n}}.
\label{lorenz}
\end{equation}






Particularly simple analytical expressions emerge in the critical limit $\mathcal{R}\to1$, corresponding to the occurrence of the largest avalanche. In this limit, the Lorenz function reduces to \cite{gini_prl_2023,eswar,ghosh_gkp}

\begin{equation}
    L(x,n)=1-(1-x)^{1-n},\quad \text{where } n=\frac{1}{\beta-1}
    \label{eq:n_beta}
\end{equation}.

\noindent This gives $g(n)=1-2\int_{0}^{1}L(x,n)\,dx=\frac{n}{2-n}$. Furthermore, using Eq. (\ref{lorenz}) together with the defining equation of the Kolkata index, Eq. (\ref{eq:k}), one obtains $1-n=\frac{\ln k}{\ln(1-k)}$. This then gives the simpler ($n$-independent) form

\begin{equation}
g=\frac{\ln(1-k)-\ln k}{\ln(1-k)+\ln k}.
\label{gk_relation}
\end{equation}

\noindent Although Eq.~(\ref{gk_relation}) is derived in the $\mathcal{R}\to1$ limit, numerical simulations and real earthquake data show that most observations for $\mathcal{R}<1$ also lie close to this universal curve.

The Hirsch index ($h$) is introduced to characterize the distribution of avalanche sizes by simultaneously accounting for the number and sizes of large avalanche events. To define this index, the avalanche events are first ranked in descending order of their sizes. The $h$ index is then defined as the largest integer $h$ satisfying $s_{h}\geq h$. Equivalently, a system has an $h$ index equal to $h$ at least $h$ avalanche events that have sizes greater than or equal to $h$, whereas all remaining avalanche events have sizes less than $h$, as shown in Fig. \ref{fig:lorenz}(B).

\section{Inequality measures in fracture models and experiments}
\label{inequality_in_fracture_models}

\begin{figure}[h!]
\begin{center}
    \includegraphics[width=1.0\textwidth]{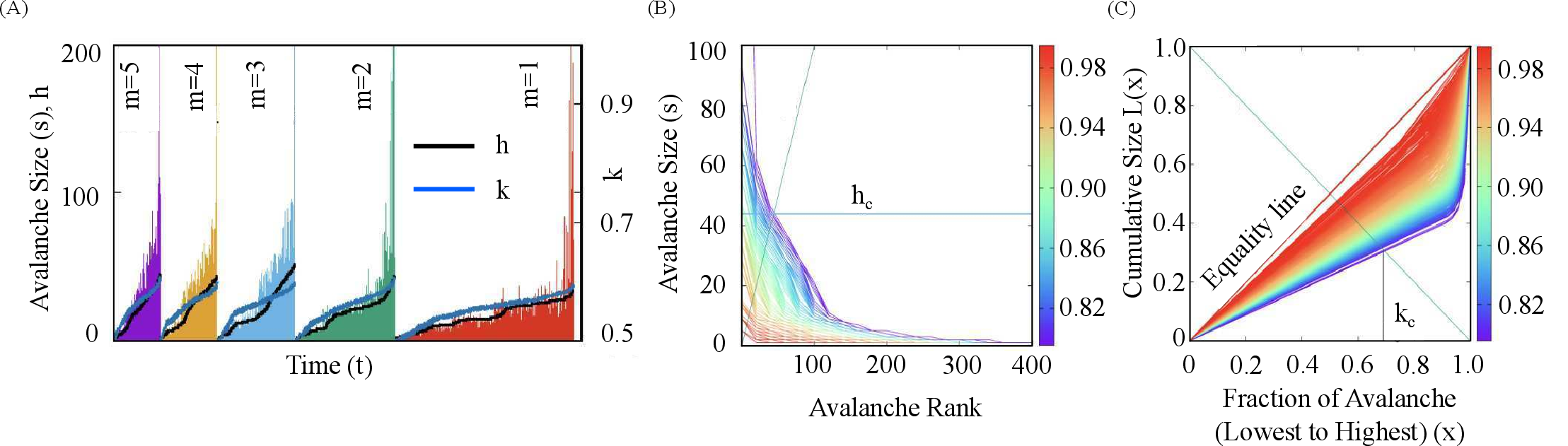}
    \end{center}
    \caption{ The measures of inequality for avalanche sizes in an FBM (with $50,000$ fibres) are shown. \textbf{(A)} The time series of avalanche sizes ($s$) are displayed for different values of Weibull modulus ($m$). Solid black and blue lines represent the $h$ and the $k$ indices, respectively. The left scale is for $s$ and the $h$ index, while the right scale is for the $k$ index. Even though different samples continue their dynamics for different amounts of time, showing different critical load values ($\sigma_c$), the final values of the $h$ and $k$ indices slightly change with $m$. \textbf{(B)} A rank plot of avalanche statistics at different stages of the failure process is shown. Colors represent the fraction of surviving fibres at the time of measurement. The $h$ index value is found where the $45^\circ$ line crosses the Lorenz curve $L(x)$. \textbf{(C)} The Lorenz curve for different stages of the failure process is shown. The percentage of surviving fibres is represented by colors, which indicate the process stage. The equality line is shown from where the Lorenz curve starts to deviate as the process continues. The critical value $k_c$ of the $k$ index is marked. [Adapted from \cite{first}]}
    \label{fig:first}
    \end{figure}

Social inequality measures were first used in studying how fractures develop by applying the Kolkata ($k$) and Hirsch ($h$) indices to the size of sudden bursts of damage, called ``avalanches", in a model called ELS-FBM (see \cite{first}). They used these inequality measures to track how damage builds up over time. This study showed that the $k$ index steadily increases during the fracture process and eventually reaches a nearly constant value, around $k_c \approx 0.62 \pm 0.03$ (as shown in Fig. \ref{fig:first}(C)). This value is not greatly affected by the type of disorder or how big the system is. As a result, the $k$ index could be an efficient indicator of the upcoming catastrophic failure because of its constancy. The $h$ index was also considered; however, it turned out to be dependent on the size of the system. In contrast to the $k$ index, its ultimate value will be determined by certain parameters such as the Weibull modulus ($m$) and the number of fibres in the bundle, which makes it not very efficient as a general indicator; see Fig. \ref{fig:first}. The current study confirmed that the metrics of social inequalities could be utilized for monitoring the propagation of damage and predicting the forthcoming failure in materials. The approach was then expanded onto other physical systems. \cite{Ghosh2022} investigated the behavior of the $g$ and $k$ indices in several models such as the kinetic wealth exchange, two-dimensional percolation, SOC sandpiles, and FBMs. It has been revealed that regardless of differences in functioning of these systems on the microlevel, the heterogeneity of wealth, clusters, or avalanches grows in the vicinity of the critical point.

\begin{figure}[h!]
\begin{center}
    \includegraphics[width=1.0\textwidth]{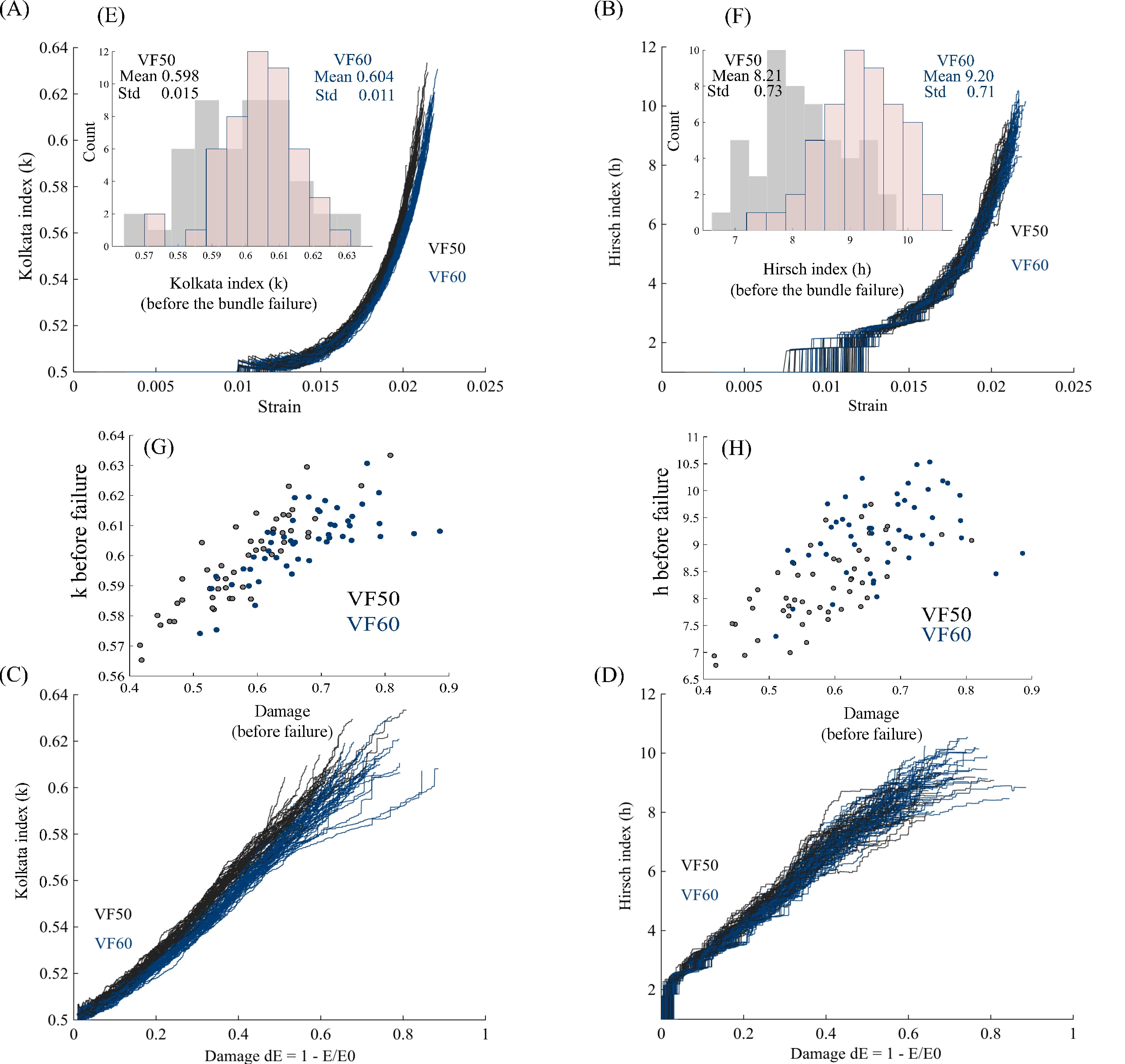}
    \end{center}
    \caption{ The Kolkata ($k$) and Hirsch ($h$) indices are shown as functions of applied strain in Figs. \textbf{(A)} and \textbf{(B)}, and as functions of damage in Figs. \textbf{(C)} and \textbf{(D)}. The Kolkata index is displayed in Figs. \textbf{(A)} and \textbf{(C)}, and the Hirsch index is shown in Figs. \textbf{(B)} and \textbf{(D)}. The Figs. \textbf{(E)} and \textbf{(F)} show the histograms of the Kolkata and Hirsch indices, respectively, at the end of the last strain increase. The Figs. \textbf{(G)} and \textbf{(H)} show the Kolkata and Hirsch indices, respectively, as they relate to the corresponding damage at the end of the last strain increase. The carbon fibre/epoxy bundles in the model had a fibre volume fraction of $50\%$ and $60\%$ (called VF50 and VF60, respectively, as mentioned in \cite{Lomov2023}). Each graph includes $50$ different examples, with the VF50 shown in black lines and the VF60 shown in blue lines. [Private communication, \cite{Lomov2023}] }
    \label{fig:lomov}
    \end{figure}

\cite{Lomov2023} tested these ideas in a more realistic fracture model called the Impregnated FBM (see e.g., \cite{lomov_ifbm_2, lomov_ifbm_1}) which includes realistic ways of redistributing load locally. Using simulations of carbon-fibre/epoxy composites, they found that both the $h$ and $k$ indices change in a similar way to those in the ELS-FBM and reach similar values as the material is about to fail, as shown in Fig. \ref{fig:lomov}. Their results confirmed that inequality indicators are still useful even in more realistic models of composite materials.

\begin{figure}[h!]
\begin{center}
    \includegraphics[width=1.0\textwidth]{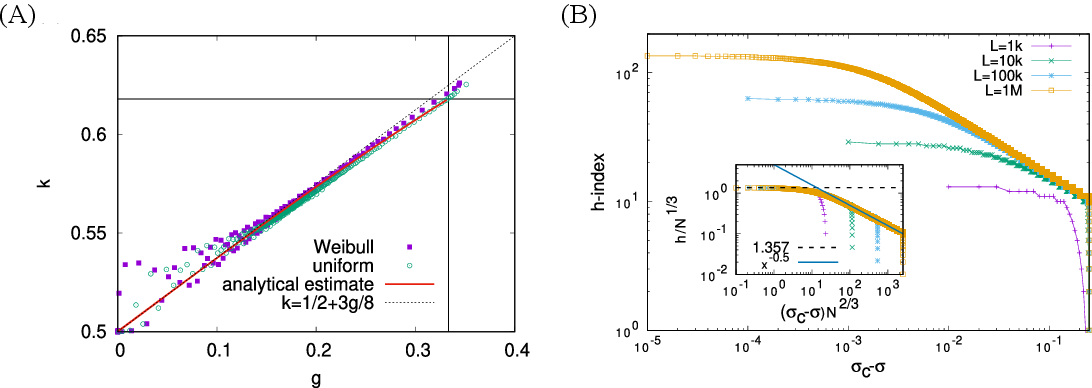}
    \end{center}
  \caption{\textbf{(A)} The Kolkata index changes depending on the Gini index in the mean-field fibre bundle model (see \cite{diksha2023inequality} for more information). Results from numerical simulations are shown for two different types of threshold distributions: uniform and Weibull. A reference line, based on an analytical formula valid for small Gini values (as in Eq.(\ref{eq:3g/8})), is also plotted for comparison. The failure points, where the system breaks down, are marked by vertical and horizontal lines: $g_f = \frac{1}{3}$ and $k_f \approx 0.618$. The close match between the analytical predictions and the simulation results shows that the critical values of the Gini and Kolkata indices are the same, no matter what type of distribution is used, as long as it's from a broad class of fibre strength threshold distributions. \textbf{(B)} The $h$ index shows a specific scaling behavior with system size in the mean-field FBM (see \cite{diksha2023inequality} for more information). The scaling collapse indicates that the $h$ index shows similar behavior for different system sizes near the critical point. The inset shows the behavior of the scaling function in the different ranges. For small values of the scaling variable, it tends to a constant value of $(\frac52)^{1/3}$ (calculated with a loading step size of $10$), while for large values, it decreases in a power-law way with an exponent of $-\frac12$. [From \cite{diksha2023inequality}].}
    \label{fig:diksha_2023}
    \end{figure}

\cite{diksha2023inequality} expanded on the theory of using inequality measures by studying the $g$, $k$, and $h$ indices in the ELS-FBM, LLS-FBM, and the Random Fuse Model (RFM). They showed how these indices change near the critical point and tested them using computer simulations. The study also showed that similar behavior is seen in the LLS-FBM and the RFM, indicating that inequality measures are useful beyond the simple average models, as shown in Fig. \ref{fig:diksha_2023}. These findings established that inequality indices are reliable signs of damage buildup and the potential for sudden failure in a wide range of fracture models.

\begin{figure}[h!]
\begin{center}
    \includegraphics[width=1.0\textwidth]{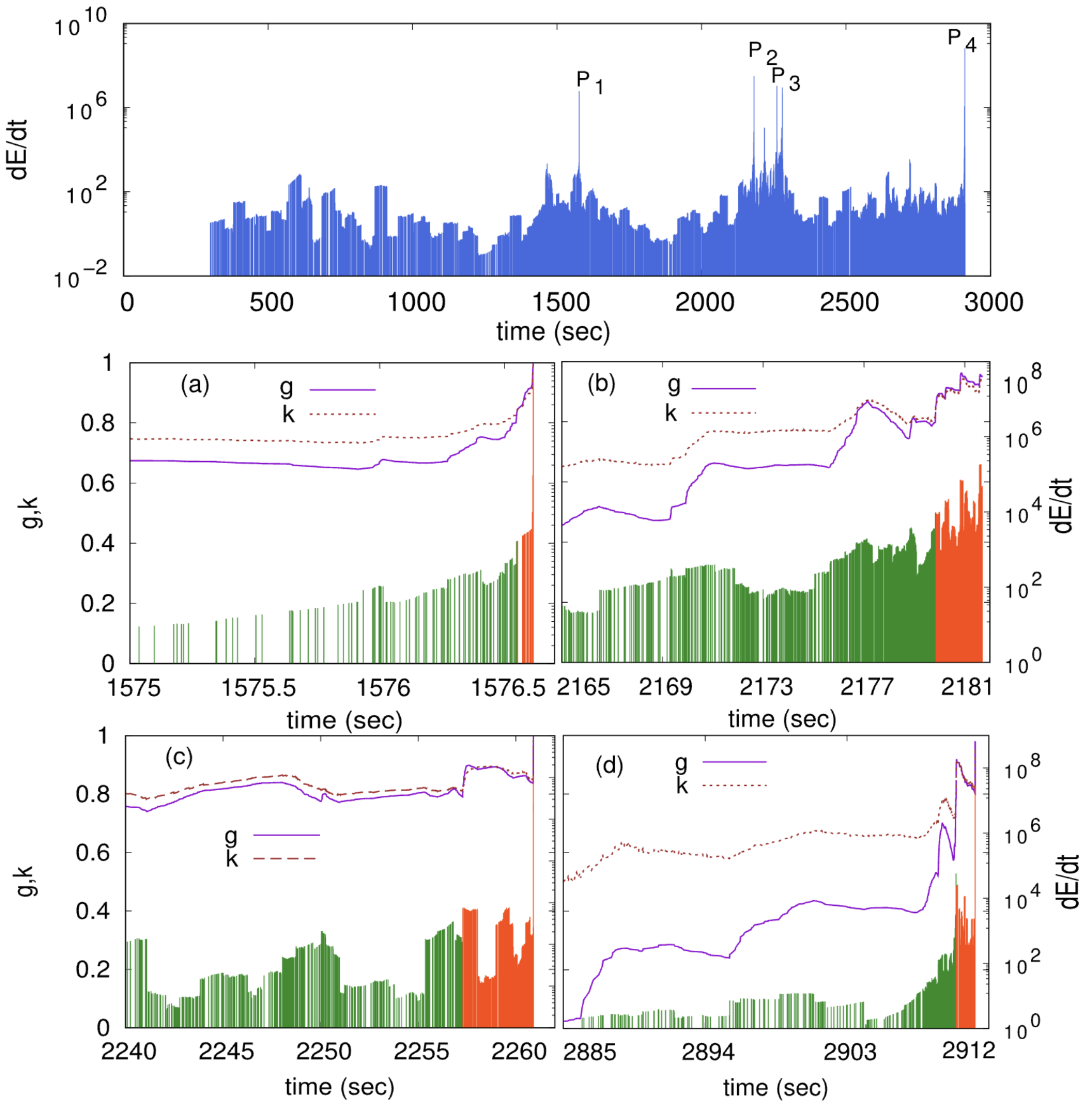}
    \end{center}
    \caption{ The time evolution of the acoustic emission (AE) energy release for the natural red sandstone sample (SR2) is shown using a window of $50$ events. Panels \textbf{(A)} to \textbf{(D)} describe the time evolution of the Gini ($g$) and Kolkata ($k$) inequality indices for the four main AE peaks (P1 to P4). The intersection of $g$ and $k$ curves occurs earlier than the major AE event for each peak. Events after the crossing are shown in red, and events before the crossing are shown in green. The fact that the $g,k$ crossing consistently precedes each major AE peak indicates that the $g,k$ crossing is a reliable precursor of impending failure. [From \cite{jordi}]}
    \label{fig:jordi}
    \end{figure}

Very recently, \cite{jordi} demonstrated experimentally the applicability of this approach for analyzing acoustic signals produced by compression of such materials as Vycor glass, Gelsil glass, and natural sandstone. Rather than focusing on the size of sudden bursts, they analyzed the non-uniformity of energy release rate $\frac{dE}{dt}$, which grows sharply prior to the formation of large fractures. In particular, they found that the crossing point of the $g$ and $k$ parameters occurred immediately prior to sharp decreases of the strain, as illustrated in Fig. \ref{fig:jordi}. Thus, it confirms that assessment of inequality of energy release rate is an effective tool for predicting large-scale avalanche events despite the absence of any indication of the emergence of catastrophes in the statistics of break sizes of such avalanches. Inequality measures applied in real-life (mostly earthquake) predictions face a number of issues in addition to those associated with fracture studies conducted in a laboratory or computational environment. Unlike controlled systems where the loading regimes, system boundaries, and times of failure can be known or fixed, tectonic systems are heterogeneous and partially observable. The earthquake catalogs may be incomplete, especially for small earthquakes, while the differences in the sensitivity of the network, its spatial heterogeneity, and aftershock clustering could affect the values of inequality indices.

\section{Earthquake statistics and models}
\label{sec:eq_stat_models}

\subsection{Fundamental laws of earthquakes}
\label{earthquake_laws}

Earthquakes are one of the most dangerous natural events, as they cause many deaths, a lot of money being damaged and long-lasting problems for communities. Scientists have been studying earthquakes for a long time, but it is still difficult to know when big earthquakes will happen. This is because the process that causes earthquakes is very complicated and happens deep in the Earth’s crust, which is made of different kinds of rock. However, there are some patterns to earthquakes that are seen over many different places and sizes. By looking at them, people have worked out some rules that help us understand how earthquakes happen. The rules form the basis of our statistical study of earthquakes. Some of the famous laws are the Gutenberg-Richter law \cite{gr1944} and the Omori-Utsu law \cite{Utsu_1961, Utsu_1995}. Other rules such as Bath's law \cite{Bath1965} and the unified scaling law \cite{Bak2002} have also been proposed, but they focus on different aspects of how earthquakes happen.

The Gutenberg-Richter (GR) law describes the frequency-magnitude distribution of earthquakes by relating the cumulative number of earthquakes, $N_e(\geq M)$, with magnitude greater than or equal to $M$ occurring within a specified region and time interval. The GR relation \cite{gr1944} is expressed as

\begin{equation}
\log N_e(\geq M)=a-bM,
\label{eq:gr}
\end{equation}

\noindent where $a$ is a measure of regional seismic activity and $b$ is the GR exponent, which characterizes the relative occurrence of small and large earthquakes. Although the $b$ value varies by region, it typically falls between $0.8\leq b\leq 1.06$ for small earthquakes and $1.23\leq b\leq 1.54$ for large earthquakes \cite{Pacheco_1992_nature, Frohlich1993}. Since the seismic energy released by an earthquake scales exponentially with magnitude, the cumulative GR relation can be expressed in terms of the seismic energy as

\begin{equation*}
N_e(\geq E)\propto E^{-2b/3}.
\end{equation*}

\noindent On the other side, the Omori-Utsu law states that the aftershock decay rate $n_e(t)$ over time (elapsed since the main shock) $t$ is given by \cite{omori,Utsu_1961,Utsu_1995}

\begin{equation*}
    n_e(t)=\frac{\mathcal{C}}{(t+c)^{p_o}},
    \label{eq:omori}
\end{equation*}

\noindent where $\mathcal{C}$ is the aftershock productivity that depends on the main shock magnitude, $c$ is the time offset parameter that prevents the divergence of the aftershock rate at the time of the main shock, $(t=0)$ and $p_o$ is the Omori-Utsu exponent. For most real aftershock sequences, the exponent $p_o\neq1$, however it is close to $1$ (as mentioned in \cite{Rundle_Turcotte_complexity_and_eq_2015}).

Earthquakes happen when existing tectonic plates or faults slip and rub against each other. This is not just a geophysical event but a major topic in material science and understanding how materials break. Stick-slip friction, where friction suddenly releases stored potential energy, has been shown to be related to earthquakes in various studies \cite{Brace1966, rice_1993, Ben_zion_2008, scholz2019_3rd}. Constitutive laws, or rules, explain how rocks slide and break and help explain how earthquakes start, spread and how often they happen. These rules are important for understanding the seismic activity \cite{Dieterich1992, rice_1993, Lapusta2003, Rubin2005, scholz2019_3rd}. Among the theories, the rate-and-state friction (RSF) law is widely used. It performs well in steady sliding conditions \cite{Marone1998, Leeman_lab_slow_eq_2016, Acosta_dynamic_weaking_2018,Ji_fault_2022,Rubino_DynamicFriction_2017}. But if we do not include the factors like frictional melting, thermal pressurization, flash heating, silica-gel lubrication, and granular flow in fault gouge, the RSF model does not fully explain fast seismic slip. In the current scenario, seismologists and earth scientists build some complex models where they apply these processes \cite{Sibson1973, Shimamoto_1986, DiToro2011, Rice2006, Garagash_FractureMechanics_2021,Shi_HowFrictionalSlipEvolves_2023}.

Complex and diverse fault systems, difficult to reproduce in the laboratory, are the sites of natural earthquakes, so simplified models have become indispensable in the study of earthquakes. The models emphasize important features like the accumulation of stress, frictional instability, migration of stress, and rupture and are still simple enough to be handled mathematically and computationally. They are not exhaustive of the details of real faults, but they allow understanding the general behavior of earthquakes and are broadly used in statistical physics approaches for studying earthquake mechanisms.

\subsection{Earthquake models}
\label{earthquake_model}

Earthquake models can be classified into some categories. The first are physics-based dynamical models that rely on equations of motion and rules about how friction works. These models include the Burridge–Knopoff spring-block model \cite{bkmodel}, which considers faults as continuous surfaces and other mechanical descriptions of the fault motion. The second kind are statistical models that describe earthquakes as random events based on observations. A well-known example is the ``Epidemic-Type Aftershock Sequence" (ETAS) model \cite{ogata_ETAS_1988, ogata_ETAS_1998} which is a superposition of the general earthquake activity and the aftershock sequences. These events are consistent with the Gutenberg–Richter magnitude distribution and with the modified Omori–Utsu law. The third type are simpler statistical physics models, which are interested in the behavior of groups of events together and exhibit patterns like scaling. The models include SOC models like the ``Olami-Feder-Christensen" (OFC) model \cite{ofc} and the two-fractal overlap model \cite{ChakrabartiStinchcombe1999} . These models, while simple, can demonstrate many facets of earthquake activity, such as the patterns of earthquake sizes, the way they happen in time, and the way they behave near critical points. These models are useful for studying the general features of how earthquakes work. In our review we focus on some of these models and their statistical behaviors.

\subsubsection{One-Dimensional Burridge–Knopoff Model:}
\label{subsec:bk}

\begin{figure}[h!]
\begin{center}
    \includegraphics[width=0.55\textwidth]{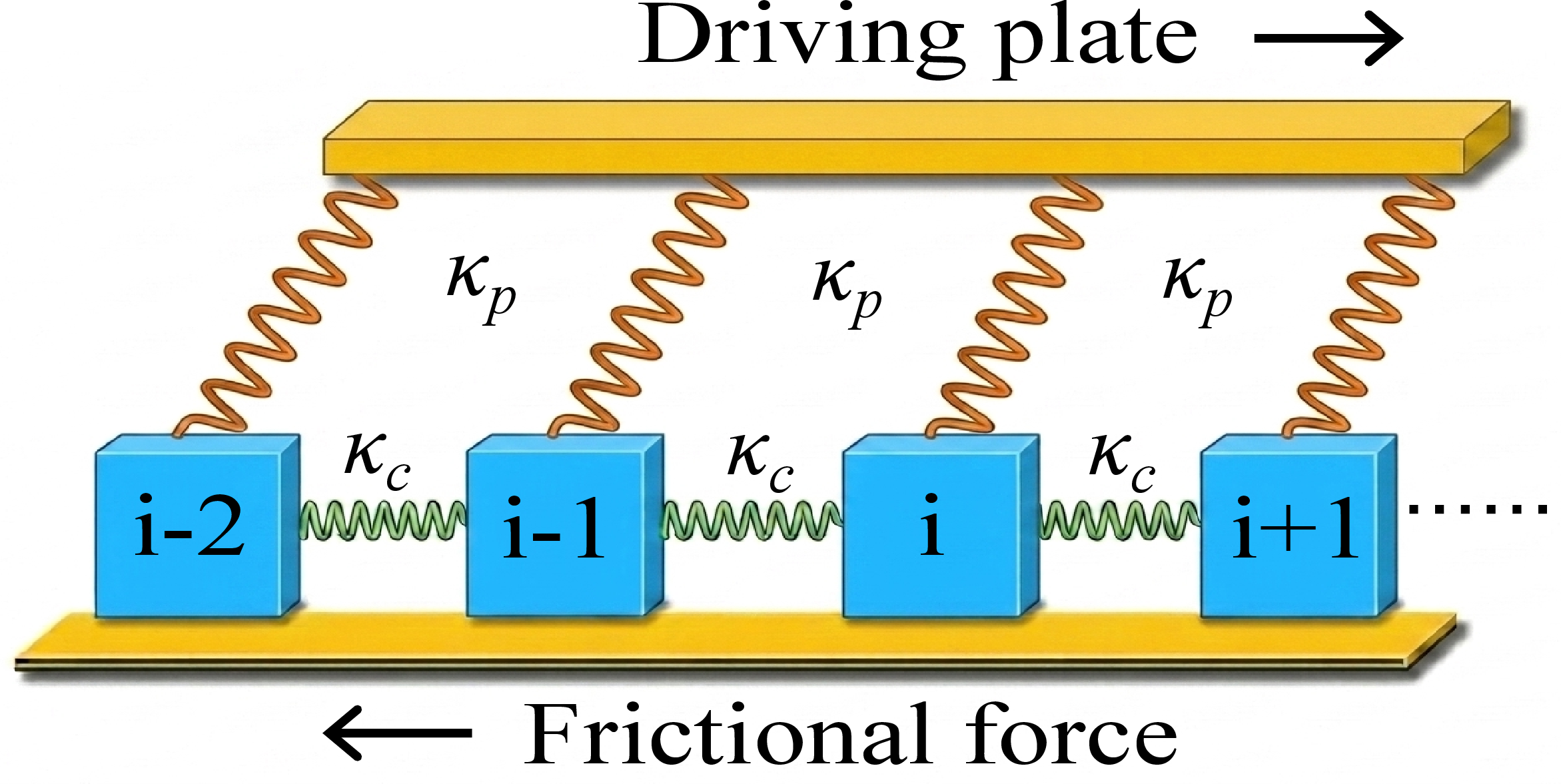}
    \end{center}
    \caption{Schematic representation of the one-dimensional Burridge-Knopoff model.}
    \label{fig:1d_bk}
    \end{figure}

The Burridge-Knopoff (BK) model is one of the most widely studied mechanical models used to understand earthquake dynamics through the lens of statistical physics. The model treats a fault as a collection of blocks that are elastic, coupled, and situated on a surface that is moving at a very low velocity. Interactions among the blocks and also their interaction with the surface produce stick-slip behavior that resembles most of the observed properties of earthquakes, including the sudden release of stress and clustering of events, as well as earthquake magnitude distribution. As a result of its simplicity and ability to capture major characteristics of faults, the BK model has been extensively employed in studying statistical characteristics of earthquakes. The one-dimensional version of the BK model is made up of a line of identical blocks, each with mass $\mathcal{M}$, spaced a distance $a$ apart. Each block is connected to the blocks next to it using springs that are stiff with a stiffness $\mathcal{K}_c$, and to a moving plate at the bottom using springs that have a stiffness $\mathcal{K}_p$. The movement of this plate happens at a constant speed $V$. The blocks also interact with the rough surface beneath them through a frictional force, $\phi$, which causes the stick-slip behavior that leads to events like earthquakes (see Fig.~\ref{fig:1d_bk}). The distance each block moves from its original position at any given time $t$ is described by $U_i(t)$. The motion of the $i^{\mathrm{th}}$ block is explained by the equation of motion \cite{Carlson_1989_PRL,Carlson_1989_PRA,kawamura2012}

\begin{equation}
\mathcal{M}\ddot{U}_i(t)
=
\mathcal{K}_c
\left[
U_{i+1}(t)-2U_i(t)+U_{i-1}(t)
\right]
-
\mathcal{K}_p
\left[
U_i(t)-Vt
\right]
-
\phi(\dot{U}_i),
\label{eq:bk}
\end{equation}

\noindent The first part on the right side shows how blocks push against each other elastically. The second part is the force that pulls the blocks back due to the loader plate moving steadily. The last part is the resistance from friction between the blocks and the surface. The balance between the elastic push and the friction that changes with speed causes blocks to sometimes stick and then suddenly slip. The friction force is given by \cite{Carlson_1989_PRA, kawamura2012}

\begin{equation*}
\phi(\dot{U})=
\begin{cases}
(-\infty,\,1], & \dot{U}\leq 0,\\[2ex]
\dfrac{1-\delta_f}
{1+\dfrac{2\alpha\dot{U}}{1-\delta_f}},
& \dot{U}>0,
\end{cases}
\label{eq:friction}
\end{equation*}

\noindent In the above relation $\delta_f$ and $\alpha$ denote friction drop parameter and the velocity weakening parameter. $\delta_f$ parameter determines how fast the friction force drops right after sliding starts and $\alpha$ controls how fast the friction decreases if we increase the sliding speed. Together, these two parameters control the stick-slip behavior and the statistical features of the earthquake sequences created by the model. For numerical simulations, we can rewrite equation (\ref{eq:bk}) in a dimensionless form by using the characteristic frequency $\omega_p = \sqrt{\mathcal{K}_p / \mathcal{M}}$, the characteristic displacement $\mathcal{D} = \phi_0 / \mathcal{K}_p$, the dimensionless displacement $u_i = U_i / \mathcal{D}$, the dimensionless time $t' = \omega_p t$, the dimensionless loading rate $\mathcal{V} = V / (\mathcal{D} \omega_p)$, and the dimensionless stiffness parameter $l = \sqrt{\mathcal{K}_c / \mathcal{K}_p} = \xi / a$ (see \cite{Carlson_1989_PRA, Carlson_1989_PRL, kawamura2012} for more details).

\begin{figure}[h!]
\begin{center}
    \includegraphics[width=1.0\textwidth]{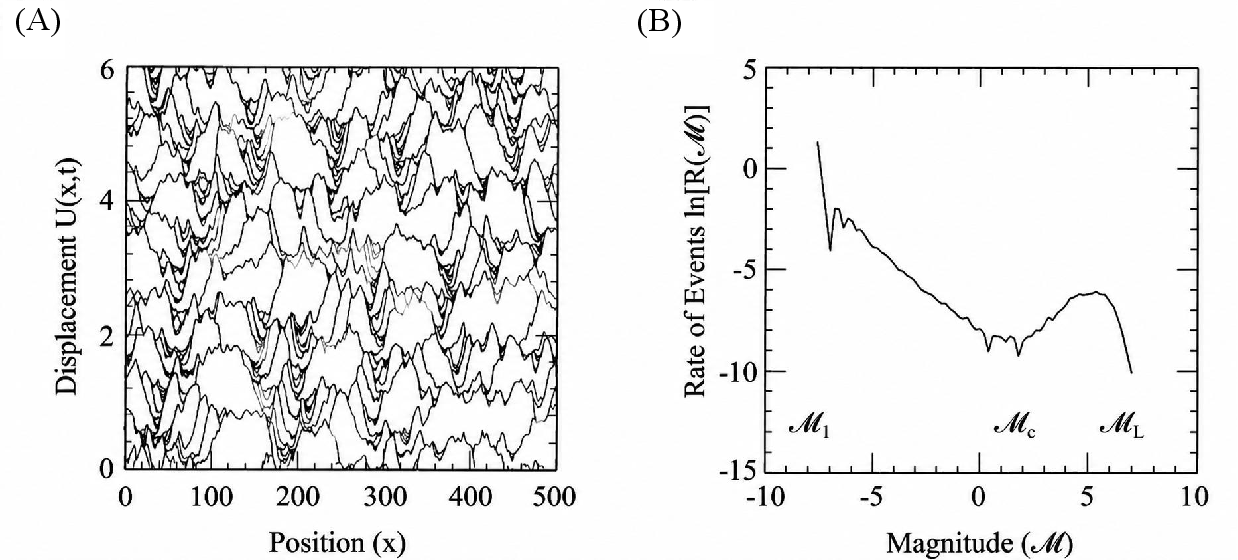}
    \end{center}
    \caption{\textbf{(A)} The displacement $U(x,t)$ for a sequence of events is shown right after each event, as a function of position $x$ along the fault. The lowest curve shows the earliest configuration after the initial phase, when the system has settled into a steady state. The following curves show the different stuck configurations. The seismic moment of each event is defined by the area between two consecutive curves. Even though the model is spatially uniform, the behavior is very complex. Small clusters occur frequently in local depressions in the displacement distribution, forming regions where larger clusters can form in the future. \textbf{(B)} The magnitude-frequency distribution for the uniform BK model exhibits a clear distinction between small and large earthquakes characterized by the crossover magnitude $\mathcal{M}_c$, which represents the maximal magnitude of the small clusters. Closer examination of the data reveals that the magnitude of the largest earthquake depends on the wavelength cut-off. In particular, when $N$ is sufficiently large, the crossover and largest magnitudes $\mathcal{M}_c$ and $\mathcal{M}_L$ do not depend on the system size anymore. However, $\mathcal{M}_c$ does not depend on the spacing between blocks $a$, but the largest-event magnitude increases as the mesh becomes finer, and it scales roughly as $\mathcal{M}_L \sim \ln (\xi^2/a)$.  The lower limit of the small-event range is marked by $\mathcal{M}_1$. [Adapted from \cite{Carlson1994}]}
    \label{fig:carlson}
    \end{figure}
   
Figure \ref{fig:carlson}(A) illustrates the evolution of fault displacement following each earthquake event in the BK model. The distance between two adjacent curves represents the energy released during each earthquake. It can be observed from the figure that frequent small quakes tend to smooth out small dips in the fault line, whereas rare large earthquakes result in huge displacements and account for majority of the slip along the fault line. Furthermore, it indicates that large earthquakes initiate at locations where small earthquakes are common, indicating that there exists some relationship between the movement of the fault at various points. Figure \ref{fig:carlson}(B) represents size and recurrence of earthquakes from the BK model, considering a very large system over a long duration of time. Smaller earthquakes exhibit the power-law behavior described by GR over a broad range of magnitudes, where the exponent $b$ equals roughly $1$ if the coefficient of friction $\alpha$ is large. Beyond the crossover magnitude $\mathcal{M}_c$ point, there are more larger events than expected from the GR law, implying that these earthquakes are characteristic earthquake events. The maximum magnitude is $\mathcal{M}_L$, which is independent of the size of the system once the system size is large enough.

Over the years, there are many improvements made to the BK model to make it more realistic. These improvements include two-dimensional BK models that help study how ruptures spread and how faults interact \cite{Carlson_1989_PRL, Carlson_1989_PRA, Carlson_time_interval_1991,Carlson_BK_2D_1991, Myers_1993, Carlson1994, Shaw_1994, Myers_1996, Kawamura_2018_2d_BK}; models that use more general elastic and nonlinear properties with vector and self-similar elasticity \cite{Schmittbuhl_1996, AKISHIN_2000_BK}; viscoelastic and slider-block models that include standard linear solid (SLS) elements and other effects to show aftershocks and delayed stress release \cite{rundle_2003_bk, Petrillo_2020, Sliderblock_Shcherbavok_2023}; and rate-and-state friction models used to understand how earthquakes start, how ruptures spread, and the statistical behavior of seismic events \cite{xia_2005_BK, Mori_kawamura_2006_bk, Kawamura_2008, Ueda_2015_BK, Kawamura_2018_2d_BK}; as well as generalized friction laws and spring-block models that better represent real friction behavior \cite{Jagla_2010_BK, Braun_2018_BK_generalised}. For more detailed reviews on spring-block and BK models, earthquake physics, and the statistical properties of earthquakes, see \cite{Pelletier_2000_BK, Ben_zion_2008, kawamura2012, Rundle_Turcotte_complexity_and_eq_2015, Hainzl_BK_2000}.

\subsubsection{Olami-Feder-Christensen (OFC) model:}
\label{subsec:ofc_model}

\begin{figure}[h!]
\begin{center}
    \includegraphics[width=1.0\textwidth]{ofc.eps}
    \end{center}
    \caption{\textbf{(A)} Schematic representation of the two-dimensional Burridge–Knopoff (BK) spring-block model, showing a block at a position \(U_{i,j}\) connected to its four nearest neighbors by coupling springs. The black dots indicate the initial stable positions of the blocks before the application of the driving force. This spring-block representation provides a mechanical framework for understanding the stress redistribution in the OFC dynamics \textbf{(B)} Avalanche-size distribution and the corresponding exponent for different values of the OFC stress-transfer parameter \(\alpha_t\). For \(\alpha_t=0.25\), the bulk redistribution rule is conservative, whereas for \(\alpha_t<0.25\), it is non-conservative. [Adapted from \cite{Kawamura_2010_ofc}] }
    \label{fig:ofc}
    \end{figure}
    
The Olami–Feder–Christensen (OFC) model \cite{ofc} is a cellular automaton model introduced to study earthquake dynamics and self-organized criticality (SOC). In this model, a two-dimensional lattice of sites evolves through slow driving and rapid relaxation events. When the stress at a site reaches a prescribed threshold, the site fails and redistributes stress to its nearest neighbors. Depending on the stress-transfer parameter, this redistribution can be conservative or non-conservative, allowing the model to investigate the emergence of scale-invariant avalanche statistics. The OFC model can be related to the two-dimensional Burridge–Knopoff (BK) spring-block model, which provides a mechanical framework for understanding the stress redistribution underlying the OFC dynamics \cite{Carlson_BK_2D_1991,Kawamura_2018_2d_BK}.

To establish the connection between the OFC model and the 2D BK model of earthquake dynamics, we consider a square lattice of blocks indexed by $(i,j)$. Each block is connected to its four nearest neighbors through springs with elastic constants $\mathcal{K}_1$ in the horizontal direction and $\mathcal{K}_2$ in the vertical direction, as illustrated in Fig. \ref{fig:ofc}(A). Each block is also connected to a uniformly driven plate by a loader spring of stiffness $\mathcal{K}_L$. Let $U_{i,j}$ denote the displacement of the block $(i,j)$ from its relaxed position. The elastic-force expression for the standard 2D BK model is given by \cite{ofc}

\begin{equation*}
F_{i,j}=
\mathcal{K}_1\bigl(2U_{i,j}-U_{i-1,j}-U_{i+1,j}\bigr)+\mathcal{K}_2\bigl(2U_{i,j}-U_{i,j-1}-U_{i,j+1}\bigr)
+\mathcal{K}_L U_{i,j}.
\label{eq:2D_BK_force}
\end{equation*}

\noindent A block slips when its force reaches or exceeds the static-friction threshold. Suppose the block $(i,j)$ becomes unstable and slips by a positive displacement magnitude $\Delta U_{i,j}$ in the direction that reduces its force. Assuming that the neighboring blocks remain fixed during this instantaneous slip, the magnitude of the force reduction is $\Delta F_{i,j}=-\bigl(2\mathcal{K}_1+2\mathcal{K}_2+\mathcal{K}_L\bigr) \Delta U_{i,j}$. Under the assumption that the slipping block relaxes to zero force, we have $F_{i,j}+\Delta F_{i,j}=0$. Consequently, the slip magnitude is $\Delta U_{i,j}=\frac{F_{i,j}}{2\mathcal{K}_1+2\mathcal{K}_2+\mathcal{K}_L}$. The slip changes the extensions of the springs connecting the failing block to its nearest neighbors, thereby redistributing the elastic forces across the lattice. For a horizontal neighbor $(i\pm1,j)$, the magnitude of the force transferred is

\begin{equation*}
\begin{aligned}
\Delta F_{i\pm1,j}
&=\mathcal{K}_1\Delta U_{i,j}=\frac{\mathcal{K}_1}
{2\mathcal{K}_1+2\mathcal{K}_2+\mathcal{K}_L}
F_{i,j}.
\end{aligned}
\label{eq:horizontal_transfer}
\end{equation*}

\noindent Similarly, for a vertical neighbor $(i,j\pm1)$, the transferred force is

\begin{equation*}
\begin{aligned}
\Delta F_{i,j\pm1}
&=\mathcal{K}_2\Delta U_{i,j}=\frac{\mathcal{K}_2}
{2\mathcal{K}_1+2\mathcal{K}_2+\mathcal{K}_L}
F_{i,j}.
\end{aligned}
\label{eq:vertical_transfer}
\end{equation*}

\noindent We therefore define the dimensionless redistribution parameters

\begin{equation*}
\alpha_{1t}
=\frac{\mathcal{K}_1}
{2\mathcal{K}_1+2\mathcal{K}_2+\mathcal{K}_L},
\qquad
\alpha_{2t}
=\frac{\mathcal{K}_2}
{2\mathcal{K}_1+2\mathcal{K}_2+\mathcal{K}_L}.
\label{eq:alpha12}
\end{equation*}

\noindent The corresponding OFC relaxation rules can then be expressed as follows: (i) the force on the failing block is reset to zero, $F_{i,j}\rightarrow0$; (ii) each horizontal neighbor receives a fraction $\alpha_{1t}$ of the force released by the failing block, so that $F_{i\pm1,j}\rightarrow F_{i\pm1,j}+\alpha_{1t}F_{i,j}$; and (iii) each vertical neighbor receives a fraction $\alpha_{2t}$, so that $F_{i,j\pm1}\rightarrow F_{i,j\pm1}+\alpha_{2t}F_{i,j}$. These rules provide the connection between the elastic stress redistribution in the idealized BK model and the threshold-based dynamics of the OFC model.

For the isotropic case, in which $\mathcal{K}_1=\mathcal{K}_2=\mathcal{K}_c$, the two redistribution parameters become equal, $\alpha_{1t}=\alpha_{2t}=\alpha_t$, giving

\begin{equation}
\alpha_t=\frac{\mathcal{K}_c}{4\mathcal{K}_c+\mathcal{K}_L}.
\label{eq:alpha_t}
\end{equation}

\noindent The total fraction of force redistributed to the four nearest neighbors is $4\alpha_t$, while the fraction not transferred to these neighbors is $1-4\alpha_t=\frac{\mathcal{K}_L}{4\mathcal{K}_c+\mathcal{K}_L}$. Using Eq. \ref{eq:alpha_t}; in the limit $\mathcal{K}_L=0$, we obtain $\alpha_t=0.25$, corresponding to conservative bulk redistribution; for $\mathcal{K}_L>0$, we have $4\alpha_t<1$, and the redistribution becomes non-conservative; when $\mathcal{K}_c=\mathcal{K}_L$, the redistribution parameter is $\alpha_t=0.20$, and the fraction not transferred to the nearest neighbors is $1-4\alpha_t=0.20$. Thus, the loader-spring stiffness controls the degree of non-conservation in the effective OFC relaxation rule.

The avalanche of size ($s$) in the OFC model is described as the chain of block ruptures that starts when a block becomes unstable and ends once all blocks return to their stable state. The same block can rupture more than once within one avalanche, and each of the ruptures is considered in determining the avalanche size. The original OFC paper describes the statistics of released earthquake energy through the exponent $B$, which is dependent on the redistribution parameter and therefore on the conservation parameter \cite{ofc}. The scaling for the cumulative energy distribution is then, $N(E_0>E)\sim E^{-B}$ while for the probability density function it is $P(E)\sim E^{-(1+B)}$. The connection between these exponents and the Gutenberg-Richter $b$ value is conditioned by the magnitude-energy scaling relation. More details on the OFC model can be found in \cite{Hargarten_2002_ofc, Helmstetter_ofc_2004, Kotani_ofc_2008, Kawamura_2010_ofc}.

\subsubsection{One-Dimensional Train Model:}
\label{subsec:train_model}

The one-dimensional train model, which is a modified version of the one-dimensional BK model \cite{vieira} consists of an array of blocks of identical mass, arranged on a discrete lattice and connected by identical Hookean springs of spring constant $\mathcal{K}$. The system is driven quasistatically by pulling the block at one end of the chain (here, the rightmost block). The driving continues until the system becomes unstable, at which point the external pulling is halted while the resulting relaxation process takes place. Once all block movements cease and the system reaches a stable configuration, the pulling resumes. Repetition of this loading–relaxation cycle produces intermittent stick-slip dynamics after an initial transient period. Each interior block is subjected to three forces: elastic forces exerted by the two neighboring springs and a frictional force arising from its contact with the underlying surface.

\begin{figure}[h!]
\begin{center}
    \includegraphics[width=0.60\textwidth]{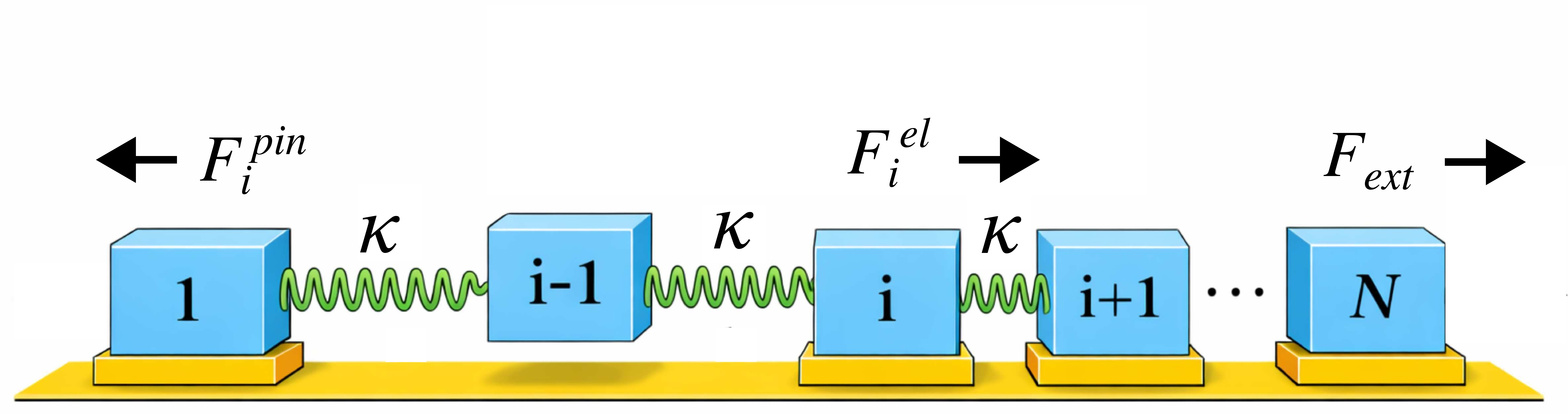}
    \end{center}
    \caption{Schematic representation of the modified one-dimensional train model. Random pinning force (chosen between $0$ to $1$) acts on all blocks in contact with the lattice sites (indicated by yellow plates), except for the $(i-1)$-th block, which is not in contact with any of the  lattice sites.}
    \label{fig:1d_train}
    \end{figure}

In the modified version of the train model considered here (see \cite{biswas2013}), the frictional force is replaced by a random pinning force. This simplification is motivated by the fact that the friction between two contacting surfaces depends on their local microscopic properties and therefore changes whenever a block establishes a new contact. Accordingly, each time a block slides to a new lattice site, it is assigned a new pinning force, chosen randomly from a uniform distribution between 0 and 1 (see Fig. \ref{fig:1d_train}). Unlike an earlier approach \cite{Chianca2009}, where the friction force was determined explicitly from the roughness of the contacting surfaces. The dynamics of the train model are governed by the competition between the elastic restoring forces and the local pinning force. Let $r_i(t)$ the position of the $i^{th}$ block be at time $t$. The total force acting on this block is given by \cite{biswas2013}

\begin{equation*}
F_i^{tot}(t)=F_i^{el}(t)+C\{r_i(t)\}F_i^{pin},
\label{eq:train_total}
\end{equation*}

\noindent where $F_i^{el}(t)$ is the net elastic force arising from the two neighboring springs, $F_i^{pin}$ is the local pinning force that opposes the direction of motion and $C\{r_i(t)\}$ is a binary function that determines whether the block is in contact with a pinning site.

\begin{equation*}
C\{(r_i)\}=
\begin{cases}
1, & \text{if a pinning site exists beneath the block,} \\
0, & \text{otherwise.}
\end{cases}
\end{equation*}

\noindent Thus, the pinning force acts only when the block is in contact with the lower chain. The elastic force is determined by the extensions and compressions of the neighboring springs and is given by \cite{biswas2013}

\begin{equation*}
F_i^{el}(t)=\mathcal{K}\left[r_{i+1}(t)-r_i(t)-R_{i,i+1}^{0}\right]
+\mathcal{K}\left[R_{i-1,i}^{0}- r_i(t)-r_{i-1}(t))\right],
\label{eq:train_elastic_force}
\end{equation*}

\noindent where $\mathcal{K}$ is the spring constant and $R_{i,i+1}^{0}$ denotes the equilibrium separation between the $i^{th}$ and $(i+1)^{th}$ blocks. At each update, the total force on a block is checked. If the total force is greater than zero, the block moves forward by one unit, but only if the next position is empty. After each move, a new random pinning force is given to the block from a range between $0$ and $1$, showing the block has made a new connection with the ground. The blocks that are located on the sides behave differently. While the leftmost block is fixed with only one spring, thus having force reduced, the rightmost one is the driving mechanism, which moves slowly. Forces will be added to the mechanism until it destabilizes, after which the driving will be stopped and the release of energy takes place through the sudden displacement of parts. When all actions stop and equilibrium is established, the slow loading will begin again.

\begin{figure}[h!]
\begin{center}
\includegraphics[width=1.0\textwidth]{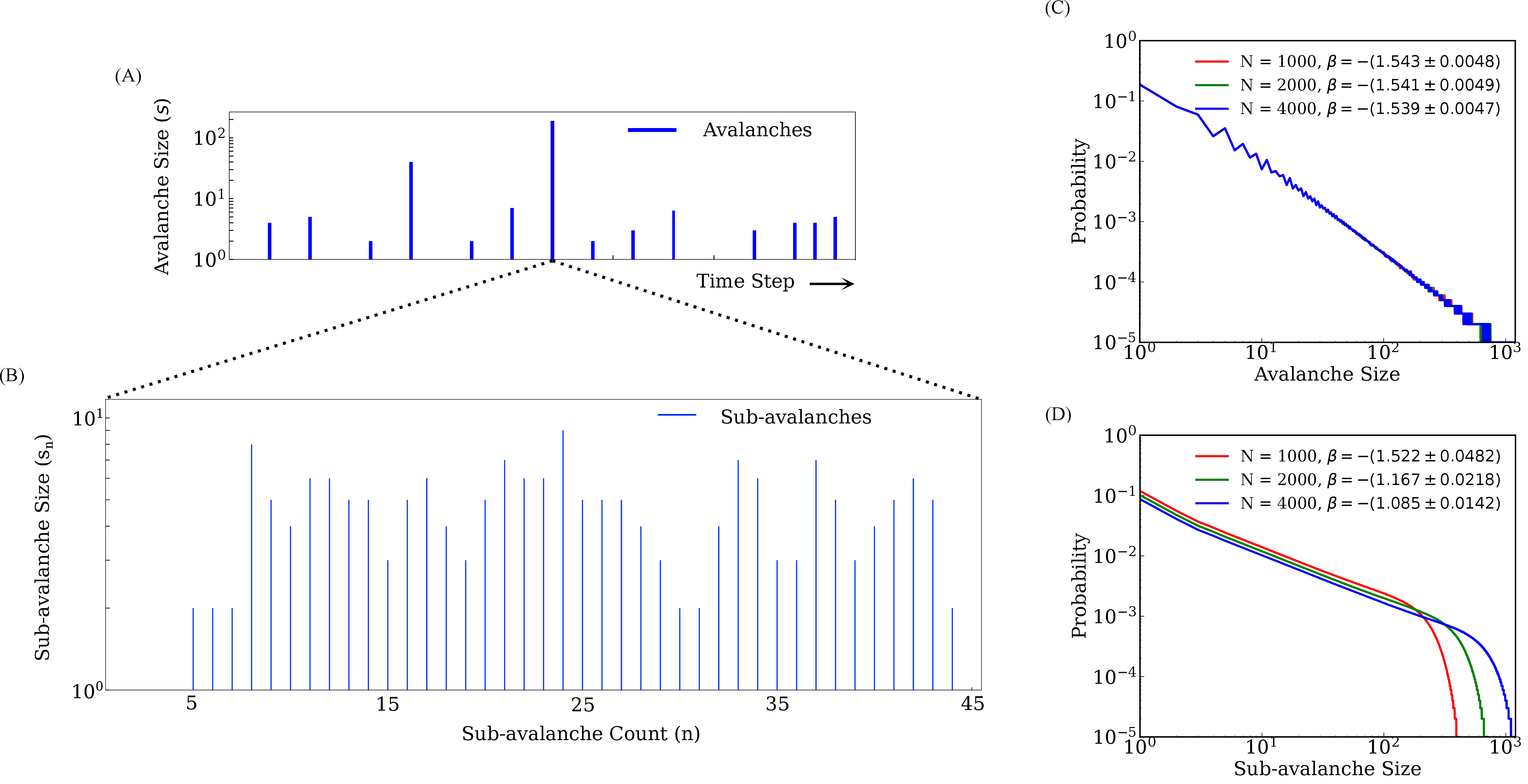}
\end{center}
\caption{\textbf{(A)} Schematic of avalanche size time series generated using the 1D train model with modification, depicting a small portion of the time series rather than the full time series. \textbf{(B)} Schematic of sub-avalanche size time series associated with the maximum-sized avalanche depicted in \textbf{(A)}, indicating that an avalanche is comprised of several sub-avalanches. \textbf{(C)} and \textbf{(D)} Avalanche and sub-avalanche size distributions for varying system sizes $N$, displaying power law scaling with the exponent $\beta$ as discussed in Eq. (\ref{eq:n_beta}).}
\label{fig:train_dist}
\end{figure}

In the train model, an ``avalanche" ($s$) starts when an external force $F_{ext}$ is applied on the rightmost block. This causes a chain reaction of blocks sliding and moving. The sliding continues until the system reaches a stable state. The size of the avalanche ($s$) is the total number of blocks that move up to the stable state. The avalanche is made up of smaller parts called ``sub-avalanches" ($s_n$), each showing a single step or wave of movement. Simply, the total avalanche size is the sum of the sizes of all sub-avalanches $(s=\sum_n s_n)$, as shown in Figs. \ref{fig:train_dist}(A) and (B). The distributions of avalanche and sub-avalanche sizes are shown in Figs. \ref{fig:train_dist}(C) and (D), respectively.

In \cite{biswas2013}, it was shown that the modified train model is mathematically the same as the boundary-driven Edwards-Wilkinson (EW) model for how interfaces grow or move (or depin).
This link shows that earthquake behavior can be connected to how interfaces move through rough materials.

\subsubsection{Two-Fractal Overlap (CS) Model:}
\label{subsec:CS_model}

\begin{figure}[h!]
\begin{center}
\includegraphics[width=1.0\textwidth]{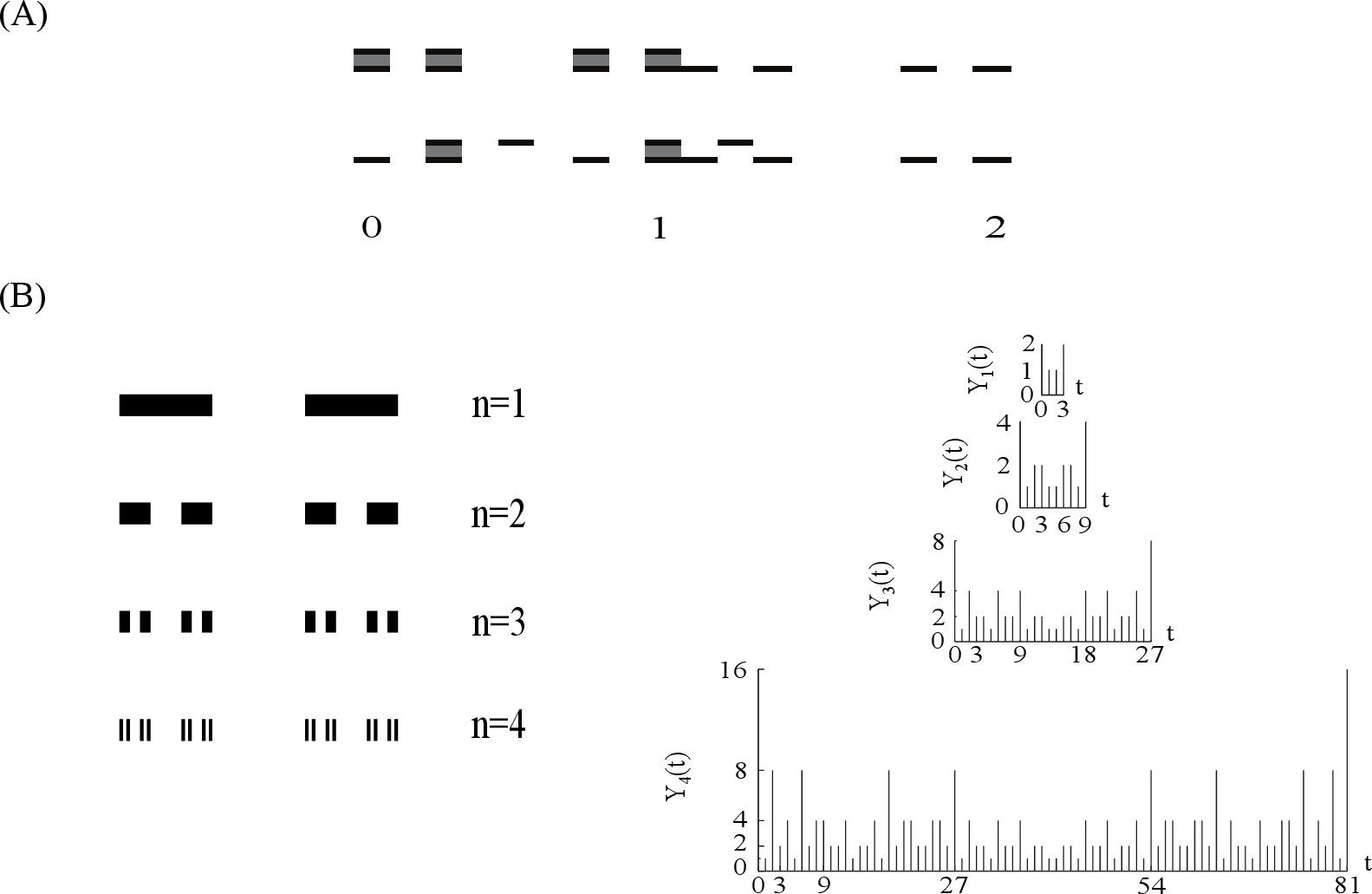}
\end{center}
\caption{\textbf{(A)} Recursive structure of the overlap time series for the first four generations of the two-fractal overlap model. The overlap time series of every lower generation is embedded within that of the higher generation. \textbf{(B)} Illustration of the second-generation Cantor sets at $t=0$ and $t=2$. The overlapping intervals are shaded in gray. The lower Cantor set is periodically repeated to implement periodic boundary conditions, while the upper Cantor set slides uniformly over it. [Adapted from \cite{Bhattacharya_2011}]}
\label{fig:cantor}
\end{figure}

The two-fractal overlap model (also known as the Chakrabarti-Stinchcombe (CS) model \cite{ChakrabartiStinchcombe1999}) provides a simple geometric description of earthquake dynamics based on the fractal nature of fault surfaces (see, e.g., \cite{Bhattacharya_2011}). In this model, the fractured surfaces of the earth's crust and the underlying tectonic plate are represented by two identical Cantor sets of generation $n$. One Cantor set slides uniformly over its replica under periodic boundary conditions, thereby generating a time series of overlaps $Y_n(t)$ (see Fig.~\ref{fig:cantor}). The overlap between the two fractals represents the contact area between the fault surfaces and is assumed to be proportional to the elastic energy accumulated during the sticking phase and released during the subsequent slip event (earthquake). At the $n$-th generation, the overlap magnitude is defined as the number of overlapping occupied intervals between the two Cantor sets. Since the Cantor set contains $2^n$ occupied intervals, the overlap can assume only discrete values belonging to the geometric sequence \cite{bhattacharyya2005}

\begin{equation*}
    Y_n(t)=2^{\,n-q}, \qquad q=0,1,\ldots,n.
\end{equation*}

\noindent Initially, all occupied intervals overlap, giving the maximum overlap $Y_n(0)=2^n$. The motion is discretized by taking one unit of time as the time required to translate the upper Cantor set by one elementary interval of length $\frac{1}{3^n}$. Owing to the periodic boundary conditions and the mirror symmetry of the finite-generation Cantor set, the overlap time series satisfies

\begin{equation*}
    Y_n(t)=Y_n(3^n-t), \qquad 0\le t\le3^n.
\end{equation*}

\noindent One key aspect of the overlap dynamics is how it works recursively. The overlap time series for generation $n$ includes, as part of it, the full overlap time series from all earlier generations. This layered structure shows the self-similar pattern of the Cantor set and creates a sequence of overlapping sizes that are nested within each other. By using this recursive feature along with basic counting methods, the exact probability of different overlap sizes can be calculated, as shown in \cite{Bhattacharya_2011}. Since the overlap size is thought to be related to the energy released during an earthquake, a measure similar to earthquake magnitude is defined as $M=\log_2(Y_n)=n-q$, where $q=0,1,\ldots,n$. The probability distribution for these earthquake magnitudes follows a binomial distribution

\begin{equation}
\mathcal{B} (M)=
\binom{n}{M}
\left(\frac{1}{3}\right)^M
\left(\frac{2}{3}\right)^{n-M},
\label{eq:binomial}
\end{equation}

\noindent which follows from the recursive overlap statistics of the two Cantor sets. For a large generation number $n$, the binomial distribution approaches a Gaussian distribution through the central limit theorem. Integrating this Gaussian distribution to obtain the cumulative frequency of events and employing the asymptotic expansion of the complementary error function for large magnitudes lead to the cumulative frequency-magnitude relation \cite{Bhattacharya_2011}

\begin{equation}
\log N(\ge M)=A-\frac{3}{4}M-\log\left(M-\frac{n}{3}\right),
\label{eq:cantor}
\end{equation}

\noindent where $N(\ge M)$ denotes the cumulative number of earthquakes with magnitude greater than or equal to $M$, and $A$ is a constant that depends only on the generation number of the Cantor set. For sufficiently large magnitudes, the logarithmic correction varies slowly compared with the linear term, yielding \cite{Bhattacharya_2011}

\begin{equation}
    \log N(\ge M)\simeq A-\frac{3}{4}M,
    \label{eq:cantor_2}
\end{equation}

\noindent which has the same functional form as the empirical GR relation as mentioned in Eq. (\ref{eq:gr}). Thus, the CS model is naturally consistent with the GR scaling and the theoretical $b$ exponent value $b=\frac{3}{4}$. In general, the theoretical value of the parameter $b$ depends on the fractal dimension of fault surfaces in question, which implies that the regional differences in the GR exponent can be due to the fault surface geometry. The constant $A$ in Eqs.(\ref{eq:cantor}) and (\ref{eq:cantor_2}) depends on the generation number $n$ of the Cantor set. Given a constant fractal dimension, the larger the generation number, $n$ the more overlap events occur and, consequently, the greater the number of seismicity episodes generated by the model. Mathematically, the constant $A$ works similarly to the parameter $a$ in the GR law as given in Eq.(\ref{eq:gr}), and it shows the level of seismic activity in the system \cite{Bhattacharya_2011}.


\section{The Gutenberg–Richter Exponent as a Potential Earthquake Precursor and its limitations}
\label{sec:forecasting_with_b}

\begin{figure}[h!]
\begin{center}
\includegraphics[width=0.85\textwidth]{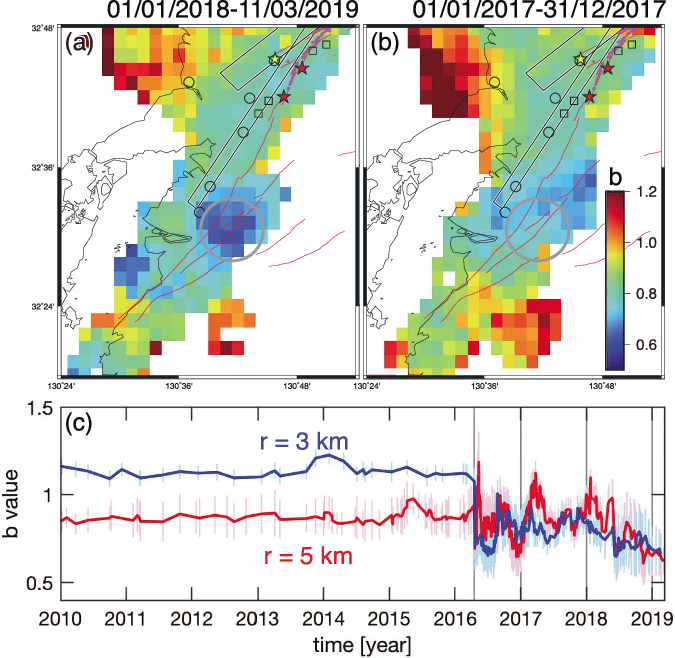}
\end{center}

\caption{\textbf{(A)} Represents a spatial map of the $b$ value distribution using the earthquake data from 1 January 2018 to 11 March 2019 of the region around the southern part of the northern Futagawa‐Hinagu fault zone in Japan. Clear localized low $b$ value anomalies are seen near the southern end of the activated northern section of the Hinagu fault. The low $b$ value region becomes much more concentrated spatially in the 2018 onward map compared with the preceding years. \textbf{(B)} It shows the corresponding $b$ value map for 2017 (covering 1 January to 31 December). In contrast to panel \textbf{(A)}, the low $b$ values are distributed over a bordered region. The comparison of panels \textbf{(A)} and \textbf{(B)} shows a clear spatial localization of the low $b$ value with time. The circles marked in the two maps denote the selected region for the temporal analysis presented in Fig. \textbf{(C)} where the time-dependent $b$ value analysis for the earthquakes taking place in a cylindrical region of radius $r=5$ km (denoted in red) and $r=3$ km (denoted in blue) is done. Here, the $b$ value has been computed using a window of $100$ consecutive earthquakes, and each calculated $b$ value has been plotted against the time corresponding to the last earthquake in the respective window. The vertical bars correspond to the uncertainty of the $b$ values. As seen from the figure, at the onset of the 2016 Kumamoto earthquake sequence, denoted by the first vertical gray line, the temporal behavior of $b$ value shows clear variations due to the presence of many aftershocks. The second, third, and fourth gray lines represent 1 January of 2017, 2018, and 2019, respectively, and serve as a guide for the two periods taken for the spatial $b$ value maps. After showing sharp variations due to the 2016 sequence, the $b$ value starts decreasing with time and attains the value of $b\sim 0.6$. [From \cite{nanjo_2019}]}
\label{fig:nanjo_2019}
\end{figure}

Among the various conventional approaches to earthquake forecasting, the spatio-temporal variation of the Gutenberg-Richter (GR) exponent ($b$ value) is one of the most widely studied statistical precursors. A physical basis for the use of $b$ value in earthquake forecasting was first established in the work of Scholz \cite{scholz} by demonstrating that laboratory micro-fractures obey a frequency-magnitude relation analogous to the GR law. This result demonstrated that the $b$ value depends on the stress conditions, with its variations being primarily related to the applied stress. Further studies examining the relation between the earthquake frequency distribution and crustal stress \cite{wyss_1973} demonstrated that lower $b$ values are associated with higher stress values (see also \cite{Schorlemmer_2005}). The possibility that the temporal variation of $b$ value could further give information about an impending earthquake was explicitly explored in early precursor studies. In \cite{Smith_1981}, they examined the earthquake catalogs from New Zealand together with the 1967 Caracus and 1971 San Fernando earthquake sequences and found systematic temporal changes in the $b$ value preceding several large earthquakes. They found that the $b$ value increased somewhat from its background level and then decreased back to the background level before the large events, which suggests that the evolution of the frequency-magnitude distribution could have some information about the temporal evolution of the earthquake cycle. Further, other works \cite{imoto_1991,gao_2002,hirose_2002,nuannin_2005} reinforced the interest in the possibility that a decrease in the $b$ value might be associated with stress accumulation before large events. In the work \cite{nanjo_2012} this hypothesis was reinforced by the reporting of a marked $b$ value decrease on a decade scale in the vicinity of the hypocenters of both the Sumatra earthquake (2004) and the Tohoku earthquake (2011). The similarity of the observed temporal evolution in these independent regions provided was interpreted as evidence that the $b$ value could provide an indicator of the approach to the large event. Another recent study related to the Kumamoto earthquake sequences in Japan \cite{nanjo_2019} showed that after the strong fluctuations after the 2016 sequence, the $b$ value progressively decreases with time (as shown in Fig. \ref{fig:nanjo_2019}). Some other studies in different regions and fault zones (see \cite{huang_2015, nanjo_2016, yoshida_2017,nanjo_2017,nanjo_2021}) also gave the same observations. These observations motivated the development of $b$ value-based approaches to real-time earthquake forecasting.

Recently, a foreshock traffic-light system (FTLS) based on changes in the Gutenberg-Richter $b$ value following moderate-to-large earthquakes \cite{Gulia_2019}. The method exploits the observation that ordinary aftershock sequences often exhibit an increase in $b$ relative to the background value, whereas sequences subsequently followed by a larger earthquake may exhibit a smaller increase or a decrease. Nonetheless, these variations should not be taken as direct evidence for an increased level of stress or a coming large earthquake. The validity of the $b$-values depends on many factors, including catalog completeness, number of samples, magnitude span, estimation methods, and spatial and temporal scale choices \cite{Marzocchi2020}. In addition, the magnitude-frequency distribution can be different even among nearby seismicity patches and can change differently in various parts of the earthquake sequence \cite{Herrmann2022}. It is important to confirm first the statistical significance of any variation to draw conclusions about its physical or predictive nature. For instance, according to some study, it was shown that there are significant spatial but not significant temporal variations in the Parkfield region, despite a visible temporal trend \cite{Mirwald2024}. For the prospective study of CSEP-Italy using five-year forecasting models that were driven by instrumental seismicity and spatial changes in the $b$ value, performance was low overall, even though the performance varied for individual target earthquakes \cite{Taroni2018}. However, these results relate to long-term spatial prediction and are different from the FTLS, which uses temporal changes in the $b$ value in evolving aftershock sequences. Specifically, in the FTLS method, aftershock sequences are classified in accordance with the change in $b$ relative to the background value of $b$, where a reduction of at least 10\% leads to the red color of the alert, and an increase of at least 10\% leads to the green color of the alert \cite{Gulia_2019}. For the retrospective testing using 58 aftershock sequences, the classification accuracy reached 95\%. For the pseudo-prospective study using the 2019 Ridgecrest sequence, $b$ decreased by about 23\% after the $M_w 6.4$ earthquake and led to the red alert and then increased by about 26\% after the $M_w 7.1$ earthquake and led to the green alert \cite{Gulia_2020}.

In general, the estimation of $b$ represents a compromise between stability and resolution; while smaller windows are able to detect local changes, they will produce imprecise estimates in cases where there are few events. Larger windows will produce more stable estimates at the risk of missing any local or short-term variations. Furthermore, by increasing the number of events that are being included in each estimate, we may unintentionally wash out the changes that take place on a smaller time scale. A proper analysis needs to address how sensitive the results are to the amount of completeness, sample size, and choice of the window, determine the uncertainty of the estimates, and check whether the changes are statistically significant \cite{Biswas_lucas_2019_mapping,Marzocchi2020,Mirwald2024}.

\section{Other earthquake forecasting approaches and their limitations}
\label{sec:modern_forecasting}

In the past few decades, there have been numerous advancements in the area of earthquake prediction using physical modeling, laboratory experiments, statistics, and machine learning approaches that augment observations of seismicity \cite{Kagan_1997_review,Jordan_2011_review,Mignan_2014_review,Johnson_review_2021,Hardebeck_2024_review,Kubo_review_2024,Mizrahi_review}. All these methods address various dimensions of earthquake dynamics, such as damage accumulation, variations in the rates of seismic activity, earthquake triggering, temporal dependencies, and development of seismic sequences.

Acoustic Emission (AE) and damage localization indicators have been studied extensively within laboratory tests of rock fracture and friction. As the specimen approaches failure, the rates of AE might evolve, and event hypocenters might localize around the incipient fault \cite{scholz,Lockner_1993,Lei_2000,Goebel_2013}. Continuous machine learning analysis of acoustic emission and fault slip data has also proven the potential of predicting time-to-failure of laboratory specimens \cite{RouetLeduc_2017,Johnson_review_2021,Kubo_review_2024,Mizrahi_review}. Laboratory tests provide both high-resolution temporal and, in cases where there are several sensors present, also spatial information about damage evolution processes. Due to the controllable loading and failure process, it is possible to study statistical properties across many failures. However, scaling these results up to tectonic earthquakes is the biggest problem here. Natural faults vary in terms of geometry, heterogeneity, loading history, and monitoring. In some laboratory failures there is no recognizable foreshock activity at all \cite{Goebel_2023}.

The class of statistical forecasting methods includes the use of the "Epidemic-Type Aftershock Sequence" (ETAS) model as a probabilistic method. The ETAS model considers earthquake occurrences to be a self-exciting process, where earthquakes have the potential to cause further earthquakes while the seismic background activity continues. ETAS models have proven useful for predicting aftershocks and estimating short-term earthquake rates because of their explicit consideration of earthquake triggering and clustering \cite{ogata_ETAS_1988,ogata_ETAS_1998,Jordan_2011_review,Mizrahi_review}. These models usually use earthquake time and magnitude as input parameters. ETAS models have been extensively studied via both retrospective and prospective analysis, with certain forms of them having succeeded in various short-term forecast experiments \cite{Helmstetter_2006, Tsuruoka_2012, Taroni2018, Mizrahi_review}. However, their success is contingent on correct parameter estimation, completeness of the catalog used, and the specification of the model itself. For instance, short-term aftershock incompleteness following major events can lead to erroneous estimation of parameters and thus the resulting forecast \cite{Hainzl_2016}. Besides that, while ETAS successfully models the statistics of earthquake occurrence and triggering, it does not monitor the dynamic state of a fault.

Another significant advancement in earthquake forecasting is ``Operational Earthquake Forecasting'' (OEF). This approach aims at generating time-dependent probabilistic forecasts of future seismicity based on continually collected data. Contrary to deterministic approaches, the OEF approach evaluates the time dependence of earthquake probabilities in specific regions for certain magnitude ranges and time periods. OEF can include statistical, physical, and machine learning models, which means that this is not just a particular forecast model but a forecasting approach \cite{Jordan_2011_review,Hardebeck_2024_review}. It is important to be able to objectively assess the forecasting performance. That is why ``Collaboratory for the Study of Earthquake Predictability'' (CSEP) was founded for that purpose \cite{Jordan_2011_review,Schorlemmer_2010,Zechar_2010,Schorlemmer_CSEP_2018}. These experiments rely on target areas, magnitude thresholds, and time windows, combined with statistical tests like the N-test, L-test, S-test, and M-test, as applicable. Information gain measures and likelihood ratio tests could also be utilized to benchmark forecasts against reference models. CSEP experiments have been carried out using earthquake catalogs from California, Japan, Italy, New Zealand, and elsewhere. Systems have also been implemented in various nations, with the Italian system generating probabilistic forecasts in connection with the 2016--2017 Amatrice-Norcia earthquake sequence \cite{Marzocchi_2017,Mizrahi_review}. However, probabilities of occurrence assigned to extreme earthquakes are typically small, making their practical implications difficult to ascertain. Data quality and reporting practices also vary regionally \cite{Zechar_2016_OEF,Mizrahi_review}, making consistent implementation of OEF problematic. Nonetheless, OEF serves as an experimental methodology and communication standard for earthquake forecasts, although individual model performance should be established independently.

The next methodology is ``Natural Time Analysis,'' which was introduced by Varotsos and his collaborators. This approach describes earthquake sequences in terms of a new time scale, which takes into account the order of events rather than traditional clock time. By combining the sequence of events with their energy, natural time analysis uses parameters such as variance and entropy to study the evolution of seismic dynamics and possible transitions to critical states \cite{Varotsos_2005,Varotsos_2011,Varotsos_2011_NATURAL_TIME,Varotsos_2013,Varo}. These kinds of measures can be obtained from earthquake catalogs that include event times and magnitudes or energies, with the resolution depending on the number of events used. Retroactive studies have studied natural time indicators with ROC analysis, including the minimum of order parameter fluctuations before major earthquakes \cite{Sarlis_2013}. But there is an important distinction that needs to be made between this retrospective study of discrimination and independent prospective verification. As far as the related VAN \cite{VAN_1,VAN_2} method based on seismic electric signals is concerned, the controversy regarding its statistical significance, lack of clarity of the prediction rules, and the design of the validation process \cite{Geller_1996, Mulargia_Gasperini_1996,Geller_1997} is well known. These controversies emphasize the need for advance definition and testing of the performance of candidate signals and alarm rules using independent data.

Earthquake nowcasting offers another related but different way to investigate the process of evolution of earthquake cycles. As opposed to estimating the exact time of the future earthquake, nowcasting attempts to determine how mature the current earthquake cycle is based on the statistics of smaller earthquakes that happen between large earthquakes. One such indicator is called the Earthquake Potential Score (EPS), which determines where the current number of events is positioned compared to the distribution of numbers of events in previous earthquake cycles \cite{Rundle_2016_NOWCASTING,Rundle_2018_NOWCASTING_NATURAL_TIME,Rundle_2021_NOWCASTING,Rundle_code_NOWCASTING}. The technique becomes more interesting when there are enough numbers of historical cycles in the catalog, and the rule of thumb suggests that there should be at least 20 or more cycles. Nowcasting has been used in various tectonic environments like induced seismicity and the Bay of Bengal \cite{Luginbuhl_2018,Pasari_2019} and even information-theoretic extensions have been proposed \cite{Rundle_2019_entropy}. Statistical stability of the method depends on the quality of the catalog and its cycles' representation. Moreover, estimation of the maturity of the cycle is not a prediction of the future time of a strong event. Therefore, assessment of the technique should be done according to the purpose and with reference models.

Another option for studying the earthquake catalogs and geophysical variables is to use machine learning algorithms. Supervised and unsupervised learning, deep neural networks, graph neural networks, and physics-informed models can be applied to detect complicated patterns that would be hard to capture using standard statistics \cite{Johnson_review_2021,Kubo_review_2024,Mizrahi_review}. Depending on the application, these algorithms would require large labeled datasets, continuous seismogram data, or a combination of seismic and geophysical data. These algorithms have high resolution for sufficiently detailed data, but they rely on the representativeness of the training dataset and the distinction between the training and testing periods. The lack of a very large number of major earthquakes gives rise to the class-imbalance issue, whereas overfitting, data leakage, and regional discrepancies cause overoptimistic retrospective results. For example, a deep neural network developed to estimate aftershock locations was subsequently shown to have comparable performance to a logistic-regression model with only two parameters \cite{DeVries_2018,Mignan_Broccardo_2019}. A further assessment of neural-network earthquake-prediction studies identified widespread concerns about suitable baselines and out-of-sample testing \cite{Mignan_Broccardo_2020}. Therefore, the complexity of a machine-learning model should not be taken as evidence of superior forecasting skill. Independent temporal and regional testing, comparison with simpler alternatives such as ETAS, and prospective evaluation are necessary before operational performance can be established.

Other statistical measures of clustering, intermittency, and energy release provide additional descriptions of earthquake sequences. Cluster analysis and the scaling of the inter-event times have been studied for describing the time organization \cite{Kagan_Jackson_1991, Corral_2004} whereas nearest-neighbor methods determine clusters based on the time, space, and size dependencies between events \cite{Zaliapin_2013, Zaliapin_BenZion_2013}. Network representations have also been used to characterize the structure of earthquake triggering and interactions \cite{Baiesi_Paczuski_2004}. These methods are able to detect statistical patterns that cannot be detected using only the event rate approach, but they all have issues related to the completeness of the catalogs, the choice of events, and the windows used for the analyses. They are mostly diagnostic methods rather than predictive models. Their predictive usefulness will depend on whether the patterns detected offer any predictive information compared to a null model.

One common problem that is inherent in the above-discussed approaches is the false-alarm issue. An approach can predict periods before certain large earthquakes but will trigger alarms for many other periods where no earthquake happens. However, only the correlation between associations and ROC curves cannot fully measure such tradeoff without specifying the forecast period, alarm period, or an independent test period. For probabilistic predictions, likelihood scores and information gain can measure their performance based on reference models which have already incorporated the effects of seismic activity and clustering \cite{Schorlemmer_2010,Zechar_2010,Schorlemmer_CSEP_2018}. In the case of alarm-based methods, the Molchan diagram serves as an alternative measure by comparing the proportion of missed target events to the proportion of space-time covered by alarms. Area skill score in this case is based on comparison with a reference method \cite{Molchan_1990,Molchan_1997,Molchan_KeilisBorok_2008,Zechar_Jordan_2008,Kagan_2009}. Improvement in the trade-off between missing and false alarms will only be possible if analysis parameters and thresholds for alarms are chosen before evaluation, baseline models appropriate for the problem are used, and performance is estimated on independent data. In some cases, probabilistic forecasts might even provide more valuable information than alarms because they capture the uncertainty associated with rare events \cite{Molchan_1997,Zechar_2016_OEF,Mizrahi_review}. At present, obtaining a good hit rate without having a high false-alarm rate is one of the main problems in forecasting large earthquakes.

Given the broad range of existing forecasting approaches, it is worthwhile to investigate whether inequality measures of seismic energy release can provide complementary information about earthquake dynamics.

\section{Inequality measures of earthquake energy release}
\label{inequality_measure_in_EQ}

The Gini ($g$) and Kolkata ($k$) inequality indices can be calculated directly from the time series of seismic energy or seismic moment associated with earthquakes, without requiring the estimation of a power-law exponent from earthquake catalog data. Recent studies combining model simulations with real earthquake data suggest that large earthquakes may occur following periods of highly unequal energy release, potentially indicating changes in the state of the system \cite{g_eq, PRE_2026}. These findings motivate further investigation into whether changes in the inequality indices can provide useful information for earthquake forecasting.

\begin{figure}[h!]
\begin{center}
    \includegraphics[width=1.0\textwidth]{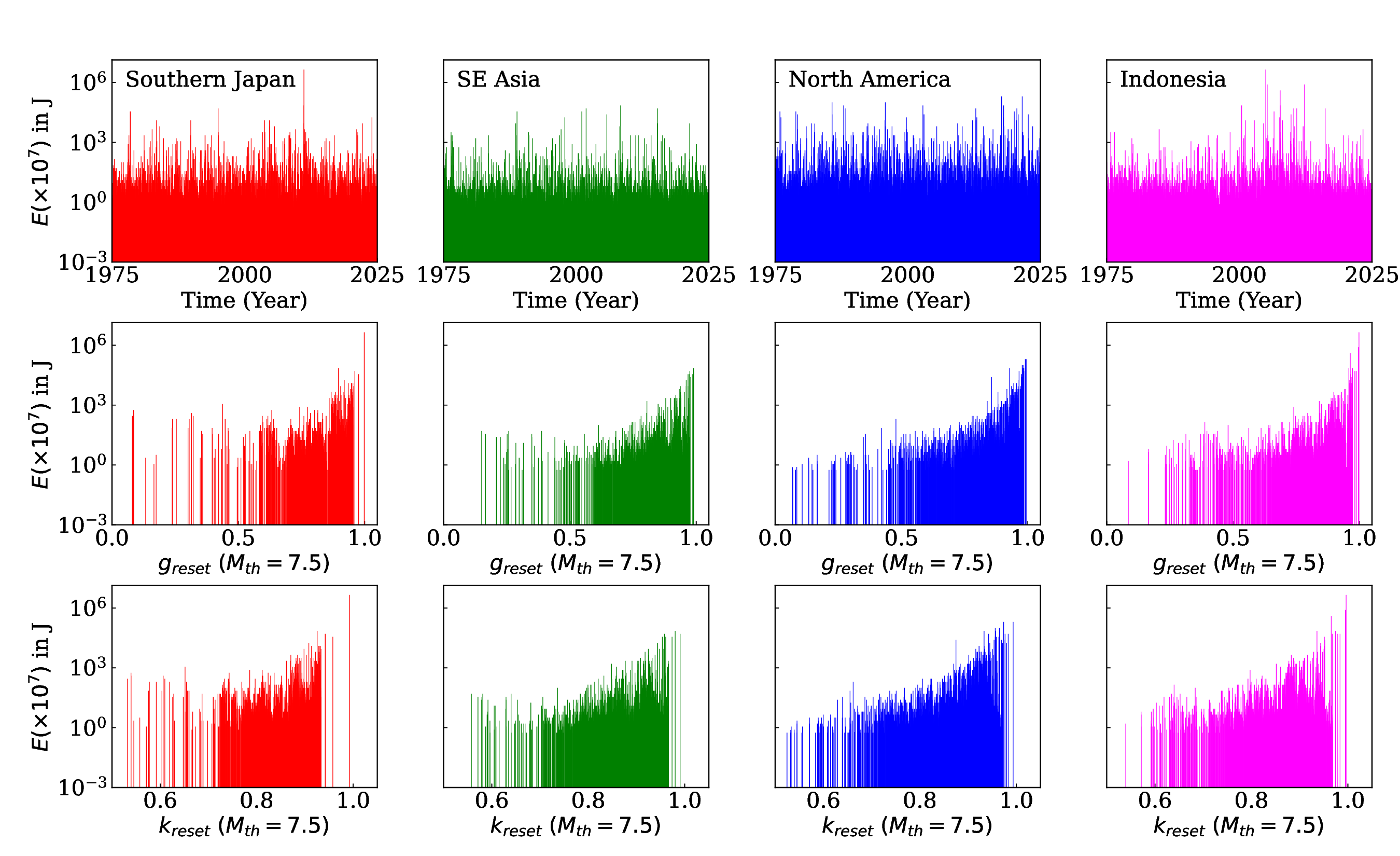}
    \end{center}
    \caption{The top-row panels show earthquake records from different areas around the world that are active in terms of earthquakes, from the year 1975 up to 2025 (with a lower magnitude cutoff $4.5$) arranged in the order they actually happened. The middle and bottom panels show the same earthquake records but sorted based on their assigned Gini ($g$) and Kolkata ($k$) values calculated using the ``dynamic reset" method with reset threshold $M_\mathrm{th}\geq 7.5$ , respectively. It's clear that bigger earthquakes tend to happen more often in periods where there's higher inequality in energy release across all the regions. While some smaller earthquakes also happen during times of high inequality value, there are no large earthquakes when the inequality indices are low. [From \cite{PRE_2026}]}
    \label{fig:real_reset_order}
    \end{figure}
    
If we consider a time series of \(N\) earthquake events, indexed by \(i=1,2,\ldots,N\), with occurrence times \(t_1<t_2<\cdots<t_N\) and magnitudes \(M_i\). An event \(i\) is classified as large if \(M_i\geq M_{\mathrm{th}}\), where \(M_{\mathrm{th}}\) is a prescribed magnitude threshold. On the other hand, the energy criterion can be written as \(E_i \geq E_{\mathrm{th}}\). The inequality measures \(g_i\) and \(k_i\) associated with the event \(i\) are computed using all previous events in the current time window except for event \(i\). Let $n$ denote the index of the first event in the current window. For an event \(i>n\), the indices are calculated using the events \(j=n,n+1,\ldots,i-1\). For the first event of each window (\(i=n\)), no preceding events are available; therefore, the indices are initialized by convention as \(g_n=0\) and \(k_n=0.5\). For every event \(i\) where \(M_i\geq M_{\mathrm{th}}\) holds, its indices are firstly determined by using the events preceding it in the same window, \(j=n,\ldots,i-1\). Then, the window is set again for the next event, \(i+1\), which is considered the first event in the new window. In this way, \(g_{i+1}=0\) and \(k_{i+1}=0.5\) holds. Indices of the successive events are determined by cumulative use of the events starting from the resetting process to the end of the window without any contribution from the previous windows. This process continues after each large event which is defined as ``dynamic reset". As the number of events in a window increases, the inequality indices may evolve and cross one another. These crossings may be analyzed as possible precursors for future large earthquakes, but the predictive significance of the crossings will have to be proven by means of statistical analysis. In view of the fact that estimates from a few occurrences may not be statistically reliable, the early events in each time window may be discarded using a certain minimum event criterion.

For comparison, the indices can also be calculated using a ``sliding window method" with a fixed number of events in the window. In the earthquake catalogs examined here, this approach yields a weaker association with subsequent large earthquakes. One possible explanation is that a large event can continue to influence the indices while it remains within the sliding window, including in windows following its occurrence. This persistence may affect the temporal evolution of the indices and complicate the interpretation of their association with subsequent seismic activity. A sliding-window approach may be appropriate for systems approaching terminal catastrophic failure, such as a rock specimen undergoing fracture (as discussed in Sec. \ref{inequality_in_fracture_models}), where the time series terminates at failure. However, in systems with statistically stationary behavior, which includes SOC models in their stationary regime, one should take into account the persistence of large events in sequential windows.

In the recent study \cite{PRE_2026}, the analysis of earthquake catalogs for the period of time January 1975--October 2025 in four regions has been performed: Southern Japan (30.145°--41.278°N, 129.792°--142.119°E); Southeast Asia (20.633°--35.747°N, 67.588°--99.000°E); North America (12.383°--61.143°N, 145.547°--46.318°W); Indonesia (7.537°S--5.703°N, 95.142°--107.051°E) (data was taken from \cite{usgs}). The common lower magnitude cutoff value \(M_{c}^{lower}=4.5\) was used for all the catalogs without any focal depths restriction for earthquakes. Earthquake magnitudes were transformed to seismic energies via the empirical relation \(\log_{10} E = 1.5M+C\) \cite{moment}. Also the additive constant \(C\) was not used (as \(C\) depends on region and depth). Catalog completeness was assessed by examining the magnitude–frequency distributions and their consistency with the Gutenberg–Richter law.

\begin{figure}[h!]
\begin{center}
    \includegraphics[width=0.85\textwidth]{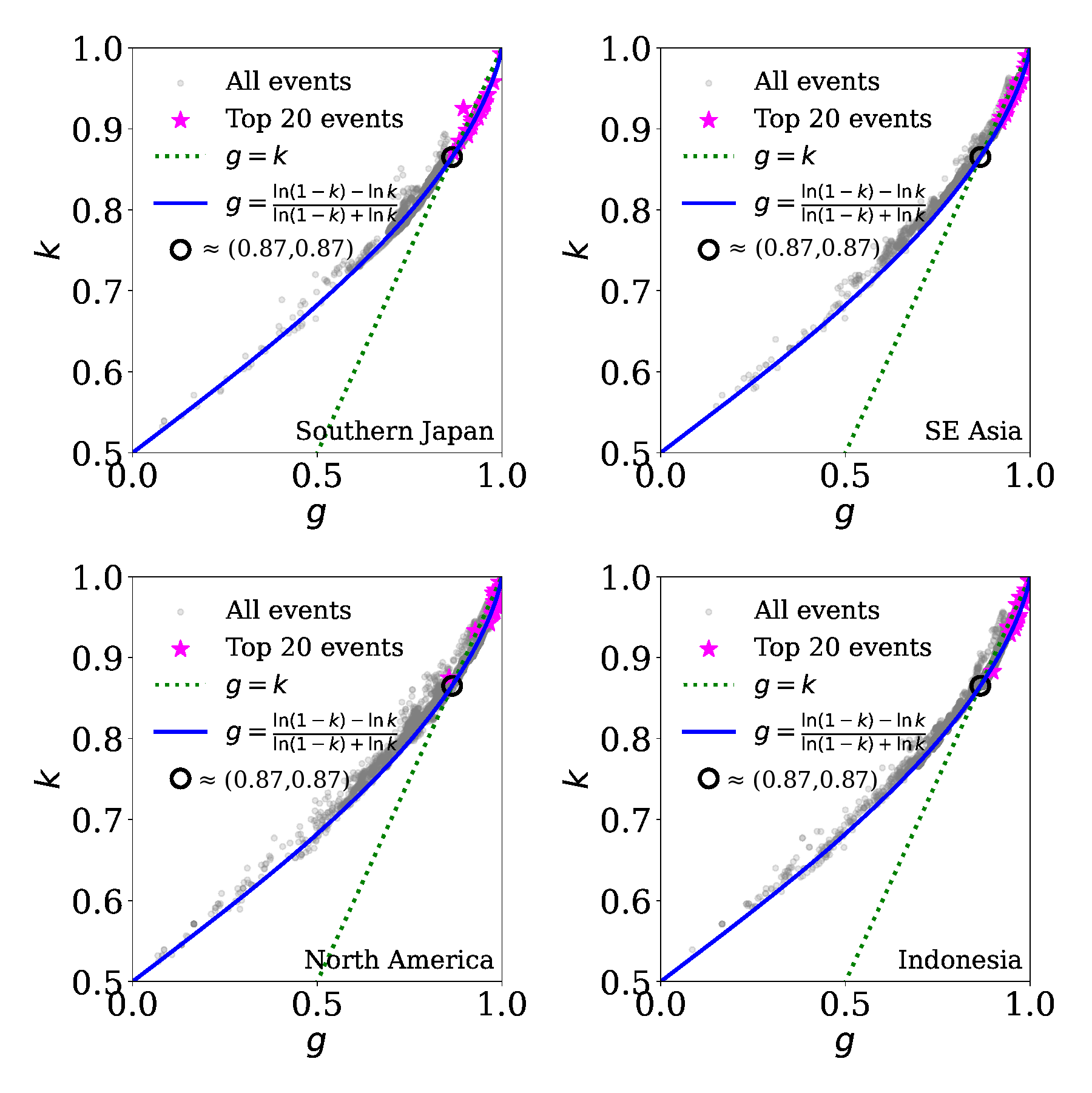}
    \end{center}
    \caption{ For earthquake records from various seismically active areas, the Gini ($g$) and Kolkata ($k$) inequality indices are plotted using the ``dynamic reset" method for the reset threshold $M_\mathrm{th}\geq 7.5$, with the top 20 largest earthquakes highlighted. The green dashed line shows where $g = k$, and the blue solid line shows the relationship given by Eq.(\ref{gk_relation}). From these plots, it is clear that most of the biggest earthquakes happen near $g\approx k\approx 0.87$ or at even higher values of these inequality measures. This suggests that high values of the Gini and Kolkata indices might be early signs that a big earthquake is about to happen. [Adapted from \cite{PRE_2026}] }
    \label{fig:gvk_real}
    \end{figure}

As shown in Fig. \ref{fig:real_reset_order}, the earthquake catalogs from the selected regions indicate that many large earthquakes occur after the inequality indices reach high values. Although smaller earthquakes also occur during periods of high inequality, the analyzed catalogs show that many large events are associated with the high-inequality regime. These observations suggest that elevated inequality may serve as a ``retrospective precursor indicator or a statistical discriminator" of periods preceding large earthquakes. A similar pattern has been observed in laboratory rock-fracture experiments \cite{jordi}. Further evidence is provided by the $g$-$k$ diagrams in Fig. \ref{fig:gvk_real}, where many large earthquakes occur near the crossing region, with both indices approaching approximately $0.87$. Similar near-universal values have been reported for certain self-organized criticality (SOC) models approaching their stationary critical states \cite{manna_2022}.

However, this behavior should not be generalized to all fracture models. In particular, the conventional ELS-FBM, which undergoes catastrophic breakdown, is distinct from SOC models in its standard formulation. In such fiber bundle models, the terminal values of the inequality indices depend on the avalanche-size distribution and its associated power-law exponent \cite{eswar}. For the SOC models, the crossing near $g=k\approx0.87$ is associated with their respective avalanche statistics. By contrast, steeper avalanche-size distributions, such as those with exponents $\theta=\frac52$ or $\theta=3$ in relevant ELS-FBM regimes (as mentioned in Subsec. \ref{universility_in_ELSFBM}), need not exhibit the same crossing behavior (as discussed in Sec. \ref{inequality_in_fracture_models}). Thus, the occurrence and location of a $g,k$ crossing depend on the event-size distribution and the underlying model dynamics. The similarity between the crossing values observed in earthquake catalogs and selected SOC models therefore does not, by itself, establish a common underlying mechanism or demonstrate operational earthquake forecasting capability.

\begin{figure}[h!]
\begin{center}
    \includegraphics[width=1.0\textwidth]{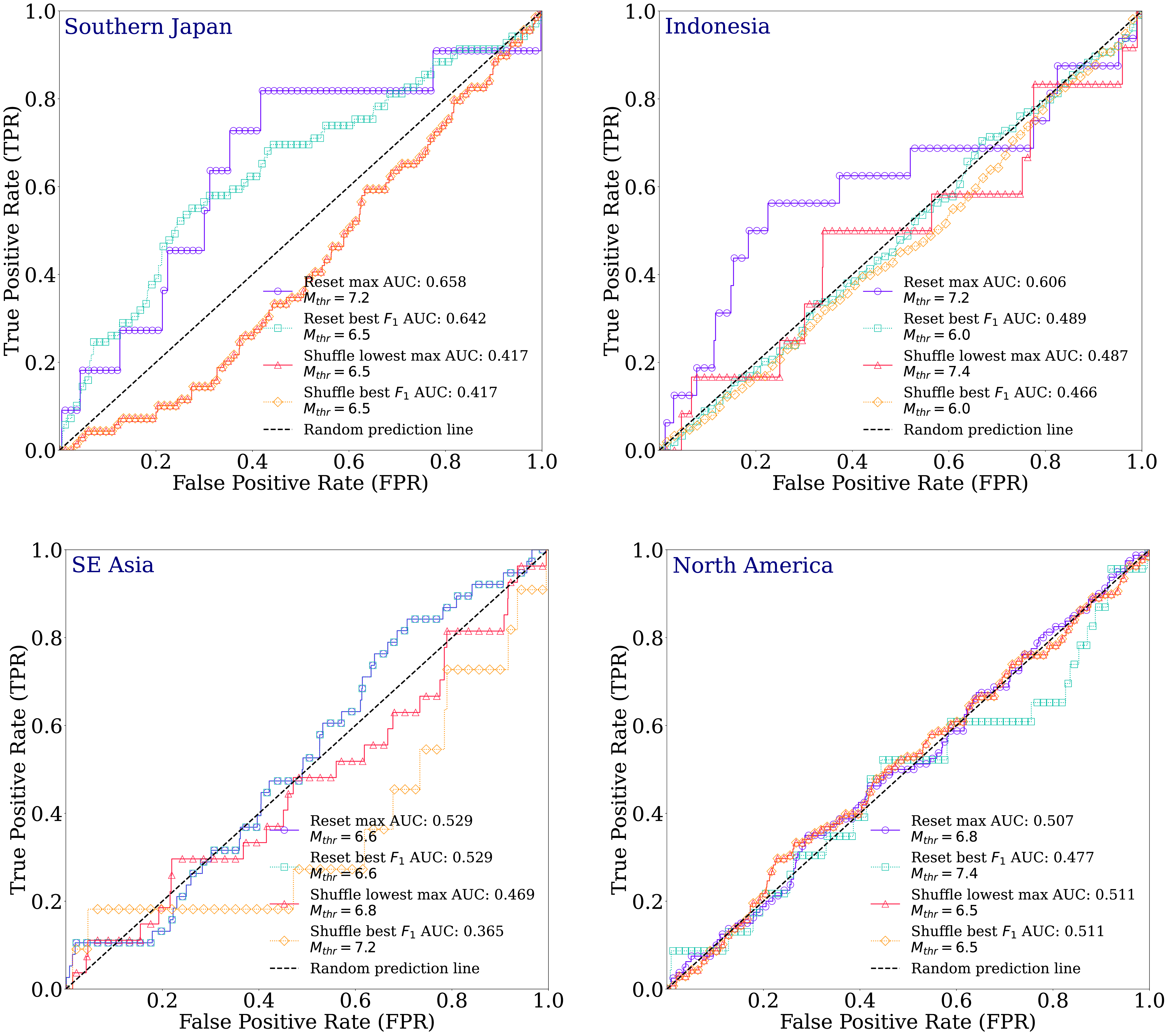}
    \end{center}
    \caption{Receiver Operating Characteristic (ROC) curves and corresponding area under the curve (AUC) values were made for different seismic areas using the inequality indices, which are calculated via ``dynamic reset" method, and the reset magnitude for each of the chosen regions varies from $6.0$ to $7.5$. The ROC curve shows the True Positive Rate (TPR), which is the proportion of large earthquakes that are correctly predicted, compared to the False Positive Rate (FPR), which is the proportion of non-large earthquakes that are wrongly predicted as large events. [From \cite{PRE_2026}]}
    \label{fig:ROC}
    \end{figure} 

The ability of the proposed method to discriminate between earthquake occurrence and non-occurrence was evaluated using receiver operating characteristic (ROC) analysis, as shown in Fig. \ref{fig:ROC}. ROC analysis is a common technique used to evaluate the performance of a binary classification model. We defined a threshold earthquake magnitude ($M_\mathrm{th}$) to classify an event as ``large". Also, the threshold values of $g$ and $k$ are then varied systematically as $g_\mathrm{th}$ and $k_\mathrm{th}$. If any event has a magnitude (avalanche size), $M\geq M_\mathrm{th}$ then it is classified as ``Actual large event", and if the magnitude (avalanche size) $M<M_\mathrm{th}$, it is classified as ``Actual small event". To show the predictive behavior by using the inequality indices, we define ``Predicted large event" as $g\geq g_{\mathrm{th}}$ and ``Predicted small event" as $g<g_{\mathrm{th}}$. A large event is correctly predicted as large, i.e., a true positive (TP), if \(M \geq M_{\mathrm{th}}\) and $g \geq g_{\mathrm{th}}$). A large event is incorrectly predicted as small, i.e., a false negative (FN), if \(M \geq M_{\mathrm{th}}\) but \(g < g_{\mathrm{th}}\). A small event is correctly identified as small, i.e., a true negative (TN), if \(M < M_{\mathrm{th}}\) and \(g < g_{\mathrm{th}}\). Finally, a small event is incorrectly identified as large, i.e., a false positive (FP), if \(M < M_{\mathrm{th}}\) but \(g \geq g_{\mathrm{th}}\). We can also do the same with the Kolkata index. The ROC curve is obtained by plotting True Positive Rate (TPR) versus False Positive Rate (FPR), and the predictive performance is measured using the area under the curve (AUC). An AUC value of 0.5 corresponds to random prediction, while larger AUC values show better predictive capability of the inequality indices in identifying large earthquake events or avalanches. TPR and FPR are defined as 

\begin{equation*}
\mathrm{R} = \mathrm{TPR} = \frac{\mathrm{TP}}{\mathrm{TP}+\mathrm{FN}}
\qquad\qquad
\mathrm{FPR} = \frac{\mathrm{FP}}{\mathrm{FP}+\mathrm{TN}}.
\end{equation*}

\noindent We also use the \(\mathrm{F}_1\) score as an additional criterion for selecting the optimal classification threshold in the ROC analysis. The \(\mathrm{F}_1\) score is the harmonic mean of precision (\(\mathrm{P}\)) and recall (\(\mathrm{R}\)). Precision is the fraction of predicted positive events that are truly positive, while recall is the fraction of actual positive events that are correctly identified. They are defined as

\begin{equation*}
\mathrm{F_1} = \frac{2\,\mathrm{P}\,\mathrm{R}}{\mathrm{P}+\mathrm{R}}
\qquad\qquad
\mathrm{P} = \frac{\mathrm{TP}}{\mathrm{TP}+\mathrm{FP}}.
\end{equation*}

\noindent At each point on the ROC curve, the corresponding $\mathrm{F_1}$ score is calculated. We then identify the maximum $\mathrm{F_1}$ score for that ROC curve. This procedure is repeated over all event-size thresholds. The threshold whose ROC curve gives the largest $\mathrm{F_1}$ score is selected as the ``best $\mathrm{F_1}$ threshold". The AUC is then calculated from the complete ROC curve corresponding to this selected threshold. Therefore, the $\mathrm{F_1}$ score is used to select the threshold, while the AUC is still calculated from the ROC curve. The ROC analysis was conducted through varying the magnitude of the resetting from 6.0 to 7.5 for all chosen regions. Based on the findings, it is possible to state that the forecast's effectiveness in large geographic zones with various tectonic features is not high due to the differences in the seismicity of the regions, which may reduce the effectiveness of the forecasting signal \cite{Kagan_1997_review, Hardebeck_2024_review, Mizrahi_review}. However, the forecasting method shows better results when applied to small regions like Southern Japan and Indonesia. Additionally, one of the factors that can impact the forecasting effectiveness is the incompleteness of the catalogs \cite{mizrahi_2021,wang_2025}. Furthermore, a shuffled earthquake catalog, which preserves the event-size distribution but randomizes the temporal ordering of events, does not exhibit the same effects. This suggests that the observed behavior is associated with temporal correlations in event sizes rather than the size distribution alone (see \cite{PRE_2026}).

\begin{figure}[h!]
\begin{center}
    \includegraphics[width=0.85\textwidth]{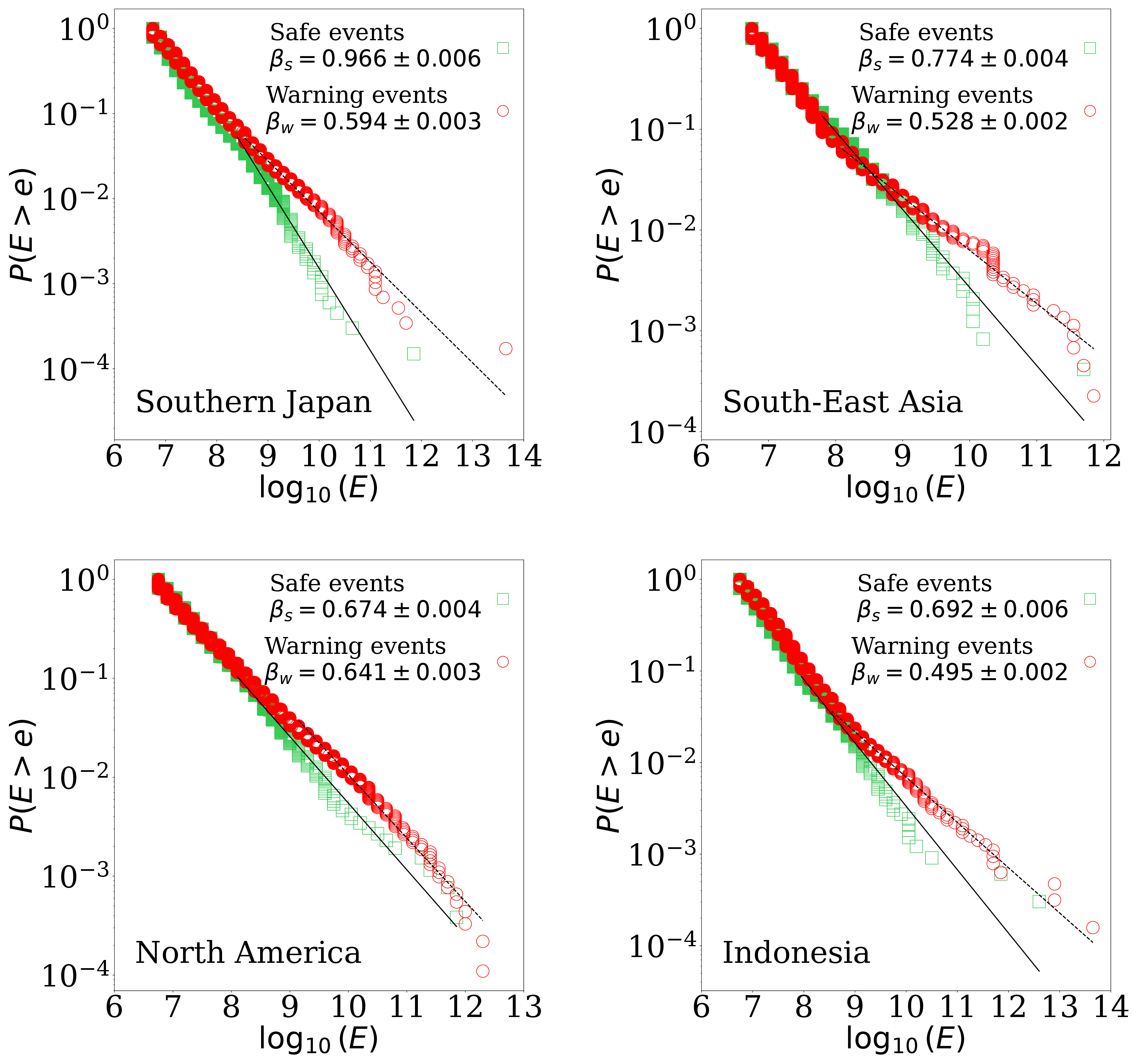}
    \end{center}
    \caption{ Cumulative frequency-energy distributions of earthquakes are shown for two different regimes: the ``safe" regime, where the value of $g\leq k$, and the ``warning" regime, where $g>k$. The given results show that the warning regime has a smaller exponent than the safe regime ($\beta_w<\beta_s$). The exponent $\beta$ is related to the relation (\ref{eq:n_beta}), which connects the GR exponent to the inequality indices, and also matches with the observation discussed in section \ref{sec:forecasting_with_b}. [From \cite{PRE_2026}]}
    \label{fig:s_w_real}
    \end{figure}

To check if this framework matches the known finding that the GR exponent (here $\beta$, as $P(E) \propto E^{-\beta}$, where $E$ denotes the seismic energy) drops before large earthquakes (as discussed in Sec. \ref{sec:forecasting_with_b}), earthquake records were split into two different groups: the ``safe" regime, where the value of $g\leq k$, and the ``warning" regime, where $g>k$. The earthquake records are split into sections, called ``reset panels" (for the reset threshold $M_\mathrm{th}\geq 7.5$). The position of this crossing can affect how many earthquakes are in each regime within a reset panel. In some cases, there may be too few earthquakes in one regime to accurately determine the power-law exponent ($\beta$) for that reset panel. To overcome this, the data from all safe zones across all reset panels are combined to create the cumulative distribution for the safe regime, and all warning zones are combined similarly for the warning regime. The given results show that the warning regime has a smaller exponent than the safe regime ($\beta_w<\beta_s$). 

The Gini ($g$) and Kolkata ($k$) inequality indices provide complementary measures for characterizing the temporal evolution of seismic activity. Unlike the Gutenberg-Richter $b$ value, which requires a sufficiently large sample of earthquake magnitudes for reliable estimation, these indices can be calculated for relatively short sequences and directly quantify the inequality in the distribution of released seismic energy or seismic moment. Nevertheless, their statistical reliability also depends on the number and distribution of available events. The results indicate that fluctuations in these indices can be associated with periods preceding major earthquakes and may correlate with previously reported decreases in the $b$ value before some large earthquakes and material failures. Yet ROC analyses show that the ability of the Gini ($g$) and Kolkata ($k$) indices to separate bad cases is weak and changes with location. Thus, the Gini and Kolkata indices may be regarded as retrospective precursor indicators and statistical discriminators of seismic activity, rather than established earthquake-prediction methods.

An important direction for future research is to determine whether inequality-based predictive models can distinguish periods preceding very large earthquakes from those in which smaller events occur despite similarly high inequality. The principal challenge is to reduce false alarms associated with the high-inequality regime without substantially increasing missed large earthquakes. This effort needs a check of how complete the catalog is, how sensitive the results are to the parameters, and a fine‑tuning of the inequality thresholds, the reset magnitude, the minimum number of events, and the forecast horizon (also mentioned in Sec. \ref{sec:modern_forecasting}). The Gini ($g$) and Kolkata ($k$) indices must also be compared with well‑known reference methods, such as seismicity rates Gutenberg–Richter $b$-value estimates and ETAS-based forecasts, to see if they add any predictive value. Evaluation should provide uncertainty estimates, rates of alarms and missed events, and validation that is independent, in time, and in region. In the end testing that uses alarm rules will be crucial to show whether inequality-based models can truly help with everyday earthquake forecasting.

\section{Summary, discussion \& perspective}
\label{sec:discussion}

\begin{table}[!htbp]
    \centering
    \caption{Comparison of the principal statistical models of fracture.}
    \label{tab:fracture_models}
    \footnotesize
    \renewcommand{\arraystretch}{1.2}
    \begin{tabularx}{\linewidth}{|>{\raggedright\arraybackslash}p{2.0cm}|>{\raggedright\arraybackslash}X|>{\raggedright\arraybackslash}X|>{\raggedright\arraybackslash}X|>{\raggedright\arraybackslash}X|}
        \hline
        \textbf{Model} & \textbf{Degrees of freedom and loading} & \textbf{Interaction or governing rule} &
        \textbf{Dissipation, critical behavior, and statistics} & \textbf{Relevance and limitations} \\
        \hline
        \textbf{ELS-FBM} (see Subsec. \ref{subsec:ELSFBM}) &
        Elastic, brittle fibres with random failure thresholds; external load applied through rigid plates. &
        The load of each failed fibre is redistributed equally among all surviving fibres. &
        Irreversible fibre failure; avalanche bursts precede catastrophic breakdown at a macroscopic critical load; avalanche statistics depend on the threshold distribution and loading protocol. &
        Analytically tractable mean-field model; neglects spatial stress correlations, damage localization, and crack propagation. \\
        \hline
        \textbf{LLS-FBM} (see Subsec. \ref{subsec:LLSFBM}) &
        Spatially arranged, elastic, brittle fibres with randomly distributed failure thresholds; quasistatic or time-dependent loading, depending on the model variant. &
        The load of a failed fibre is redistributed locally among its intact nearest neighbors. &
        Correlated damage, crack nucleation, and potentially abrupt macroscopic failure; avalanche statistics depend on the redistribution rule and interaction range. &
        Captures spatial stress redistribution and localization, but is less analytically tractable than the ELS-FBM. \\
        \hline
        \textbf{2D LLS-FBM (Models I and II)} (see Subsec. \ref{sec:soc_llsfbm}) &
        2D lattice of brittle fibres with uniformly distributed failure thresholds; quasistatic, localized loading applied initially to a central fibre. &
        \textbf{Model I:} the load of a failed fibre is shared equally among its intact nearest neighbours. \par\smallskip
        \textbf{Model II:} the load is redistributed equally among the surviving fibres along the boundary of the damaged region. &
        Self-organized critical-like avalanche dynamics in both models. \par\smallskip
        \textbf{Model I:} avalanche-size exponent $1.35\pm0.03$ and duration exponent $1.53\pm0.02$. \par\smallskip
        \textbf{Model II:} mean-field-like SOC behavior, with avalanche-size exponent $1.50\pm0.01$ and duration exponent $2.00\pm0.01$. &
        Model I retains local stress fluctuations, whereas Model II suppresses them through boundary-based redistribution, producing mean-field-like scaling. \\
        \hline
    \end{tabularx}
\end{table}

\begin{table}[!htbp]
    \centering
    \caption{Comparison of the statistical models of earthquakes.}
    \label{tab:earthquake_models}
    \footnotesize
    \renewcommand{\arraystretch}{1.2}
    \begin{tabularx}{\linewidth}{|>{\raggedright\arraybackslash}p{2.0cm}|>{\raggedright\arraybackslash}X|>{\raggedright\arraybackslash}X|>{\raggedright\arraybackslash}X|>{\raggedright\arraybackslash}X|}
        \hline
        \textbf{Model} & \textbf{Degrees of freedom and loading} & \textbf{Interaction or governing rule} &
        \textbf{Dissipation, critical behavior, and statistics} & \textbf{Relevance and limitations} \\
        \hline
        \textbf{1D BK model} (see Subsec. \ref{subsec:bk}) &
        Chain of elastically coupled blocks driven by a plate moving at constant speed. &
        Nearest-neighbor elastic coupling, loader springs, and velocity-dependent friction. &
        Stick-slip dynamics; event-size statistics can show Gutenberg-Richter-like scaling, with a possible excess of large characteristic events depending on the friction law and system parameters. &
        Mechanical model of fault slip and earthquake sequences; results depend on friction laws, elastic coupling, damping, and system size. \\
        \hline
        \textbf{OFC model} (see Subsec. \ref{subsec:ofc_model}) &
        2D lattice of stress-bearing sites under slow driving and rapid stress redistribution. &
        A site exceeding a threshold transfers a fraction of its stress to its nearest neighbors and resets. &
        Self-organized critical-like avalanche dynamics; for isotropic four-neighbor coupling, $\alpha_t=1/4$ is conservative in the bulk, whereas $\alpha_t<1/4$ introduces bulk dissipation. &
        Efficient model of collective seismicity and power-law scaling; omits realistic frictional dynamics, inertia, and detailed fault geometry. \\
        \hline
        \textbf{1D train model} (see Subsec. \ref{subsec:train_model}) &
        Quasistatically driven 1D chain of elastically connected blocks, with one end driven. &
        Elastic coupling competes with spatially varying static friction or pinning forces. &
        Stick-slip avalanches with power-law event-size statistics in certain pinning and driving protocols. &
        Connects earthquake-like dynamics to interface depinning; statistics depend sensitively on the pinning distribution and driving protocol. \\
        \hline
        \textbf{Two-fractal overlap (CS) model} (see Subsec. \ref{subsec:CS_model}) &
        Two finite-generation Cantor sets sliding uniformly past each other. &
        Overlap between the two sets represents fault contact and acts as a geometric proxy for stored/released elastic energy. &
        Overlap statistics yield Gutenberg-Richter-like power-law behavior. &
        Provides an analytically tractable link between fractal fault geometry and magnitude-frequency statistics; idealizes fault contact and does not model full elastodynamic rupture. \\
        \hline
    \end{tabularx}
\end{table}

Elastic properties of materials and their breaking mechanics or failure behavior had been the earliest topics of physical science and mechanical engineering. Even before Hooke's (1678) elastic or linear response to stresses on solid materials got established (suggesting self-averaging statistics of the elastic constant, which remains defined independent  of its volume), da Vinci reported from his experimental studies, more than 500 years back (from now), that the failure or fracture strength of the material tends to decrease with the sample volume (suggesting extreme statistics, like the Weibull or Gumbel distributions). These were discussed in sections \ref{sec:da_vinci} to \ref{sec:Fracture_exponents_for_disorder_concentration}, in the light of Griffith's crack nucleation theory and percolation theory in sections \ref{sec:crack_nucleation} to \ref{sec:Fracture_exponents_for_disorder_concentration}. Although the Fibre Bundle Model (FBM) of heterogeneous materials failure was introduced by Peirce in 1926, detailed studies for the Equal Load Sharing (ELS) models started with Daniels in 1945. We discussed analytically in subsections \ref{sec:uniform_ELSFBM} to \ref{sec:decreasing_ELSFBM} the different bundle strengths in three cases of simple distributions of fibre strengths in the bundles and the universal values of critical exponents of different failure-dynamical quantities near the respective bundle strengths. Then, in subsection \ref{sec:soc_llsfbm}, we essentially discuss the Self-Organized Critical (SOC) behavior of the two-dimensional FBM, giving the power law statistics of the broken fibre avalanches. Although these models differ in their degrees of freedom, interaction rules, loading protocols, and mechanisms of dissipation, they offer complementary perspectives on collective failure and the statistical organization of event sizes, as shown in Table \ref{tab:fracture_models}.

This review focuses on the relationship between avalanche statistics, the physical phenomena underlying avalanche dynamics, and measures of inequality. In the middle, in sections \ref{sec:inequality_measures} and \ref{inequality_in_fracture_models}, we introduce the social inequality measures and their applications for the failure avalanche statistics, where the number of large avalanches or energy released (like the number  of very rich people in any society) decreases with the corresponding magnitude of avalanche sizes or energy releases (like the magnitude of income or  wealth). In statistical physics of critical phenomena, we are concerned with the growth of statistical fluctuations, measured by the critical exponents (belonging to different universality classes), as the system approaches the  transition or critical point either by tuning the system externally or without any tuning in SOC systems. Although generally, some appropriately defined (and measurable) susceptibility captures this increase in fluctuations, which  tend to diverge as the system approaches the critical point, it is not always helpful to predict the critical point from the extrapolation of such large fluctuations. Social scientists, on the other hand, measure these fluctuations with some (bounded value) inequality indices. As discussed in Fig. \ref{fig:lorenz}(A) of section \ref{sec:inequality_measures}, the Lorenz function $L(x)$ (introduced in 1905) represents the cumulative fraction of wealth (here avalanche mass) possessed by $x$ a fraction of the poorest people (or smallest avalanches) in any country (system). The Gini index ($g$) is given by the area between the equality line given by the diagonal (connecting $(0,0)$ with $(1,1)$)  and the Lorenz function $L(x)$ (for the range $x = 0$ to $1$), normalized by the area $\frac{1}{2}$ under the equality line. Although the Gini index ($0 \le g \le 1$) has remained  extremely popular since 1921  (along with a few other indices obtained from the nonlinear Lorenz function)  in economics and sociology in quantifying the inequalities of wealth and other distributions. Another index, called the Kolkata index ($k$, introduced in 2014), has been quite useful to identify the critical points. It is based on the observation that, while $L(x)$ itself has trivial fixed points at $x = 0$ and $1$, the complementary Lorenz function $L_c(x) \equiv 1 - L(x)$ has a nontrivial fixed point at $k$: $L_c(k) = k$, $0.5 \le k \le 1$. This fixed-point value $k$ gives the fraction of wealth (or avalanche mass or released energy) possessed by the $(1-k)$ fraction of the richest population (avalanches or quakes). As discussed in this section $k = 0.80$ corresponds to Pareto’s 80-20 law (suggesting $80\%$ effects come from $20\%$ efforts or causes), while for all the sandpile models at their respective SOC points, one gets $g = k \simeq 0.87$ (suggesting approximately all $87\%$ of the avalanche masses are dissipated through only about $13\%$ of the largest avalanches). As such, the evaluation of $g$ and $k$ index values prior to major breakdown (and also some smooth crossing behavior of $g$ and $k$ as shown in Fig. \ref{fig:jordi} of section \ref{inequality_in_fracture_models} for the acoustic energy emission statistics before fracture  and in Fig. \ref{fig:gvk_real} of section \ref{inequality_measure_in_EQ}, for the seismic activity measures prior to the measured events in different regions of the world have been useful to get the indication of reliable precursors of the critical or major failure points. As mentioned in section \ref{inequality_in_fracture_models}, some such indications can also be obtained from the study of the Hirsch index ($h$, introduced in 2005), which gives the fixed point of the nonlinear citation  (again increasing with the number of publications or avalanches having those many citations or avalanche sizes/masses; see Fig. \ref{fig:lorenz}(B) of section \ref{sec:inequality_measures}). Thus the value $h$ gives the value of the number of papers of the author (here, number of avalanches of the system), each having citations (here, avalanche size or mass) not smaller than $h$. (See Fig. \ref{fig:lomov} for the damage statistics characteristics in two solid samples, before complete failure).

Earthquakes are large-scale mechanical failure phenomena, which are so common. We discussed the empirically observed Gutenberg-Richter (1944) power law for the decrease of frequency of earthquakes with the magnitude of the released elastic energies  and the Omori-Utsu law (1894) for decreasing frequency of aftershocks with the time elapsed since the main shock. In the following subsection \ref{subsec:bk}, we introduce the one-dimensional continuous space dynamical stick-slip motions  of earth crust blocks connected to each other with elastic springs on the moving tectonic plates of the earth responsible for sticking with a nonlinear friction force until slipping. The numerically observed Gutenberg-Richter (GR) type power law for the quake frequencies with their released elastic energies is discussed. In the next subsection, \ref{subsec:ofc_model}, the Olami-Feder-Christensen (OFC) cellular automaton model for earthquake dynamics and its SOC is discussed. In subsection \ref{subsec:train_model}, we introduce the simpler discrete motion train model and its self-similar avalanche statistics (obtained numerically) consistent with the GR law. Finally, in the next subsection, we introduce the contact (overlap) length statistics (interpreted as stored energy in the system) released with time as one fractal moves over the other with uniform velocity in the two-fractal-overlap model. Some simple analytical results supporting the GR law in the model are discussed. The models discussed in this review differ in their degrees of freedom, loading protocols, interaction rules, and mechanisms of dissipation or relaxation. These differences give rise to distinct forms of collective behavior, criticality, and event-size statistics. Table \ref{tab:earthquake_models} summarize the principal features, applications, and limitations of the statistical models of earthquake discussed in this manuscript, respectively.

An important question is whether statistical measures can act as clues that a failure is coming. Experiments on laboratory fracture, computer simulations, and lists of earthquakes have looked at inequality indices before big events. Still, careful consideration is required when dealing with these patterns. A relationship of some index to the occurrence of a significant future event doesn’t automatically guarantee its predictive ability. Such patterns may result from the size of the system, the threshold used for the definition of events, the nature of their distribution, the system loading, or the way the analysis window is constructed. As for earthquakes, additional complications appear due to data quality, spatial and temporal non-uniformity of distribution, clustering of events, existence of slow and low-frequency earthquakes, and the limited number of major events. For future research, the combination of the inequality indices with already tested approaches, a proper reference model, estimation of errors, and testing with unseen data should be considered. In terms of predicting, it should be assessed with target events, prediction intervals and alarm criteria, as well as false negative and false positive rates.

There are several approaches that appear particularly promising. Firstly, side-by-side testing of fracture-based and earthquake-based models will enable one to determine what parts of the inequality measures remain constant for varying rules of interaction types of disorder or loading regimes. Secondly, examining the impact of scaling of the system size and sensitivity analysis will be able to distinguish changes in the collective behavior of the system from artifacts induced by the finite size and limited observation times. Thirdly, combining the inequality indices with the information on avalanche sizes, time correlations, network-based quantities, and information-theoretical values will allow one to reconstruct the evolution of the failure process. Lastly, applying the suggested methods to earthquakes requires careful data validation and assessment of reliability and generalizability. All of these steps are needed to find out if inequality indices give insight beyond what normal earthquake statistics already show.

As mentioned already in the abstract, statistical physicists, condensed matter physicists, earth scientists, computer scientists, and even scientists from sociology or socio-physics are intensely interacting in research on the topics briefly summarized above. Also, there is no chapter in the textbooks on any subject that introduces the reader to these classic problems (phenomena, statistics, and models) of material failures (starting with the experiment of da Vinci five hundred years back) and large-scale dynamical failure like earthquakes. We introduce here the readers to the basic phenomena, concepts, and significant models developed over more than a century. We discuss their statistics, keeping in mind the researchers and newcomer readers from these wide backgrounds, who are interested in exploring the frontiers. We hope this review will help the readers from such a wide background to enjoy the basic concepts and the current research (with ample references to the appropriate literature) and eventually to contribute to the understanding of these long-standing problems.

\section{Additional Requirements}

\section*{Conflict of Interest Statement}

The authors declare that the research was conducted in the absence of any commercial or financial relationships that could be construed as a potential conflict of interest. Two of the authors (SB and BKC) are members of the Frontiers editorial board at the time of submission. Their editorial board membership did not have any impact on the peer-review  process or the final decision regarding this paper.

\section*{Author Contributions}

Conceptualization, S.B. and B.K.C.; Methodology, S.B. and B.K.C.; Data curation, Investigation, Visualization, and Writing—original draft preparation, S.S.; Funding acquisition, S.B.; Project administration and Resources, B.K.C.; Supervision, S.B. and B.K.C.; Writing—review and editing, S.S., S.B., and B.K.C. All authors have read and agreed to the published version of the manuscript.


\section*{Acknowledgments}

We are grateful to our colleagues Suchismita Banerjee, Jordi Bar\'o, Gilles Benguigui, Pathikrit Bhattacharya, Pratip Bhattacharyya, Soumyaditya Das, Diksha, Asim Ghosh, Lucas Goehring, Alex Hansen, Takahiro Hatano, Bijin Joseph, Hikaru Kawamura, Sumanta Kundu, Subhrangshu Sekhar Manna, Srutarshi Pradhan, Purusattam Ray, Subhadeep Roy, Parongama Sen, Shohini Sen, and Robin Stinchcombe for their valuable collaborations and contributions over the past four decades. We are also grateful to Stepan V Lomov for providing the IFBM figures.

\section*{Data Availability Statement}

Published data were analyzed in this work; no new data were generated or reported except for Fig. \ref{fig:train_dist}.

\bibliographystyle{apsrev4-2}
\bibliography{review_ref}

@article{Carlson1994,
  title = {Dynamics of earthquake faults},
  author = {Carlson, J. M. and Langer, J. S. and Shaw, B. E.},
  journal = {Rev. Mod. Phys.},
  volume = {66},
  issue = {2},
  pages = {657--670},
  numpages = {0},
  year = {1994},
  publisher = {American Physical Society},
  doi = {10.1103/RevModPhys.66.657}
}

@article{kawamura2012,
  title = {Statistical physics of fracture, friction, and earthquakes},
  author = {Kawamura, Hikaru and Hatano, Takahiro and Kato, Naoyuki and Biswas, Soumyajyoti and Chakrabarti, Bikas K.},
  journal = {Rev. Mod. Phys.},
  volume = {84},
  issue = {2},
  pages = {839--884},
  numpages = {0},
  year = {2012},
  publisher = {American Physical Society},
  doi = {10.1103/RevModPhys.84.839}
}

@article{gr1944,
       author = {{Gutenberg}, B. and {Richter}, C.~F.},
        title = "{Frequency of earthquakes in California}",
      journal = {Bulletin of the Seismological Society of America},
         year = {1944},
      
       volume = {34},
       number = {4},
        pages = {185-188},
          doi = {10.1785/BSSA0340040185}
      
}

@article{omori,
  author  = {Omori, Fusakichi},
  title   = {On the After-shocks of Earthquakes},
  journal = {Journal of the College of Science, Imperial University of Tokyo},
  volume  = {7},
  pages   = {111--200},
  year    = {1894}
}

@article{Utsu_1995,
  title={The centenary of the Omori formula for a decay law of aftershock activity},
  author={Utsu, Tokuji and Ogata, Yosihiko and Matsu'ura, Ritsuko S},
  journal={Journal of Physics of the Earth},
  volume={43},
  number={1},
  pages={1--33},
  year={1995},
  publisher={The Seismological Society of Japan},
  doi={10.4294/jpe1952.43.1}
}

@article{scholz,
    author = {Scholz, C. H.},
    title = {The frequency-magnitude relation of microfracturing in rock and its relation to earthquakes},
    journal = {Bulletin of the Seismological Society of America},
    volume = {58},
    number = {1},
    pages = {399-415},
    year = {1968},
    month = {02},
    issn = {0037-1106},
    doi = {10.1785/BSSA0580010399}
}

@article{schmit,
author = {Bouchon, Michel and Durand, Virginie and Marsan, David and Karabulut, Hayrullah and Schmittbuhl, Jean},
year = {2013},
pages = {},
title = {The long precursory phase of most large interplate earthquakes},
volume = {6},
journal = {Nature Geoscience},
doi = {10.1038/ngeo1770}
}

@article{Rundle_2021_NOWCASTING,
  author  = {Rundle, John B. and Stein, Seth and Donnellan, Andrea and Turcotte, Donald L. and Klein, William and Saylor, Cameron},
  title   = {The complex dynamics of earthquake fault systems: new approaches to forecasting and nowcasting of earthquakes},
  journal = {Reports on Progress in Physics},
  year    = {2021},
  volume  = {84},
  number  = {7},
  pages   = {076801},
  doi     = {10.1088/1361-6633/abf893},
  issn    = {0034-4885},
  publisher = {IOP Publishing}
}

@article{Varo,
  author  = {Varotsos, Panayiotis A. and Sarlis, Nicholas V. and Nagao, Toshiyasu},
  title   = {Complexity measure in natural time analysis identifying the accumulation of stresses before major earthquakes},
  journal = {Scientific Reports},
  year    = {2024},
  volume  = {14},
  number  = {1},
  pages   = {30828},
  doi     = {10.1038/s41598-024-81547-z}
}

@article{gini,
  author  = {Gini, Corrado},
  title   = {Measurement of Inequality of Incomes},
  journal = {The Economic Journal},
  year    = {1921},
  volume  = {31},
  number  = {121},
  pages   = {124--126},
  doi     = {10.2307/2223319}
}

@article{kolkata,
title = {Inequality in societies, academic institutions and science journals: Gini and k-indices},
journal = {Physica A: Statistical Mechanics and its Applications},
volume = {410},
pages = {30-34},
year = {2014},
issn = {0378-4371},
doi = {https://doi.org/10.1016/j.physa.2014.05.026},
author = {Asim Ghosh and Nachiketa Chattopadhyay and Bikas K. Chakrabarti},

}

@book{pareto,
  author    = {Pareto, Vilfredo},
  title     = {Cours d'{\'e}conomie politique},
  year      = {1896},
  publisher = {F. Rouge},
  address   = {Lausanne},
  note      = {Vols. I--II, p. 97}
}

@article{g_eq,
  title = {Relations among different inequality measures in complex systems: From kinetic exchange to earthquake models},
  author = {Sen, Shohini and Banerjee, Suchismita and Chakrabarti, Bikas K.},
  journal = {Phys. Rev. E},
  volume = {113},
  issue = {6},
  pages = {064308},
  numpages = {15},
  year = {2026},
  publisher = {American Physical Society},
  doi = {10.1103/vnpc-zw1v}
}

@article{ofc,
  title = {Self-organized criticality in a continuous, nonconservative cellular automaton modeling earthquakes},
  author = {Olami, Zeev and Feder, Hans Jacob S. and Christensen, Kim},
  journal = {Phys. Rev. Lett.},
  volume = {68},
  issue = {8},
  pages = {1244--1247},
  numpages = {0},
  year = {1992},
  month = {Feb},
  publisher = {American Physical Society},
  doi = {10.1103/PhysRevLett.68.1244}
}

@article{bkmodel,
  author  = {Burridge, Robert and Knopoff, Leon},
  title   = {Model and Theoretical Seismology},
  journal = {Bulletin of the Seismological Society of America},
  year    = {1967},
  volume  = {57},
  number  = {3},
  doi = {10.1785/BSSA0570030341},
  pages   = {341--371}
}

@inbook{Kun_Hidalgo_Herrmann_2006,
author = {Kun, Ferenc and Raischel, Frank and Hidalgo, Raul and Herrmann, H.J.},
year = {2006},
month = {10},
pages = {57-92},
title = {Extensions of Fibre Bundle Models},
volume = {705},
isbn = {978-3-540-35373-7},
publisher={Springer, Berlin, Heidelberg},
journal = {Lecture Notes in Physics},
doi = {10.1007/3-540-35375-5_3}
}

@book{sen_chakrabarti_sociophysics_2014,
  title={Sociophysics: An Introduction},
  author={Sen, P. and Chakrabarti, B.K.},
  isbn={9780191639456},
  url={https://books.google.co.in/books?id=deJ1AQAAQBAJ},
  year={2013},
  publisher={OUP Oxford}
}

@article{vieira,
  title = {Self-organized criticality in a deterministic mechanical model},
  author = {de Sousa Vieira, Maria},
  journal = {Phys. Rev. A},
  volume = {46},
  issue = {10},
  pages = {6288--6293},
  numpages = {0},
  year = {1992},
  publisher = {American Physical Society},
  doi = {10.1103/PhysRevA.46.6288}
}

@article{biswas2013,
  author  = {Biswas, Soumyajyoti and Ray, Purusattam and Chakrabarti, Bikas K.},
  title   = {Equivalence of the train model of earthquake and boundary driven Edwards-Wilkinson interface},
  journal = {The European Physical Journal B},
  year    = {2013},
  volume  = {86},
  number  = {9},
  pages   = {388},
  doi     = {10.1140/epjb/e2013-40637-6}
}

@misc{usgs,
author       = {{United States Geological Survey (USGS)}},
title        = {Earthquake Catalog},
year         = {2025},
howpublished = {\url{https://earthquake.usgs.gov}},
note         = {Accessed: 2025-10. Data used from January 1975 to October 2025}
}

@article{moment,
  author  = {Bormann, Peter and Di Giacomo, Domenico},
  title   = {The moment magnitude {M\textsubscript{w}} and the energy magnitude {M\textsubscript{e}}: Common roots and differences},
  journal = {Journal of Seismology},
  volume  = {15},
  number  = {2},
  pages   = {411--427},
  year    = {2011},
  doi     = {10.1007/s10950-010-9219-2}
}

@article{gini_prl_2023,
  title = {Critical Scaling through Gini Index},
  author = {Das, Soumyaditya and Biswas, Soumyajyoti},
  journal = {Phys. Rev. Lett.},
  volume = {131},
  issue = {15},
  pages = {157101},
  numpages = {6},
  year = {2023},
  publisher = {American Physical Society},
  doi = {10.1103/PhysRevLett.131.157101}
}

@article{first,
  title = {Social inequality analysis of fiber bundle model statistics and prediction of materials failure},
  author = {Biswas, Soumyajyoti and Chakrabarti, Bikas K.},
  journal = {Phys. Rev. E},
  volume = {104},
  issue = {4},
  pages = {044308},
  numpages = {7},
  year = {2021},
  publisher = {American Physical Society},
  doi = {10.1103/PhysRevE.104.044308},
}

@article{jordi,
  title = {Inequalities of energy release rates in compression of nanoporous materials predict its imminent breakdown},
  author = {Diksha and Bar\'o, Jordi and Biswas, Soumyajyoti},
  journal = {Phys. Rev. E},
  volume = {111},
  issue = {5},
  pages = {L053502},
  numpages = {6},
  year = {2025},
  month = {May},
  publisher = {American Physical Society},
  doi = {10.1103/PhysRevE.111.L053502}
}

@article{eswar,
  title = {Prediction of depinning transitions in interface models using Gini and Kolkata indices},
  author = {Diksha and Eswar, Gunnemeda and Biswas, Soumyajyoti},
  journal = {Phys. Rev. E},
  volume = {109},
  issue = {4},
  pages = {044113},
  numpages = {10},
  year = {2024},
  publisher = {American Physical Society},
  doi = {10.1103/PhysRevE.109.044113}
}

@article{Rundle_code_NOWCASTING,
author = {Rundle, John B. and Baughman, Ian and Zhang, Tianjian},
title = {Nowcasting Earthquakes With Stochastic Simulations: Information Entropy of Earthquake Catalogs},
journal = {Earth and Space Science},
volume = {11},
number = {6},
pages = {e2023EA003367},
doi = {https://doi.org/10.1029/2023EA003367},
year = {2024}
}

@article{lund_byrne_2001,
  author  = {Lund, Jay R. and Byrne, Joseph P.},
  title   = {Leonardo da Vinci's tensile strength tests: Implications for the discovery of engineering mechanics},
  journal = {Civil Engineering and Environmental Systems},
  year    = {2001},
  volume  = {18},
  number  = {3},
  pages   = {243--250},
  doi     = {10.1080/02630250108970302}
}

@book{hooke1678,
  author    = {Hooke, Robert},
  title     = {De Potentia Restitutiva, or of Spring. Explaining the Power of Springing Bodies},
  publisher = {Printed for John Martyn Printer to the Royal Society},
  address   = {London},
  year      = {1678}
}

@article{griffith1921,
  author  = {Griffith, A. A.},
  title   = {The Phenomena of Rupture and Flow in Solids},
  journal = {Philosophical Transactions of the Royal Society of London. Series A, Containing Papers of a Mathematical or Physical Character},
  year    = {1921},
  volume  = {221},
  pages   = {163--198},
  doi     = {10.1098/rsta.1921.0006}
}

@book{born1954,
  title={Dynamical Theory of Crystal Lattices},
  author={Born, M. and Huang, K.},
  isbn={9780192670083},
  lccn={99192617},
  series={International series of monographs on physics},
  url={https://books.google.co.in/books?id=FkZRAAAAMAAJ},
  year={1954},
  publisher={Clarendon Press}
}

@article{imoto_1991,
title = {Changes in the magnitude—frequency b-value prior to large (${M} \geq 6.0$) earthquakes in Japan},
journal = {Tectonophysics},
volume = {193},
number = {4},
pages = {311-325},
year = {1991},
doi = {https://doi.org/10.1016/0040-1951(91)90340-X},
author = {Masajiro Imoto},
}

@article{hirose_2002,
  author  = {Hirose, Fuyuki and Nakamura, Ayako and Hasegawa, Akira},
  title   = {b-value Variation Associated with the Rupture of Asperities—Spatial and Temporal Distributions of b-value East off NE Japan},
  journal = {Zisin (Journal of the Seismological Society of Japan. 2nd ser.)},
  volume  = {55},
  number  = {3},
  pages   = {249--260},
  year    = {2002},
  doi     = {10.4294/zisin1948.55.3_249}
}

@article{yoshida_2017,
author = {Yoshida, Keisuke and Saito, Tatsuhiko and Urata, Yumi and Asano, Youichi and Hasegawa, Akira},
title = {Temporal Changes in Stress Drop, Frictional Strength, and Earthquake Size Distribution in the 2011 Yamagata-Fukushima, NE Japan, Earthquake Swarm, Caused by Fluid Migration},
journal = {Journal of Geophysical Research: Solid Earth},
volume = {122},
number = {12},
pages = {10,379-10,397},
doi = {https://doi.org/10.1002/2017JB014334},
year = {2017}
}

@article{RayChakrabarti1985,
  author  = {Ray, Purusattam and Chakrabarti, Bikas K.},
  title   = {A Microscopic Approach to the Statistical Fracture Analysis of Disordered Brittle Solids},
  journal = {Solid State Communications},
  year    = {1985},
  volume  = {53},
  number  = {6},
  pages   = {477--479},
  doi     = {10.1016/0038-1098(85)91061-0}
}

@article{duxbury1986,
  title = {Size Effects of Electrical Breakdown in Quenched random Media},
  author = {Duxbury, P. M. and Beale, P. D. and Leath, P. L.},
  journal = {Phys. Rev. Lett.},
  volume = {57},
  issue = {8},
  pages = {1052--1055},
  year = {1986},
  publisher = {American Physical Society},
  doi = {10.1103/PhysRevLett.57.1052}
}

@book{ChakrabartiBenguigui1997,
    author = {Chakrabarti, Bikas K and Gilles Benguigui, L},
    title = {Statistical Physics of Fracture and Breakdown in Disordered Systems},
    publisher = {Oxford University Press},
    year = {1997},
    month = {08},
    isbn = {9780198520566},
    doi = {10.1093/oso/9780198520566.001.0001}
}

@book{Sahimi2003,
  author    = {Sahimi, Muhammad},
  title     = {Heterogeneous Materials II: Nonlinear and Breakdown Properties and Atomistic Modeling},
  publisher = {Springer},
  address   = {New York},
  year      = {2003},
  doi       = {10.1007/b97505}
}

@book{BiswasRayChakrabarti2015,
  author    = {Biswas, Soumyajyoti and Ray, Purusattam and Chakrabarti, Bikas K.},
  title     = {Statistical Physics of Fracture, Breakdown, and Earthquake: Effects of Disorder and Heterogeneity},
  publisher = {John Wiley \& Sons},
  address   = {Weinheim},
  year      = {2015},
  edition   = {reprint},
  pages     = {344},
  isbn      = {9783527672677},
  doi       = {10.1002/9783527672646}
}

@book{StaufferAharony2018,
  author    = {Stauffer, Dietrich and Aharony, Ammon},
  title     = {Introduction to Percolation Theory},
  edition   = {2},
  year      = {1992},
  publisher = {Taylor \& Francis},
  doi       = {10.1201/9781315274386}
}

@article{Peirce1926,
  author  = {Peirce, F. T.},
  title   = {Tensile Tests for Cotton Yarns, Part V: The Weakest Link},
  journal = {Journal of the Textile Institute Transactions},
  year    = {1926},
  volume  = {17},
  number  = {7},
  pages   = {T355--T368},
  doi     = {10.1080/19447027.1926.10599953}
}

@article{Daniels1945,
    author = {Daniels, Henry Ellis},
    title = {The statistical theory of the strength of bundles of threads. I},
    journal = {Proceedings of the Royal Society of London. A. Mathematical and Physical Sciences},
    volume = {183},
    number = {995},
    pages = {405-435},
    year = {1945},
    month = {06},
    issn = {0080-4630},
    doi = {10.1098/rspa.1945.0011},
}

@incollection{Irwin_1961,
    author = {Irwin, G. R.},
    isbn = {978-0-8031-6111-5},
    title = {Effects of Size and Shape on Fracture of Solids},
    booktitle = {Properties of Crystalline Solids},
    publisher = {ASTM International},
    year = {1961},
    month = {01},
    doi = {10.1520/STP45060S}
}

@book{weibull_1939,
  title={A Statistical Theory of the Strength of Materials},
  author={Weibull, W.},
  lccn={ac39002528},
  series={Handlingar / Ingeni{\"o}rsvetenskapsakademien},
  url={https://books.google.co.in/books?id=otVRAQAAIAAJ},
  year={1939},
  publisher={Generalstabens litografiska anstalts f{\"o}rlag}
}

@article{McMahon,
author = {Thomas McMahon},
title = {Size and Shape in Biology},
journal = {Science},
volume = {179},
number = {4079},
pages = {1201-1204},
year = {1973},
doi = {10.1126/science.179.4079.1201},
}

@article{Koch2004,
  author  = {Koch, George W. and Sillett, Stephen C. and
             Jennings, Gregory M. and Davis, Stephen D.},
  title   = {The limits to tree height},
  journal = {Nature},
  year    = {2004},
  volume  = {428},
  number  = {6985},
  pages   = {851--854},
  doi     = {10.1038/nature02417},
  url     = {https://doi.org/10.1038/nature02417}
}

@article{Bazant_1993,
author = {Zdeněk P. Bažant},
title = {Scaling Laws in Mechanics of Failure},
journal = {Journal of Engineering Mechanics},
volume = {119},
number = {9},
pages = {1828-1844},
year = {1993},
doi = {10.1061/(ASCE)0733-9399(1993)119:9(1828)},
}

@article{Wiseman_1995,
  title = {Lack of self-averaging in critical disordered systems},
  author = {Wiseman, Shai and Domany, Eytan},
  journal = {Phys. Rev. E},
  volume = {52},
  issue = {4},
  pages = {3469--3484},
  numpages = {0},
  year = {1995},
  month = {Oct},
  publisher = {American Physical Society},
  doi = {10.1103/PhysRevE.52.3469},
  url = {https://link.aps.org/doi/10.1103/PhysRevE.52.3469}
}

@article{Niklas_2004,
  author  = {Niklas, Karl J. and Spatz, Hanns-Christof},
  title   = {Growth and hydraulic (not mechanical) constraints govern the scaling of tree height and mass},
  journal = {Proceedings of the National Academy of Sciences},
  year    = {2004},
  volume  = {101},
  number  = {44},
  pages   = {15661--15663},
  month   = {nov},
  doi     = {10.1073/pnas.0405857101},
}

@article{HemmerPradhan2007,
  title = {Failure avalanches in fiber bundles for discrete load increase},
  author = {Hemmer, Per C. and Pradhan, Srutarshi},
  journal = {Phys. Rev. E},
  volume = {75},
  issue = {4},
  pages = {046101},
  numpages = {4},
  year = {2007},
  month = {Apr},
  publisher = {American Physical Society},
  doi = {10.1103/PhysRevE.75.046101},
}

@article{Sornette1989,
  author  = {Sornette, Didier},
  title   = {Elasticity and Failure of a Set of Elements Loaded in Parallel},
  journal = {Journal of Physics A: Mathematical and General},
  year    = {1989},
  volume  = {22},
  number  = {6},
  pages   = {L243--L250},
  doi     = {10.1088/0305-4470/22/6/010}
}

@article{HemmerHansen1992,
  author  = {Hemmer, Per Chr. and Hansen, Alex},
  title   = {The Distribution of Simultaneous Fiber Failures in Fiber Bundles},
  journal = {Journal of Applied Mechanics},
  year    = {1992},
  volume  = {59},
  number  = {4},
  pages   = {909--914},
  doi     = {10.1115/1.2894060}
}

@article{Pradhan2010,
  title = {Failure processes in elastic fiber bundles},
  author = {Pradhan, Srutarshi and Hansen, Alex and Chakrabarti, Bikas K.},
  journal = {Rev. Mod. Phys.},
  volume = {82},
  issue = {1},
  pages = {499--555},
  numpages = {0},
  year = {2010},
  publisher = {American Physical Society},
  doi = {10.1103/RevModPhys.82.499}
}

@book{HansenHemmerPradhan2015,
author = {Hansen, Alex and Hemmer, P.C. and Pradhan, Srutarshi},
publisher = {John Wiley \& Sons},
address   = {Weinheim},
year = {2015},
month = {09},
pages = {1-236},
title = {The Fiber Bundle Model: Modeling Failure in Materials},
isbn = {9783527412143},
doi = {10.1002/9783527671960}
}

@article{ChakrabartiStinchcombe1999,
  author  = {Chakrabarti, Bikas K. and Stinchcombe, Robin B.},
  title   = {Stick-Slip Statistics for Two Fractal Surfaces: A Model for Earthquakes},
  journal = {Physica A: Statistical Mechanics and its Applications},
  year    = {1999},
  volume  = {270},
  number  = {1--2},
  pages   = {27--34},
  doi     = {10.1016/S0378-4371(99)00146-6}
}

@article{manna_2022,
  author  = {Manna, S. S. and Biswas, Soumyajyoti and Chakrabarti, Bikas K.},
  title   = {Near universal values of social inequality indices in self-organized critical models},
  journal = {Physica A: Statistical Mechanics and its Applications},
  volume  = {596},
  pages   = {127121},
  year    = {2022},
  issn    = {0378-4371},
  doi     = {10.1016/j.physa.2022.127121}
}

@article{Taroni2018,
  author  = {Taroni, Matteo and Marzocchi, Warner and Schorlemmer, Danijel
             and Werner, Maximilian J. and Wiemer, Stefan
             and Zechar, Jeremy D. and Heiniger, Lukas and Euchner, Fabian},
  title   = {Prospective {CSEP} Evaluation of 1-Day, 3-Month, and 5-Yr
             Earthquake Forecasts for Italy},
  journal = {Seismological Research Letters},
  year    = {2018},
  volume  = {89},
  number  = {4},
  pages   = {1251--1261},
  doi     = {10.1785/0220180031}
}

@article{Gulia_2020,
  author  = {Gulia, Laura and Wiemer, Stefan and Vannucci, Gianfranco},
  title   = {Pseudoprospective Evaluation of the Foreshock
             Traffic-Light System in Ridgecrest and Implications
             for Aftershock Hazard Assessment},
  journal = {Seismological Research Letters},
  year    = {2020},
  volume  = {91},
  number  = {5},
  pages   = {2828--2842},
  doi     = {10.1785/0220190307}
}

@article{Marzocchi2020,
  author  = {Marzocchi, Warner and Spassiani, Ilaria and
             Stallone, Antonella and Taroni, Matteo},
  title   = {How to be Fooled Searching for Significant
             Variations of the b-Value},
  journal = {Geophysical Journal International},
  year    = {2020},
  volume  = {220},
  number  = {3},
  pages   = {1845--1856},
  doi     = {10.1093/gji/ggz541}
}

@article{Herrmann2022,
  author  = {Herrmann, Marcus and Piegari, Ester and
             Marzocchi, Warner},
  title   = {Revealing the Spatiotemporal Complexity of the
             Magnitude Distribution and b-Value during an
             Earthquake Sequence},
  journal = {Nature Communications},
  year    = {2022},
  volume  = {13},
  pages   = {5087},
  doi     = {10.1038/s41467-022-32755-6}
}

@article{Mirwald2024,
  author  = {Mirwald, Aron and Mizrahi, Leila and Wiemer, Stefan},
  title   = {How to b-Significant When Analyzing b-Value Variations},
  journal = {Seismological Research Letters},
  year    = {2024},
  volume  = {95},
  number  = {6},
  pages   = {3343--3359},
  doi     = {10.1785/0220240190}
}

@article{Lockner_1993,
  author  = {Lockner, David},
  title   = {The Role of Acoustic Emission in the Study of
             Rock Fracture},
  journal = {International Journal of Rock Mechanics and
             Mining Sciences \& Geomechanics Abstracts},
  year    = {1993},
  volume  = {30},
  number  = {7},
  pages   = {883--899},
  doi     = {10.1016/0148-9062(93)90041-B}
}

@article{Lei_2000,
  author  = {Lei, Xinglin and Kusunose, Kinichiro and
             Nishizawa, Osamu and Cho, Akira and Satoh, Takashi},
  title   = {On the Spatio-Temporal Distribution of Acoustic
             Emissions in Two Granitic Rocks under Triaxial
             Compression: The Role of Pre-Existing Cracks},
  journal = {Geophysical Research Letters},
  year    = {2000},
  volume  = {27},
  number  = {13},
  pages   = {1997--2000},
  doi     = {10.1029/1999GL011190}
}

@article{Goebel_2013,
  author  = {Goebel, T. H. W. and Schorlemmer, D. and Becker, T. W.
             and Dresen, G. and Sammis, C. G.},
  title   = {Acoustic Emissions Document Stress Changes over
             Many Seismic Cycles in Stick-Slip Experiments},
  journal = {Geophysical Research Letters},
  year    = {2013},
  volume  = {40},
  number  = {10},
  pages   = {2049--2054},
  doi     = {10.1002/grl.50507}
}

@article{RouetLeduc_2017,
  author  = {Rouet-Leduc, Bertrand and Hulbert, Claudia and
             Lubbers, Nicholas and Barros, Kipton and
             Humphreys, Colin J. and Johnson, Paul A.},
  title   = {Machine Learning Predicts Laboratory Earthquakes},
  journal = {Geophysical Research Letters},
  year    = {2017},
  volume  = {44},
  number  = {18},
  pages   = {9276--9282},
  doi     = {10.1002/2017GL074677}
}

@article{Goebel_2023,
  author  = {Goebel, Thomas H. W. and Brodsky, Emily E. and
             Dresen, Georg},
  title   = {Fault Roughness Promotes Earthquake-Like Aftershock
             Clustering in the Lab},
  journal = {Geophysical Research Letters},
  year    = {2023},
  volume  = {50},
  number  = {8},
  pages   = {e2022GL101241},
  doi     = {10.1029/2022GL101241}
}

@article{Helmstetter_2006,
  author  = {Helmstetter, Anne and Kagan, Yan Y. and Jackson, David D.},
  title   = {Comparison of Short-Term and Time-Independent Earthquake Forecast Models for Southern California},
  journal = {Bulletin of the Seismological Society of America},
  year    = {2006},
  volume  = {96},
  number  = {1},
  pages   = {90--106},
  doi     = {10.1785/0120050067}
}

@article{Hainzl_2016,
  author  = {Hainzl, Sebastian},
  title   = {Apparent Triggering Function of Aftershocks Resulting from Rate-Dependent Incompleteness of Earthquake Catalogs},
  journal = {Journal of Geophysical Research: Solid Earth},
  year    = {2016},
  volume  = {121},
  number  = {9},
  pages   = {6499--6509},
  doi     = {10.1002/2016JB013319}
}

@article{Kagan_Jackson_1991,
  author  = {Kagan, Yan Y. and Jackson, David D.},
  title   = {Long-Term Earthquake Clustering},
  journal = {Geophysical Journal International},
  year    = {1991},
  volume  = {104},
  number  = {1},
  pages   = {117--133},
  doi     = {10.1111/j.1365-246X.1991.tb02498.x}
}

@article{Zaliapin_2013,
  author  = {Zaliapin, Ilya and Ben-Zion, Yehuda},
  title   = {Earthquake Clusters in Southern California, I: Identification and Stability},
  journal = {Journal of Geophysical Research: Solid Earth},
  year    = {2013},
  volume  = {118},
  number  = {6},
  pages   = {2847--2864},
  doi     = {10.1002/jgrb.50179}
}

@article{Zaliapin_BenZion_2013,
  author  = {Zaliapin, Ilya and Ben-Zion, Yehuda},
  title   = {Earthquake Clusters in Southern California, II: Classification and Relation to Physical Properties of Lithosphere},
  journal = {Journal of Geophysical Research: Solid Earth},
  year    = {2013},
  volume  = {118},
  number  = {6},
  pages   = {2865--2877},
  doi     = {10.1002/jgrb.50178}
}

@article{Tsuruoka_2012,
  author  = {Tsuruoka, Hiroshi and Hirata, Naoshi and Schorlemmer, Danijel and Euchner, Fabian and Nanjo, Kazuyoshi Z. and Jordan, Thomas H.},
  title   = {{CSEP} Testing Center and the First Results of the Earthquake Forecast Testing Experiment in Japan},
  journal = {Earth, Planets and Space},
  year    = {2012},
  volume  = {64},
  number  = {8},
  pages   = {661--671},
  doi     = {10.5047/eps.2012.06.007}
}

@article{Corral_2004,
  author  = {Corral, {\'A}lvaro},
  title   = {Long-term clustering, scaling, and universality in the temporal occurrence of earthquakes},
  journal = {Physical Review Letters},
  volume  = {92},
  number  = {10},
  pages   = {108501},
  year    = {2004},
  doi     = {10.1103/PhysRevLett.92.108501}
}

@article{Mignan_Broccardo_2020,
  author  = {Mignan, Arnaud and Broccardo, Marco},
  title   = {Neural Network Applications in Earthquake Prediction
             (1994--2019): Meta-Analytic and Statistical Insights
             on Their Limitations},
  journal = {Seismological Research Letters},
  volume  = {91},
  number  = {4},
  pages   = {2330--2342},
  year    = {2020},
  doi     = {10.1785/0220200021}
}

@article{Marzocchi_2017,
  author  = {Marzocchi, Warner and Taroni, Matteo and Falcone, Giuseppe},
  title   = {Earthquake Forecasting during the Complex Amatrice-Norcia Seismic Sequence},
  journal = {Science Advances},
  year    = {2017},
  volume  = {3},
  number  = {9},
  pages   = {e1701239},
  doi     = {10.1126/sciadv.1701239}
}

@article{Sarlis_2013,
  author  = {Sarlis, Nicholas V. and Skordas, Efthimios S. and
             Varotsos, Panayiotis A. and Nagao, Toshiyasu and
             Kamogawa, Masashi and Tanaka, Haruo and Uyeda, Seiya},
  title   = {Minimum of the order parameter fluctuations of seismicity
             before major earthquakes in Japan},
  journal = {Proceedings of the National Academy of Sciences},
  volume  = {110},
  number  = {34},
  pages   = {13734--13738},
  year    = {2013},
  doi     = {10.1073/pnas.1312740110}
}

@article{Luginbuhl_2018,
  author  = {Luginbuhl, Molly and Rundle, John B. and Turcotte, Donald L.},
  title   = {Natural time and nowcasting induced seismicity at the Groningen gas field in the Netherlands},
  journal = {Geophysical Journal International},
  volume  = {215},
  number  = {2},
  pages   = {753--759},
  year    = {2018},
  doi     = {10.1093/gji/ggy315}
}

@article{Pasari_2019,
  author  = {Pasari, Sumanta},
  title   = {Nowcasting earthquakes in the Bay of Bengal region},
  journal = {Pure and Applied Geophysics},
  volume  = {176},
  number  = {4},
  pages   = {1417--1432},
  year    = {2019},
  doi     = {10.1007/s00024-018-2037-0}
}

@article{Zechar_2016_OEF,
  author  = {Zechar, J. Douglas and Marzocchi, Warner and Wiemer, Stefan},
  title   = {Operational Earthquake Forecasting in Europe: Progress, Despite Challenges},
  journal = {Bulletin of Earthquake Engineering},
  year    = {2016},
  volume  = {14},
  number  = {9},
  pages   = {2459--2469},
  doi     = {10.1007/s10518-016-9930-7}
}

@article{Geller_1996,
  author  = {Geller, Robert J.},
  title   = {Debate on evaluation of the {VAN} Method: Editor's introduction},
  journal = {Geophysical Research Letters},
  volume  = {23},
  number  = {11},
  pages   = {1291--1293},
  year    = {1996},
  doi     = {10.1029/96GL00742}
}

@article{Geller_1997,
  author  = {Geller, Robert J.},
  title   = {Earthquake prediction: a critical review},
  journal = {Geophysical Journal International},
  volume  = {131},
  number  = {3},
  pages   = {425--450},
  year    = {1997},
  doi     = {10.1111/j.1365-246X.1997.tb06588.x}
}

@article{Mulargia_Gasperini_1996,
  author  = {Mulargia, Francesco and Gasperini, Paolo},
  title   = {Precursor candidacy and validation: The {VAN} case so far},
  journal = {Geophysical Research Letters},
  volume  = {23},
  number  = {11},
  pages   = {1323--1326},
  year    = {1996},
  doi     = {10.1029/95GL03456}
}

@article{Rundle_2019_entropy,
  author  = {Rundle, John B. and Giguere, Alexis and Turcotte, Donald L.
             and Crutchfield, James P. and Donnellan, Andrea},
  title   = {Global seismic nowcasting with {Shannon} information entropy},
  journal = {Earth and Space Science},
  volume  = {6},
  number  = {1},
  pages   = {191--197},
  year    = {2019},
  doi     = {10.1029/2018EA000464}
}

@article{DeVries_2018,
  author  = {DeVries, Phoebe M. R. and Vi{\'e}gas, Fernanda and
             Wattenberg, Martin and Meade, Brendan J.},
  title   = {Deep learning of aftershock patterns following large earthquakes},
  journal = {Nature},
  volume  = {560},
  pages   = {632--634},
  year    = {2018},
  doi     = {10.1038/s41586-018-0438-y}
}

@article{Mignan_Broccardo_2019,
  author  = {Mignan, Arnaud and Broccardo, Marco},
  title   = {One neuron versus deep learning in aftershock prediction},
  journal = {Nature},
  volume  = {574},
  pages   = {E1--E3},
  year    = {2019},
  doi     = {10.1038/s41586-019-1582-8}
}

@article{Baiesi_Paczuski_2004,
  author  = {Baiesi, Marco and Paczuski, Maya},
  title   = {Scale-free networks of earthquakes and aftershocks},
  journal = {Physical Review E},
  volume  = {69},
  number  = {6},
  pages   = {066106},
  year    = {2004},
  doi     = {10.1103/PhysRevE.69.066106}
}

@article{Zechar_Jordan_2008,
  author  = {Zechar, J. Douglas and Jordan, Thomas H.},
  title   = {Testing alarm-based earthquake predictions},
  journal = {Geophysical Journal International},
  volume  = {172},
  number  = {2},
  pages   = {715--724},
  year    = {2008},
  doi     = {10.1111/j.1365-246X.2007.03676.x}
}

@article{Molchan_1990,
  author  = {Molchan, G. M.},
  title   = {Strategies in strong earthquake prediction},
  journal = {Physics of the Earth and Planetary Interiors},
  volume  = {61},
  number  = {1--2},
  pages   = {84--98},
  year    = {1990},
  doi     = {10.1016/0031-9201(90)90097-H}
}

@article{Molchan_1997,
  author  = {Molchan, G. M.},
  title   = {Earthquake prediction as a decision-making problem},
  journal = {Pure and Applied Geophysics},
  volume  = {149},
  pages   = {233--247},
  year    = {1997},
  doi     = {10.1007/BF00945169}
}

@article{Molchan_KeilisBorok_2008,
  author  = {Molchan, G. and Keilis-Borok, V.},
  title   = {Earthquake prediction: Probabilistic aspect},
  journal = {Geophysical Journal International},
  volume  = {173},
  number  = {3},
  pages   = {1012--1017},
  year    = {2008},
  month   = {jun},
  doi     = {10.1111/j.1365-246X.2008.03785.x}
}

@article{VAN_1,
  author  = {Varotsos, P. and Alexopoulos, K. and Nomicos, K.},
  title   = {Seismic electric currents},
  journal = {Proceedings of the Academy of Athens},
  volume  = {56},
  pages   = {277--286},
  year    = {1981}
}

@article{VAN_2,
  author  = {Varotsos, P. and Alexopoulos, K. and Nomicos, K.},
  title   = {Seven-hour precursors to earthquakes determined from telluric currents},
  journal = {Praktika of the Academy of Athens},
  volume  = {56},
  pages   = {417--433},
  year    = {1981}
}

@article{lomov_ifbm_1,
title = {Detailed experimental validation and benchmarking of six models for longitudinal tensile failure of unidirectional composites},
journal = {Composite Structures},
volume = {279},
pages = {114828},
year = {2022},
issn = {0263-8223},
doi = {https://doi.org/10.1016/j.compstruct.2021.114828},
url = {https://www.sciencedirect.com/science/article/pii/S026382232101271X},
author = {C. Breite and A. Melnikov and A. Turon and A.B. {de Morais} and C. {Le Bourlot} and E. Maire and E. Schöberl and F. Otero and F. Mesquita and I. Sinclair and J. Costa and J.A. Mayugo and J.M. Guerrero and L. Gorbatikh and L.N. McCartney and M. Hajikazemi and M. Mehdikhani and M.N. Mavrogordato and P.P. Camanho and R. Tavares and S.M. Spearing and S.V. Lomov and S. Pimenta and W. {Van Paepegem} and Y. Swolfs},
}

@article{lomov_ifbm_2,
title = {Blind benchmarking of seven longitudinal tensile failure models for two virtual unidirectional composites},
journal = {Composites Science and Technology},
volume = {202},
pages = {108555},
year = {2021},
issn = {0266-3538},
doi = {https://doi.org/10.1016/j.compscitech.2020.108555},
url = {https://www.sciencedirect.com/science/article/pii/S0266353820323472},
author = {C. Breite and A. Melnikov and A. Turon and A.B. {de Morais} and F. Otero and F. Mesquita and J. Costa and J.A. Mayugo and J.M. Guerrero and L. Gorbatikh and L.N. McCartney and M. Hajikazemi and P.P. Camanho and R.P. Tavares and S.V. Lomov and S. Pimenta and W. {Van Paepegem} and Y. Swolfs},
}

@article{Kagan_2009,
    author = {Kagan, Yan Y.},
    title = {Testing long-term earthquake forecasts: Likelihood methods and error diagrams},
    journal = {Geophysical Journal International},
    volume = {177},
    number = {2},
    pages = {532-542},
    year = {2009},
    month = {05},
    doi = {10.1111/j.1365-246X.2008.04064.x},
}

@article{diksha2023inequality,
  title = {Inequality of avalanche sizes in models of fracture},
  author = {Diksha and Kundu, Sumanta and Chakrabarti, Bikas K. and Biswas, Soumyajyoti},
  journal = {Phys. Rev. E},
  volume = {108},
  issue = {1},
  pages = {014103},
  numpages = {9},
  year = {2023},
  month = {Jul},
  publisher = {American Physical Society},
  doi = {10.1103/PhysRevE.108.014103}
}

@article{lorenz_1905,
  author  = {Lorenz, M. O.},
  title   = {Methods of Measuring the Concentration of Wealth},
  journal = {Publications of the American Statistical Association},
  volume  = {9},
  number  = {70},
  pages   = {209--219},
  year    = {1905},
  doi     = {10.2307/2276207},
  publisher = {American Statistical Association}
}

@article{PRE_2026,
  title = {Large earthquakes follow highly unequal ones},
  author = {Sarkar, Sudip and Biswas, Soumyajyoti},
  journal = {Phys. Rev. E},
  volume = {114},
  issue = {1},
  pages = {014143},
  numpages = {9},
  year = {2026},
  month = {Jul},
  publisher = {American Physical Society},
  doi = {10.1103/x98d-gtg4}
}

@book{Stauffer2003,
  title={Introduction To Percolation Theory},
  author={Aharony, A. and Stauffer, D.},
  isbn={9781135747831},
  url={https://books.google.co.in/books?id=Dph5AgAAQBAJ},
  year={2003},
  publisher={Taylor \& Francis}
}

@article{gennes1976,
  author  = {de Gennes, P. G.},
  title   = {On a Relation Between Percolation Theory and the Elasticity of Gels},
  journal = {Journal de Physique Lettres},
  volume  = {37},
  number  = {1},
  pages   = {L1--L2},
  year    = {1976},
  doi     = {10.1051/jphyslet:019760037010100}
}

@article{ray1988,
  title = {Strength of disordered solids},
  author = {Ray, P. and Chakrabarti, B. K.},
  journal = {Phys. Rev. B},
  volume = {38},
  issue = {1},
  pages = {715--719},
  numpages = {0},
  year = {1988},
  publisher = {American Physical Society},
  doi = {10.1103/PhysRevB.38.715},
}

@article{Pradhan2003,
  author  = {Pradhan, Srutarshi and Chakrabarti, Bikas K.},
  title   = {{Failure} properties of fiber bundle models},
  journal = {International Journal of Modern Physics B},
  volume  = {17},
  number  = {29},
  pages   = {5565--5581},
  year    = {2003},
  doi     = {10.1142/S0217979203023264},
}

@article{Pradhan2001PRE,
  title = {Precursors of catastrophe in the Bak-Tang-Wiesenfeld, Manna, and random-fiber-bundle models of failure},
  author = {Pradhan, Srutarshi and Chakrabarti, Bikas K.},
  journal = {Phys. Rev. E},
  volume = {65},
  issue = {1},
  pages = {016113},
  numpages = {7},
  year = {2001},
  publisher = {American Physical Society},
  doi = {10.1103/PhysRevE.65.016113}
}

@article{Biswas_flory_2020,
  title = {Flory-like statistics of fracture in the fiber bundle model as obtained via Kolmogorov dispersion for turbulence: A conjecture},
  author = {Biswas, Soumyajyoti and Chakrabarti, Bikas K.},
  journal = {Phys. Rev. E},
  volume = {102},
  issue = {1},
  pages = {012113},
  numpages = {6},
  year = {2020},
  publisher = {American Physical Society},
  doi = {10.1103/PhysRevE.102.012113}
}

@book{Stanley1971,
  title={Introduction to Phase Transitions and Critical Phenomena},
  author={Stanley, H.E.},
  isbn={9780195053166},
  lccn={87012357},
  series={International series of monographs on physics},
  url={https://books.google.co.in/books?id=C3BzcUxoaNkC},
  year={1971},
  publisher={Oxford University Press}
}

@article{Biswas_LLSFBM_2013,
  title = {Self-organized dynamics in local load-sharing fiber bundle models},
  author = {Biswas, Soumyajyoti and Chakrabarti, Bikas K.},
  journal = {Phys. Rev. E},
  volume = {88},
  issue = {4},
  pages = {042112},
  numpages = {7},
  year = {2013},
  publisher = {American Physical Society},
  doi = {10.1103/PhysRevE.88.042112}
}

@article{Mott1948,
  title={ Fracture of metals: theoretical considerations},
  author={Mott, NF},
  journal={Engineering},
  volume={165},
  pages={16--18},
  year={1948}
}

@article{Lomov2023,
  author    = {Stepan V. Lomov and Sergey G. Abaimov and Christian Breite and Yentl Swolfs},
  title     = {Inequality Indices Applied to Statistical Physics of Criticality in an Impregnated Fiber Bundle Model},
  journal   = {Mechanics of Composite Materials},
  volume    = {59},
  number    = {5},
  pages     = {841--846},
  year      = {2023},
  doi       = {10.1007/s11029-023-10137-3}
}

@article{Ghosh2022,
  author       = {A. Ghosh and S. Biswas and B. K. Chakrabarti},
  title        = {Success of Social Inequality Measures in Predicting Critical or Failure Points in Some Models of Physical Systems},
  journal      = {Frontiers in Physics},
  volume       = {10},
  pages        = {990278},
  year         = {2022},
  doi          = {10.3389/fphy.2022.990278}
}

@article{Biswas_nucleation_2015,
  title = {Nucleation versus percolation: Scaling criterion for failure in disordered solids},
  author = {Biswas, Soumyajyoti and Roy, Subhadeep and Ray, Purusattam},
  journal = {Phys. Rev. E},
  volume = {91},
  issue = {5},
  pages = {050105(R)},
  numpages = {4},
  year = {2015},
  publisher = {American Physical Society},
  doi = {10.1103/PhysRevE.91.050105},

}

@article{h_index,
  author    = {Hirsch, Jorge E.},
  title     = {An Index to Quantify an Individual's Scientific Research Output},
  journal   = {Proceedings of the National Academy of Sciences of the United States of America},
  volume    = {102},
  number    = {46},
  pages     = {16569--16572},
  year      = {2005},
  doi       = {10.1073/pnas.0507655102},
  pmid      = {16275915},
  pmcid     = {PMC1283832},
  publisher = {National Academy of Sciences}
}

@article{Bath1965,
  author = {Markus Båth},
  title   = {Lateral Inhomogeneities of the Upper Mantle},
  journal = {Tectonophysics},
  volume  = {2},
  number  = {6},
  pages   = {483--514},
  year    = {1965},
  issn = {0040-1951},
  doi = {https://doi.org/10.1016/0040-1951(65)90003-X}
}

@article{Bak2002,
  author  = {Per Bak and Kim Christensen and Leon Danon and Tim Scanlon},
  title   = {Unified Scaling Law for Earthquakes},
  journal = {Physical Review Letters},
  volume  = {88},
  number  = {17},
  pages   = {178501},
  year    = {2002},
  doi     = {10.1103/PhysRevLett.88.178501}
}

@article{rice_1993,
author = {Rice, James R.},
title = {Spatio-temporal complexity of slip on a fault},
journal = {Journal of Geophysical Research: Solid Earth},
volume = {98},
number = {B6},
pages = {9885-9907},
doi = {https://doi.org/10.1029/93JB00191},
year = {1993}
}

@article{manna_1991,
  author  = {Manna, S. S.},
  title   = {Two-state model of self-organized criticality},
  journal = {Journal of Physics A: Mathematical and General},
  year    = {1991},
  volume  = {24},
  number  = {7},
  pages   = {L363},
  month   = apr,
  doi     = {10.1088/0305-4470/24/7/009}
}

@article{Chianca2009,
  author  = {C. V. Chianca and J. S. Sa Martins and P. M. C. de Oliveira},
  title   = {Mapping the train model for earthquakes onto the stochastic sandpile model},
  journal = {European Physical Journal B},
  volume  = {68},
  number  = {4},
  pages   = {549--555},
  year    = {2009},
  month   = apr,
  doi     = {10.1140/epjb/e2009-00122-7}
}

@inbook{Rundle_Turcotte_complexity_and_eq_2015,
  author    = {Shcherbakov, Robert and Turcotte, Donald L. and Rundle, John B.},
  title     = {Complexity and Earthquakes},
  booktitle = {Treatise on Geophysics},
  year      = {2015},
  pages     = {627--653},
  publisher = {Elsevier},
  isbn      = {978-0-444-53803-1},
  doi       = {10.1016/B978-0-444-53802-4.00094-4}
}

@article{Sliderblock_Shcherbavok_2023,
  author    = {Charlotte A. Motuzas and Robert Shcherbakov},
  title     = {Viscoelastic Slider Blocks as a Model for a Seismogenic Fault},
  journal   = {Entropy},
  year      = {2023},
  volume    = {25},
  number    = {10},
  pages     = {1419},
  article-number = {1419},
  doi       = {10.3390/e25101419},
  issn      = {1099-4300},
  publisher = {MDPI}
}

@article{Petrillo_2020,
  author  = {Giuseppe Petrillo and Eugenio Lippiello and François P. Landes and Alberto Rosso},
  title   = {The influence of the brittle-ductile transition zone on aftershock and foreshock occurrence},
  journal = {Nature Communications},
  volume  = {11},
  number  = {1},
  pages    = {3010},
  year     = {2020},
  month    = jun,
  doi      = {10.1038/s41467-020-16811-7},
  issn     = {2041-1723}
}

@article{wyss_1973,
author = {Wyss, Max},
title = {Towards a Physical Understanding of the Earthquake Frequency Distribution},
journal = {Geophysical Journal of the Royal Astronomical Society},
volume = {31},
number = {4},
pages = {341-359},
doi = {https://doi.org/10.1111/j.1365-246X.1973.tb06506.x},
year = {1973}
}

@article{nanjo_2019,
author = {Nanjo, K.Z. and Izutsu, J. and Orihara, Y. and Kamogawa, M. and Nagao, T.},
title = {Changes in Seismicity Pattern Due to the 2016 Kumamoto Earthquakes Identify a Highly Stressed Area on the Hinagu Fault Zone},
journal = {Geophysical Research Letters},
volume = {46},
number = {16},
pages = {9489-9496},
doi = {https://doi.org/10.1029/2019GL083463},
year = {2019}
}

@article{gao_2002,
author = {Cao, Aimin and Gao, Stephen S.},
title = {Temporal variation of seismic b-values beneath northeastern Japan island arc},
journal = {Geophysical Research Letters},
volume = {29},
number = {9},
pages = {48-1-48-3},
doi = {https://doi.org/10.1029/2001GL013775},
year = {2002}

}

@article{nuannin_2005,
author = {Nuannin, Paiboon and Kulhanek, Ota and Persson, Leif},
title = {Spatial and temporal b value anomalies preceding the devastating off coast of NW Sumatra earthquake of December 26, 2004},
journal = {Geophysical Research Letters},
volume = {32},
number = {11},
pages = {},
doi = {https://doi.org/10.1029/2005GL022679},
year = {2005}
}

@article{huang_2015,
author = {Huang, Yihe and Beroza, Gregory C.},
title = {Temporal variation in the magnitude-frequency distribution during the Guy-Greenbrier earthquake sequence},
journal = {Geophysical Research Letters},
volume = {42},
number = {16},
pages = {6639-6646},
doi = {https://doi.org/10.1002/2015GL065170},
year = {2015}
}

@article{bhattacharyya2005,
title = {Of overlapping Cantor sets and earthquakes: analysis of the discrete Chakrabarti–Stinchcombe model},
journal = {Physica A: Statistical Mechanics and its Applications},
volume = {348},
pages = {199-215},
year = {2005},
issn = {0378-4371},
doi = {https://doi.org/10.1016/j.physa.2004.09.014},
author = {Pratip Bhattacharyya},
}

@article{ogata_ETAS_1988,
  author = {Ogata, Y.},
  title = {Statistical Models for Earthquake Occurrences and Residual Analysis for Point Processes},
  journal = {Journal of the American Statistical Association},
  volume = {83},
  number = {401},
  pages = {9--27},
  year = {1988},
  doi = {10.1080/01621459.1988.10478560}
}

@article{ogata_ETAS_1998,
  author = {Ogata, Y.},
  title = {Space-Time Point-Process Models for Earthquake Occurrences},
  journal = {Annals of the Institute of Statistical Mathematics},
  volume = {50},
  number = {2},
  pages = {379--402},
  year = {1998},
  doi = {10.1023/A:1003403601725}
}

@article{Carlson_time_interval_1991,
author = {Carlson, J. M.},
title = {Time intervals between characteristic earthquakes and correlations with smaller events: An analysis based on a mechanical model of a fault},
journal = {Journal of Geophysical Research: Solid Earth},
volume = {96},
number = {B3},
pages = {4255-4267},
doi = {https://doi.org/10.1029/90JB02474},
year = {1991}
}

@article{Carlson_BK_2D_1991,
  title = {Two-dimensional model of a fault},
  author = {Carlson, J. M.},
  journal = {Phys. Rev. A},
  volume = {44},
  issue = {10},
  pages = {6226--6232},
  numpages = {0},
  year = {1991},
  month = {Nov},
  publisher = {American Physical Society},
  doi = {10.1103/PhysRevA.44.6226},
}

@article{ray_2018,
  author  = {Ray, Purusattam},
  title   = {Statistical physics perspective of fracture in brittle and quasi-brittle materials},
  journal = {Philosophical Transactions of the Royal Society A: Mathematical, Physical and Engineering Sciences},
  volume  = {377},
  number  = {2136},
  pages   = {20170396},
  year    = {2019},
  doi     = {10.1098/rsta.2017.0396}
}

@article{Ben_zion_2008,
author = {Ben-Zion, Yehuda},
title = {Collective behavior of earthquakes and faults: Continuum-discrete transitions, progressive evolutionary changes, and different dynamic regimes},
journal = {Reviews of Geophysics},
volume = {46},
number = {4},
pages = {},
doi = {https://doi.org/10.1029/2008RG000260},
year = {2008}
}

@article{rundle_2003_bk,
author = {Rundle, John B. and Turcotte, Donald L. and Shcherbakov, Robert and Klein, William and Sammis, Charles},
title = {Statistical physics approach to understanding the multiscale dynamics of earthquake fault systems},
journal = {Reviews of Geophysics},
volume = {41},
number = {4},
pages = {},
doi = {https://doi.org/10.1029/2003RG000135},
year = {2003}
}

@article{Brace1966,
  author  = {William F. Brace and J. D. Byerlee},
  title   = {Stick-Slip as a Mechanism for Earthquakes},
  journal = {Science},
  volume  = {153},
  number  = {3739},
  pages   = {990--992},
  year    = {1966},
  doi     = {10.1126/science.153.3739.990}
}

@article{Marone1998,
  author  = {Christopher Marone},
  title   = {Laboratory-Derived Friction Laws and Their Application to Seismic Faulting},
  journal = {Annual Review of Earth and Planetary Sciences},
  volume  = {26},
  pages   = {643--696},
  year    = {1998},
  doi     = {10.1146/annurev.earth.26.1.643}
}

@book{Scholz2019_3rd,
  author    = {Christopher H. Scholz},
  title     = {The Mechanics of Earthquakes and Faulting},
  edition   = {3},
  publisher = {Cambridge University Press},
  address   = {Cambridge},
  year      = {2019},
  doi       = {10.1017/9781316681473}
}

@article{Dieterich1992,
  author  = {James H. Dieterich},
  title   = {Earthquake Nucleation on Faults with Rate- and State-Dependent Strength},
  journal = {Tectonophysics},
  volume  = {211},
  number  = {1--4},
  pages   = {115--134},
  year    = {1992},
  doi     = {10.1016/0040-1951(92)90055-B}
}

@article{Rubin2005,
  author  = {Allan M. Rubin and Jean-Paul Ampuero},
  title   = {Earthquake Nucleation on (Aging) Rate and State Faults},
  journal = {Journal of Geophysical Research},
  volume  = {110},
  pages   = {B11312},
  year    = {2005},
  doi     = {10.1029/2005JB003686}
}

@article{Lapusta2003,
  author  = {Nadia Lapusta and James R. Rice},
  title   = {Nucleation and Early Seismic Propagation of Small and Large Events in a Crustal Earthquake Model},
  journal = {Journal of Geophysical Research},
  volume  = {108},
  number  = {B4},
  pages   = {2205},
  year    = {2003},
  doi     = {10.1029/2001JB000793}
}

@article{DiToro2011,
  author  = {Giulio Di Toro and Richard Han and Takahiro Hirose and Nicolas De Paola and Shuji Nielsen and Kohtaro Mizoguchi and Tetsuro Shimamoto},
  title   = {Fault Lubrication During Earthquakes},
  journal = {Nature},
  volume  = {471},
  number  = {7339},
  pages   = {494--498},
  year    = {2011},
  doi     = {10.1038/nature09838}
}

@article{Rice2006,
  author  = {James R. Rice},
  title   = {Heating and Weakening of Faults During Earthquake Slip},
  journal = {Journal of Geophysical Research},
  volume  = {111},
  pages   = {B05311},
  year    = {2006},
  doi     = {10.1029/2005JB004006}
}

@article{Sibson1973,
  author  = {Richard H. Sibson},
  title   = {Interactions Between Temperature and Pore-Fluid Pressure During Earthquake Faulting and a Mechanism for Partial or Total Stress Relief},
  journal = {Nature Physical Science},
  volume  = {243},
  pages   = {66--68},
  year    = {1973},
  doi     = {10.1038/physci243066a0}
}

@article{Biswas_lucas_2019_mapping,
  author  = {Biswas, Soumyajyoti and Goehring, Lucas},
  title   = {Mapping heterogeneities through avalanche statistics},
  journal = {Philosophical Transactions of the Royal Society A:
             Mathematical, Physical and Engineering Sciences},
  volume  = {377},
  number  = {2136},
  pages   = {20170388},
  year    = {2019},
  doi     = {10.1098/rsta.2017.0388},
  pmid    = {30478200},
  pmcid   = {PMC6282404}
}

@book{kittel_2004,
  title={Introduction to Solid State Physics},
  author={Kittel, C.},
  isbn={9780471415268},
  lccn={2004042250},
  url={https://books.google.co.in/books?id=kym4QgAACAAJ},
  year={2004},
  publisher={Wiley}
}

@article{Stat_models_of_fracture_zapperi_alava_nakula,
author = {Mikko J. Alava and Phani K. V. V. Nukala and Stefano Zapperi},
title = {Statistical models of fracture},
journal = {Advances in Physics},
volume = {55},
number = {3-4},
pages = {349--476},
year = {2006},
publisher = {Taylor \& Francis},
doi = {10.1080/00018730300741518},
}

@article{Leeman_lab_slow_eq_2016,
  author    = {Leeman, J. R. and Saffer, D. M. and Scuderi, M. M. and Marone, C.},
  title     = {Laboratory observations of slow earthquakes and the spectrum of tectonic fault slip modes},
  journal   = {Nature Communications},
  volume    = {7},
  pages     = {11104},
  year      = {2016},
  doi       = {10.1038/ncomms11104},
  issn      = {2041-1723},
  publisher = {Nature Publishing Group}
}

@article{Ji_fault_2022,
title = {Laboratory experiments on fault behavior towards better understanding of injection-induced seismicity in geoenergy systems},
journal = {Earth-Science Reviews},
volume = {226},
pages = {103916},
year = {2022},
issn = {0012-8252},
doi = {https://doi.org/10.1016/j.earscirev.2021.103916},
author = {Yinlin Ji and Hannes Hofmann and Kang Duan and Arno Zang}}

@article{Acosta_dynamic_weaking_2018,
  author    = {Acosta, M. and Passel{\`e}gue, F. X. and Schubnel, A. and Violay, M.},
  title     = {Dynamic weakening during earthquakes controlled by fluid thermodynamics},
  journal   = {Nature Communications},
  volume    = {9},
  number    = {1},
  pages     = {3074},
  year      = {2018},
  doi       = {10.1038/s41467-018-05603-9},
  pmid      = {30082789},
  pmcid     = {PMC6079085},
  issn      = {2041-1723},
  publisher = {Springer Nature}
}

@article{Kawamura_2010_ofc,
  title = {Asperity characteristics of the Olami-Feder-Christensen model of earthquakes},
  author = {Kawamura, Hikaru and Yamamoto, Takumi and Kotani, Takeshi and Yoshino, Hajime},
  journal = {Phys. Rev. E},
  volume = {81},
  issue = {3},
  pages = {031119},
  numpages = {10},
  year = {2010},
  month = {Mar},
  publisher = {American Physical Society},
  doi = {10.1103/PhysRevE.81.031119}
}

@article{kun_herrmann_hidalgo_PRE_2002,
  title = {Fracture model with variable range of interaction},
  author = {Hidalgo, Raul Cruz and Moreno, Yamir and Kun, Ferenc and Herrmann, Hans J.},
  journal = {Phys. Rev. E},
  volume = {65},
  issue = {4},
  pages = {046148},
  numpages = {8},
  year = {2002},
  month = {Apr},
  publisher = {American Physical Society},
  doi = {10.1103/PhysRevE.65.046148}
}

@article{SB_PS_PRL_2015,
  title = {Maximizing the Strength of Fiber Bundles under Uniform Loading},
  author = {Biswas, Soumyajyoti and Sen, Parongama},
  journal = {Phys. Rev. Lett.},
  volume = {115},
  issue = {15},
  pages = {155501},
  numpages = {5},
  year = {2015},
  month = {Oct},
  publisher = {American Physical Society},
  doi = {10.1103/PhysRevLett.115.155501}
}

@article{Newman_Phoenix_2001,
  author  = {Newman, William I. and Phoenix, S. Leigh},
  title   = {Time-dependent fiber bundles with local load sharing},
  journal = {Physical Review E},
  volume  = {63},
  number  = {2},
  pages   = {021507},
  year    = {2001},
  doi     = {10.1103/PhysRevE.63.021507}
}

@article{Hidalgo_Kun_Herrmann_CREEP,
  author  = {Hidalgo, Raul Cruz and Kun, Ferenc and Herrmann, Hans J.},
  title   = {Creep rupture of viscoelastic fiber bundles},
  journal = {Physical Review E},
  volume  = {65},
  number  = {3},
  pages   = {032502},
  year    = {2002},
  doi     = {10.1103/PhysRevE.65.032502}
}

@article{Shimamoto_1986,
author = {Toshihiko Shimamoto },
title = {Transition Between Frictional Slip and Ductile Flow for Halite Shear Zones at Room Temperature},
journal = {Science},
volume = {231},
number = {4739},
pages = {711-714},
year = {1986},
doi = {10.1126/science.231.4739.711}
}

@article{Rubino_DynamicFriction_2017,
  author    = {Rubino, V. and Rosakis, A. J. and Lapusta, N.},
  title     = {Understanding dynamic friction through spontaneously evolving laboratory earthquakes},
  journal   = {Nature Communications},
  volume    = {8},
  pages     = {15991},
  year      = {2017},
  month     = jun,
  doi       = {10.1038/ncomms15991},
  issn      = {2041-1723},
  publisher = {Nature Publishing Group}
}

@article{Garagash_FractureMechanics_2021,
  author    = {Garagash, Dmitry I.},
  title     = {Fracture mechanics of rate-and-state faults and fluid injection induced slip},
  journal   = {Philosophical Transactions of the Royal Society A: Mathematical, Physical and Engineering Sciences},
  volume    = {379},
  number    = {2196},
  pages     = {20200129},
  year      = {2021},
  month     = may,
  doi       = {10.1098/rsta.2020.0129},
  issn      = {1364-503X},
  publisher = {The Royal Society}
}

@article{Shi_HowFrictionalSlipEvolves_2023,
  author    = {Shi, Shuo and Wang, Meng and Poles, Yoav and Svetlizky, Ilya and Fineberg, Jay},
  title     = {How frictional slip evolves},
  journal   = {Nature Communications},
  volume    = {14},
  pages     = {8291},
  year      = {2023},
  month     = dec,
  doi       = {10.1038/s41467-023-44086-1},
  issn      = {2041-1723},
  publisher = {Springer Nature}
}

@article{Zapperi_PRL_1997,
  title = {First-Order Transition in the Breakdown of Disordered Media},
  author = {Zapperi, Stefano and Ray, Purusattam and Stanley, H. Eugene and Vespignani, Alessandro},
  journal = {Phys. Rev. Lett.},
  volume = {78},
  issue = {8},
  pages = {1408--1411},
  numpages = {0},
  year = {1997},
  publisher = {American Physical Society},
  doi = {10.1103/PhysRevLett.78.1408},
  
}

@ARTICLE{ghosh_2022_h_index,
    
AUTHOR={Ghosh, Asim  and Chakrabarti, Bikas K.  and Ram, Dachepalli R. S.  and Mitra, Manipushpak  and Maiti, Raju  and Biswas, Soumyajyoti  and Banerjee, Suchismita },
           
TITLE={Scaling behavior of the Hirsch index for failure avalanches, percolation clusters, and paper citations},
          
JOURNAL={Frontiers in Physics},
          
VOLUME={Volume 10 - 2022},
  
YEAR={2022},
  
DOI={10.3389/fphy.2022.1019744},
  
ISSN={2296-424X},
}

@ARTICLE{Frohlich1993,
	author = {Frohlich, C. and Davis, S.D.},
	title = {Teleseismic b values; or, much ado about 1.0},
	year = {1993},
	journal = {Journal of Geophysical Research},
	volume = {98},
	number = {B1},
	pages = {631 – 644},
	doi = {10.1029/92JB01891}
}

@article{mizrahi_2021,
author = {Mizrahi, Leila and Nandan, Shyam and Wiemer, Stefan},
title = {Embracing Data Incompleteness for Better Earthquake Forecasting},
journal = {Journal of Geophysical Research: Solid Earth},
volume = {126},
number = {12},
pages = {e2021JB022379},
doi = {https://doi.org/10.1029/2021JB022379},
year = {2021}
}

@article{wang_2025,
author = {Wang, Xinyi and Li, Jiawei and Feng, Ao and Sornette, Didier},
title = {Estimating Magnitude Completeness in Earthquake Catalogs: A Comparative Study of Catalog-Based Methods},
journal = {Journal of Geophysical Research: Solid Earth},
volume = {130},
number = {9},
pages = {e2025JB031441},
doi = {https://doi.org/10.1029/2025JB031441},
year = {2025}
}

@article{Smith_1981,
  author    = {Warwick D. Smith},
  title     = {The b-value as an earthquake precursor},
  journal   = {Nature},
  year      = {1981},
  volume    = {289},
  number    = {5794},
  pages     = {136--139},
  doi       = {10.1038/289136a0},
  issn      = {0028-0836},
  publisher = {Nature Publishing Group}
}

@article{Schorlemmer_2005,
  author    = {Schorlemmer, Danijel and Wiemer, Stefan and Wyss, Max},
  title     = {Variations in Earthquake-Size Distribution across Different Stress Regimes},
  journal   = {Nature},
  year      = {2005},
  volume    = {437},
  number    = {7058},
  pages     = {539--542},
  doi       = {10.1038/nature04094},
  publisher = {Nature Publishing Group}
}

@article{nanjo_2012,
  author  = {Nanjo, K. Z. and Hirata, N. and Obara, K. and Kasahara, K.},
  title   = {Decade-scale Decrease in b Value Prior to the M9-Class 2011 Tohoku and 2004 Sumatra Quakes},
  journal = {Geophysical Research Letters},
  year    = {2012},
  volume  = {39},
  pages   = {L20304},
  doi     = {10.1029/2012GL052997},
}

@article{nanjo_2017,
  author  = {Nanjo, K. Z. and Yoshida, A.},
  title   = {Anomalous Decrease in Relatively Large Shocks and Increase in the p and b Values Preceding the April 16, 2016, M7.3 Earthquake in Kumamoto, Japan},
  journal = {Earth, Planets and Space},
  year    = {2017},
  volume  = {69},
  pages   = {13},
  doi     = {10.1186/s40623-017-0598-2},
}

@article{nanjo_2016,
  author  = {Nanjo, K. Z. and Izutsu, J. and Orihara, Y. and Furuse, N. and Togo, S. and Nitta, H. and Okada, T. and Tanaka, R. and Kamogawa, M. and Nagao, T.},
  title   = {Seismicity Prior to the 2016 Kumamoto Earthquakes},
  journal = {Earth, Planets and Space},
  year    = {2016},
  volume  = {68},
  pages   = {187},
  doi     = {10.1186/s40623-016-0558-2},
}

@article{Utsu_1961,
  author  = {Utsu, Tokuji},
  title   = {A Statistical Study on the Occurrence of Aftershocks},
  journal = {Geophysical Magazine},
  year    = {1961},
  volume  = {30},
  pages   = {521--605}
}

@article{nanjo_2021,
  author  = {Nanjo, K. Z. and Yoshida, A.},
  title   = {Changes in the b Value in and around the Focal Areas of the M6.9 and M6.8 Earthquakes off the Coast of Miyagi Prefecture, Japan, in 2021},
  journal = {Progress in Earth and Planetary Science},
  year    = {2021},
  volume  = {8},
  pages   = {66},
  doi     = {10.1186/s40623-021-01511-3}
}

@article{Gulia_2019,
  author    = {Laura Gulia and Stefan Wiemer},
  title     = {Real-time Discrimination of Earthquake Foreshocks and Aftershocks},
  journal   = {Nature},
  year      = {2019},
  volume    = {574},
  number    = {7777},
  pages     = {193--199},
  doi       = {10.1038/s41586-019-1606-4},
  publisher = {Springer Nature}
}

@article{Mizrahi_review,
author = {Mizrahi, Leila and Dallo, Irina and van der Elst, Nicholas J. and Christophersen, Annemarie and Spassiani, Ilaria and Werner, Maximilian J. and Iturrieta, Pablo and Bayona, José A. and Iervolino, Iunio and Schneider, Max and Page, Morgan T. and Zhuang, Jiancang and Herrmann, Marcus and Michael, Andrew J. and Falcone, Giuseppe and Marzocchi, Warner and Rhoades, David and Gerstenberger, Matt and Gulia, Laura and Schorlemmer, Danijel and Becker, Julia and Han, Marta and Kuratle, Lorena and Marti, Michèle and Wiemer, Stefan},
title = {Developing, Testing, and Communicating Earthquake Forecasts: Current Practices and Future Directions},
journal = {Reviews of Geophysics},
volume = {62},
number = {3},
pages = {e2023RG000823},
doi = {https://doi.org/10.1029/2023RG000823},
year = {2024}
}

@article{Jordan_2011_review,
  author  = {Jordan, Thomas H. and Chen, Y.-T. and Gasparini, Paolo and Madariaga, Ra{\'u}l and Main, Ian and Marzocchi, Warner and Papadopoulos, George and Sobolev, Gennady and Yamaoka, Kazuo and Zschau, Jochen},
  title   = {Operational Earthquake Forecasting: State of Knowledge and Guidelines for Utilization},
  journal = {Annals of Geophysics},
  year    = {2011},
  volume  = {54},
  number  = {4},
  pages   = {315--391},
  doi     = {10.4401/ag-5350}
}

@article{Mignan_2014_review,
  author  = {Mignan, Arnaud},
  title   = {The Debate on the Prognostic Value of Earthquake Foreshocks: A Meta-Analysis},
  journal = {Scientific Reports},
  year    = {2014},
  volume  = {4},
  pages   = {4099},
  doi     = {10.1038/srep04099}
}

@article{Kagan_1997_review,
    author = {Kagan, Yan Y.},
    title = {Are earthquakes predictable?},
    journal = {Geophysical Journal International},
    volume = {131},
    number = {3},
    pages = {505-525},
    year = {1997},
    month = {12},
    issn = {0956-540X},
    doi = {10.1111/j.1365-246X.1997.tb06595.x},
}

@article{Hardebeck_2024_review,
   author = "Hardebeck, Jeanne L. and Llenos, Andrea L. and Michael, Andrew J. and Page, Morgan T. and Schneider, Max and van der Elst, Nicholas J.",
   title = "Aftershock Forecasting", 
   journal= "Annual Review of Earth and Planetary Sciences",
   year = "2024",
   volume = "52",
   number = "Volume 52, 2024",
   pages = "61-84",
   doi = "https://doi.org/10.1146/annurev-earth-040522-102129",
   publisher = "Annual Reviews",
   issn = "1545-4495",
   type = "Journal Article",
  }

@article{Kawamura_2008,
author = {Mori, Takahiro and Kawamura, Hikaru},
title = {Simulation study of the two-dimensional Burridge-Knopoff model of earthquakes},
journal = {Journal of Geophysical Research: Solid Earth},
volume = {113},
number = {B6},
pages = {},
doi = {https://doi.org/10.1029/2007JB005219},
year = {2008}
}

@article{Johnson_review_2021,
  author = {Johnson, Paul A. and Rouet-Leduc, Bertrand and Pyrak-Nolte, Laura J. and Beroza, Gregory C. and Marone, Chris J. and Hulbert, Claudia and Howard, Addison and Singer, Philipp and Gordeev, Dmitry and Karaflos, Dimosthenis and Levinson, Corey J. and Pfeiffer, Pascal and Puk, Kin Ming and Reade, Walter},
  title = {Laboratory Earthquake Forecasting: A Machine Learning Competition},
  journal = {Proceedings of the National Academy of Sciences of the United States of America},
  year = {2021},
  volume = {118},
  number = {5},
  pages = {e2011362118},
  doi = {10.1073/pnas.2011362118}
}

@article{Kubo_review_2024,
  author    = {Kubo, Hisahiko and Naoi, Makoto and Kano, Masayuki},
  title     = {Recent Advances in Earthquake Seismology Using Machine Learning},
  journal   = {Earth, Planets and Space},
  year      = {2024},
  volume    = {76},
  number    = {1},
  pages     = {36},
  doi       = {10.1186/s40623-024-01982-0},
  issn      = {1880-5981},
  publisher = {Springer Nature}
}

@article{Schorlemmer_2010,
  author    = {Schorlemmer, Danijel and Gerstenberger, Matthew C. and Wiemer, Stefan and Jackson, David D. and Rhoades, David A.},
  title     = {Earthquake Likelihood Model Testing},
  journal   = {Seismological Research Letters},
  year      = {2010},
  volume    = {81},
  number    = {5},
  pages     = {718--723},
  doi       = {10.1785/gssrl.81.5.728}
}

@article{Zechar_2010,
  author    = {Zechar, J. Douglas and Schorlemmer, Danijel and Liukis, Maria and Yu, Jiancang and Werner, Max J. and Wiemer, Stefan},
  title     = {The Collaboratory for the Study of Earthquake Predictability Perspective on Computational Earthquake Science},
  journal   = {Contributions to Geophysics and Geodesy},
  year      = {2010},
  volume    = {40},
  number    = {3},
  pages     = {111--130},
  doi       = {10.2478/v10126-010-0007-9}
}

@article{Varotsos_2005,
  author    = {Varotsos, Panayiotis A. and Sarlis, Nicholas V. and Skordas, Efthimios S.},
  title     = {Long-range correlations in the natural time and the entropy of seismic electric signals},
  journal   = {Physical Review E},
  year      = {2005},
  volume    = {71},
  number    = {3},
  pages     = {032102},
  doi       = {10.1103/PhysRevE.71.032102}
}

@book{Varotsos_2011,
  author    = {Varotsos, Panayiotis A. and Sarlis, Nicholas V. and Skordas, Efthimios S.},
  title     = {Natural Time Analysis: The New View of Time---Precursory Seismic Electric Signals, Earthquakes and Other Complex Time Series},
  publisher = {Springer},
  address   = {Berlin, Heidelberg},
  year      = {2011},
  isbn      = {978-3-642-16448-1},
  doi       = {10.1007/978-3-642-16449-8}
}

@article{Varotsos_2013,
  author    = {Varotsos, Panayiotis A. and Sarlis, Nicholas V. and Skordas, Efthimios S. and Lazaridou-Varotsos, Maria S.},
  title     = {Seismic Electric Signals: An Additional Fact Showing Their Physical Interconnection with Seismicity},
  journal   = {Tectonophysics},
  year      = {2013},
  volume    = {589},
  pages     = {116--125},
  doi       = {10.1016/j.tecto.2012.12.020}
}

@article{Varotsos_2011_NATURAL_TIME,
author = {Panayiotis Varotsos  and Nicholas V. Sarlis  and Efthimios S. Skordas  and Seiya Uyeda  and Masashi Kamogawa },
title = {Natural time analysis of critical phenomena},
journal = {Proceedings of the National Academy of Sciences},
volume = {108},
number = {28},
pages = {11361-11364},
year = {2011},
doi = {10.1073/pnas.1108138108},
}

@article{Rundle_2016_NOWCASTING,
  author    = {Rundle, John B. and Turcotte, Donald L. and Donnellan, Andrea and Grant Ludwig, Lisa and Luginbuhl, Molly},
  title     = {Nowcasting Earthquakes},
  journal   = {Earth and Space Science},
  year      = {2016},
  volume    = {3},
  number    = {11},
  pages     = {480--486},
  doi       = {10.1002/2016EA000185}
}

@article{Hargarten_2002_ofc,
  title = {Foreshocks and Aftershocks in the Olami-Feder-Christensen Model},
  author = {Hergarten, Stefan and Neugebauer, Horst J.},
  journal = {Phys. Rev. Lett.},
  volume = {88},
  issue = {23},
  pages = {238501},
  numpages = {4},
  year = {2002},
  month = {May},
  publisher = {American Physical Society},
  doi = {10.1103/PhysRevLett.88.238501}
}

@article{Rundle_2018_NOWCASTING_NATURAL_TIME,
  author    = {Rundle, John B. and Luginbuhl, Molly and Giguere, Alexis and Turcotte, Donald L.},
  title     = {Natural Time, Nowcasting and the Physics of Earthquakes: Estimation of Seismic Risk to Global Megacities},
  journal   = {Pure and Applied Geophysics},
  year      = {2018},
  volume    = {175},
  number    = {2},
  pages     = {597--615},
  doi       = {10.1007/s00024-017-1720-x}
}

@article{Schorlemmer_CSEP_2018,
    author = {Schorlemmer, Danijel and Werner, Maximilian J. and Marzocchi, Warner and Jordan, Thomas H. and Ogata, Yosihiko and Jackson, David D. and Mak, Sum and Rhoades, David A. and Gerstenberger, Matthew C. and Hirata, Naoshi and Liukis, Maria and Maechling, Philip J. and Strader, Anne and Taroni, Matteo and Wiemer, Stefan and Zechar, Jeremy D. and Zhuang, Jiancang},
    title = {The Collaboratory for the Study of Earthquake Predictability: Achievements and Priorities},
    journal = {Seismological Research Letters},
    volume = {89},
    number = {4},
    pages = {1305-1313},
    year = {2018},
    month = {06},
    issn = {0895-0695},
    doi = {10.1785/0220180053},
}

@article{xia_2005_BK,
  title = {Simulation of the Burridge-Knopoff Model of Earthquakes with Variable Range Stress Transfer},
  author = {Xia, Junchao and Gould, Harvey and Klein, W. and Rundle, J. B.},
  journal = {Phys. Rev. Lett.},
  volume = {95},
  issue = {24},
  pages = {248501},
  numpages = {4},
  year = {2005},
  month = {Dec},
  publisher = {American Physical Society},
  doi = {10.1103/PhysRevLett.95.248501}
}

@article{Pacheco_1992_nature,
author = {Pacheco, Javier and Scholz, C. and Sykes, Lynn},
year = {1992},
month = {01},
pages = {71-73},
title = {Changes in frequency-size relationship from small to large earthquake},
volume = {355},
journal = {Nature},
doi = {10.1038/355071a0}
}

@article{Mori_kawamura_2006_bk,
author = {Mori, Takahiro and Kawamura, Hikaru},
title = {Simulation study of the one-dimensional Burridge-Knopoff model of earthquakes},
journal = {Journal of Geophysical Research: Solid Earth},
volume = {111},
number = {B7},
pages = {},
doi = {https://doi.org/10.1029/2005JB003942},
year = {2006}
}

@article{Myers_1993,
  title = {Rupture propagation, dynamical front selection, and the role of small length scales in a model of an earthquake fault},
  author = {Myers, Christopher R. and Langer, J. S.},
  journal = {Phys. Rev. E},
  volume = {47},
  issue = {5},
  pages = {3048--3056},
  numpages = {0},
  year = {1993},
  month = {May},
  publisher = {American Physical Society},
  doi = {10.1103/PhysRevE.47.3048}
}

@article{Myers_1996,
  title = {Slip Complexity in a Crustal-Plane Model of an Earthquake Fault},
  author = {Myers, Christopher R. and Shaw, Bruce E. and Langer, J. S.},
  journal = {Phys. Rev. Lett.},
  volume = {77},
  issue = {5},
  pages = {972--975},
  numpages = {0},
  year = {1996},
  month = {Jul},
  publisher = {American Physical Society},
  doi = {10.1103/PhysRevLett.77.972},
}

@article{Schmittbuhl_1996,
author = {Schmittbuhl, Jean and Vilotte, Jean-Pierre and Roux, Stéphane},
title = {A dissipation-based analysis of an earthquake fault model},
journal = {Journal of Geophysical Research: Solid Earth},
volume = {101},
number = {B12},
pages = {27741-27764},
doi = {https://doi.org/10.1029/96JB02294},
year = {1996}
}

@article{Shaw_1994,
author = {Shaw, Bruce E.},
title = {Complexity in a spatially uniform continuum fault model},
journal = {Geophysical Research Letters},
volume = {21},
number = {18},
pages = {1983-1986},
doi = {https://doi.org/10.1029/94GL01685},
year = {1994}
}

@inbook{Pelletier_2000_BK,
author = {Pelletier, Jon D.},
publisher = {American Geophysical Union (AGU)},
isbn = {9781118668375},
title = {Spring-Block Models of Seismicity: Review and Analysis of a Structurally Heterogeneous Model Coupled to a Viscous Asthenosphere},
booktitle = {Geocomplexity and the Physics of Earthquakes},
chapter = {},
pages = {27-42},
doi = {https://doi.org/10.1029/GM120p0027},
year = {2000},
}

@article{Kawamura_2018_2d_BK,
    author = {Kawamura, Hikaru and Yoshimura, Koji and Kakui, Shingo},
    title = {Nature of the high-speed rupture of the two-dimensional Burridge–Knopoff model of earthquakes},
    journal = {Philosophical Transactions of the Royal Society A: Mathematical, Physical and Engineering Sciences},
    volume = {377},
    number = {2136},
    pages = {20170391},
    year = {2018},
    month = {11},
    issn = {1364-503X},
    doi = {10.1098/rsta.2017.0391},
}

@article{Ueda_2015_BK,
  author  = {Y. Ueda and S. Morimoto and S. Kakui and T. Yamamoto and H. Kawamura},
  title   = {Dynamics of earthquake nucleation process represented by the Burridge--Knopoff model},
  journal = {The European Physical Journal B},
  volume  = {88},
  pages   = {235},
  year    = {2015},
  doi     = {10.1140/epjb/e2015-60499-0}
}

@article{Braun_2018_BK_generalised,
    author = {Braun, Oleg M and Peyrard, Michel},
    title = {Seismic quiescence in a frictional earthquake model},
    journal = {Geophysical Journal International},
    volume = {213},
    number = {1},
    pages = {676-683},
    year = {2018},
    month = {04},
    issn = {0956-540X},
    doi = {10.1093/gji/ggy008},
}

@article{Jagla_2010_BK,
author = {Jagla, E. A. and Kolton, A. B.},
title = {A mechanism for spatial and temporal earthquake clustering},
journal = {Journal of Geophysical Research: Solid Earth},
volume = {115},
number = {B5},
pages = {},
doi = {https://doi.org/10.1029/2009JB006974},
year = {2010}
}

@article{AKISHIN_2000_BK,
title = {Burridge–Knopoff model and self-similarity},
journal = {Chaos, Solitons \& Fractals},
volume = {11},
number = {1},
pages = {207-222},
year = {2000},
issn = {0960-0779},
doi = {https://doi.org/10.1016/S0960-0779(98)00285-9},
author = {P.G. Akishin and M.V. Altaisky and I. Antoniou and A.D. Budnik and V.V. Ivanov},
}

@Article{Hainzl_BK_2000,
AUTHOR = {Hainzl, S. and Z\"oller, G. and Kurths, J.},
TITLE = {Self-organization of spatio-temporal earthquake clusters},
JOURNAL = {Nonlinear Processes in Geophysics},
VOLUME = {7},
YEAR = {2000},
NUMBER = {1/2},
PAGES = {21--29},
DOI = {10.5194/npg-7-21-2000}
}

@article{Carlson_1989_PRL,
  title = {Properties of earthquakes generated by fault dynamics},
  author = {Carlson, J. M. and Langer, J. S.},
  journal = {Phys. Rev. Lett.},
  volume = {62},
  issue = {22},
  pages = {2632--2635},
  numpages = {0},
  year = {1989},
  month = {May},
  publisher = {American Physical Society},
  doi = {10.1103/PhysRevLett.62.2632}
}

@article{Carlson_1989_PRA,
  title = {Mechanical model of an earthquake fault},
  author = {Carlson, J. M. and Langer, J. S.},
  journal = {Phys. Rev. A},
  volume = {40},
  issue = {11},
  pages = {6470--6484},
  numpages = {0},
  year = {1989},
  month = {Dec},
  publisher = {American Physical Society},
  doi = {10.1103/PhysRevA.40.6470}
}

@book{Herrmann_2014,
  title     = {Statistical Models for the Fracture of Disordered Media},
  editor    = {Hans J. Herrmann and St{\'e}phane Roux},
  year      = {2014},
  publisher = {Elsevier},
  address   = {Amsterdam},
  isbn      = {9781483296128},
  doi       = {10.1016/C2009-0-14278-2}
}

@article{joseph_2022,
title = {Variation of Gini and Kolkata indices with saving propensity in the Kinetic Exchange model of wealth distribution: An analytical study},
journal = {Physica A: Statistical Mechanics and its Applications},
volume = {594},
pages = {127051},
year = {2022},
issn = {0378-4371},
doi = {https://doi.org/10.1016/j.physa.2022.127051},
author = {Bijin Joseph and Bikas K. Chakrabarti},
}

@article{ghosh_gkp,
  title = {Relations between the inequality indices Gini, Pietra, and Kolkata: Theory and data analysis},
  author = {Ghosh, Asim and Chakrabarti, Bikas K.},
  journal = {Phys. Rev. E},
  volume = {114},
  issue = {2},
  pages = {024302},
  numpages = {9},
  year = {2026},
  month = {Aug},
  publisher = {American Physical Society},
  doi = {10.1103/wb7q-78gb}}

@article{BTW,
  title = {Self-organized criticality: An explanation of the 1/f noise},
  author = {Bak, Per and Tang, Chao and Wiesenfeld, Kurt},
  journal = {Phys. Rev. Lett.},
  volume = {59},
  issue = {4},
  pages = {381--384},
  numpages = {0},
  year = {1987},
  month = {Jul},
  publisher = {American Physical Society},
  doi = {10.1103/PhysRevLett.59.381},
 
}

@article{Bertalan_2014,
  title = {Fracture Strength: Stress Concentration, Extreme Value Statistics, and the Fate of the Weibull Distribution},
  author = {Bertalan, Zsolt and Shekhawat, Ashivni and Sethna, James P. and Zapperi, Stefano},
  journal = {Phys. Rev. Appl.},
  volume = {2},
  issue = {3},
  pages = {034008},
  numpages = {8},
  year = {2014},
  month = {Sep},
  publisher = {American Physical Society},
  doi = {10.1103/PhysRevApplied.2.034008}
}

@article{Bhattacharya_2011,
  author  = {Bhattacharya, Pathikrit and Chakrabarti, Bikas K. and Kamal},
  title   = {A Fractal Model of Earthquake Occurrence: Theory, Simulations and Comparisons with the Aftershock Data},
  journal = {Journal of Physics: Conference Series},
  year    = {2011},
  volume  = {319},
  number  = {1},
  pages   = {012004},
  doi     = {10.1088/1742-6596/319/1/012004}
}

@article{Helmstetter_ofc_2004,
  author  = {Helmstetter, Agn{\`e}s and Hergarten, Stefan and Sornette, Didier},
  title   = {Properties of foreshocks and aftershocks of the nonconservative self-organized critical Olami-Feder-Christensen model},
  journal = {Physical Review E},
  volume  = {70},
  number  = {4},
  pages   = {046120},
  year    = {2004},
  month   = {October},
  doi     = {10.1103/PhysRevE.70.046120}
}

@article{Kotani_ofc_2008,
  title = {Periodicity and criticality in the Olami-Feder-Christensen model of earthquakes},
  author = {Kotani, Takeshi and Yoshino, Hajime and Kawamura, Hikaru},
  journal = {Phys. Rev. E},
  volume = {77},
  issue = {1},
  pages = {010102(R)},
  numpages = {4},
  year = {2008},
  month = {Jan},
  publisher = {American Physical Society},
  doi = {10.1103/PhysRevE.77.010102}
}

\end{document}